\documentclass[12pt,a4paper]{article}

\RequirePackage{graphicx}
\RequirePackage{mathptmx}      
\RequirePackage[numbers,sort&compress]{natbib}
\RequirePackage[colorlinks,citecolor=blue,urlcolor=blue,linkcolor=blue]{hyperref}
\RequirePackage{amsmath,amssymb,color}
\RequirePackage{bm}
\RequirePackage{bbm}
\RequirePackage{youngtab}
\RequirePackage{caption}
\definecolor{dark-green}{rgb}{0,0.7,0}
\definecolor{dark-blue}{rgb}{0,0.2,0.5}
\definecolor{med-blue}{rgb}{0,0.7,1}
\definecolor{mblue}{rgb}{0,0.2,1}
\definecolor{cnc}{rgb}{0.8,0,0}
\definecolor{light-red}{rgb}{1,0.8,0.8}
\definecolor{dark-yellow}{rgb}{1,0.8,0}
\definecolor{light-blue}{rgb}{0.8,0.9,1}
\definecolor{verylight-blue}{rgb}{0.93,0.95,1}
\definecolor{light-yellow}{rgb}{1,0.9,0.8}
\definecolor{grey}{gray}{0.88}

\def\a{\alpha}
\def\b{\beta}
\def\c{\gamma}
\def\r{\rho}
\def\d{\delta}
\def\g{\gamma}
 \def\m{\mu}
\def\n{\nu}
\def\eps{\epsilon}
\def\ve{\varepsilon}
\def\vt{\vartheta}

\def\stareq{\stackrel{*}{=}}

\def\ANT{\nearrow\hspace{-10.5pt}A}

\def\vt{\vartheta}

\def\s{\sigma}

\newcommand{\tvect}[2]{
  \ensuremath{\Bigl(\negthinspace\begin{smallmatrix}#1\\#2\end{smallmatrix}
    \negthinspace\Bigr)}}

\begin{document}

\title{Premetric teleparallel theory of gravity and\\ its local and linear
  constitutive law} 

\author{Yakov Itin$^{1,\star}$, Yuri N. Obukhov$^{2,\diamond}$,
  Jens Boos$^{3,\ast}$, Friedrich W.\ Hehl$^{4,\dagger}$\\
  \small $^1$The Hebrew University and Jerusalem College of Technology, Jerusalem 91160, Israel\\
  \small $^2$Institute for Nuclear Safety, Russian Academy of Sciences, 115191 Moscow, Russia\\
  \small $^3$Theor. Phys. Inst., Univ. Alberta, Edmonton, AB T6G 2E1, Canada\\
  \small $^4$Inst.\ Theor.\ Physics, Univ.\ of Cologne, 50923 K\"oln, Germany\\
  \small $^{\star}$itin@math.huji.ac.il\qquad $^{\diamond}$obukhov@ibrae.ac.ru\\
  \small $^{\ast}$boos@ualberta.ca  \qquad $^{\dagger}$hehl@thp.uni-koeln.de}

\date{\begin{footnotesize}{\it file teleGRchi43arXiv.tex, 06 Nov 2018}
  \end{footnotesize}}
\maketitle

\begin{abstract}
  We continue to investigate the premetric teleparallel theory of
  gravity (TG) with the coframe (tetrad) as gravitational
  potential. We start from the field equations and a local and linear
  constitutive law. We create a Tonti diagram of TG in order to
  disclose the structure of TG.  Subsequently we irreducibly decompose
  the 6th order constitutive tensor under the linear group.  Moreover,
  we construct the most general constitutive tensors from the metric
  and the totally antisymmetric Levi-Civita symbol, and we demonstrate
  that they encompass nontrivial axion and skewon type pieces. Using
  these tools, we derive for TG in the geometric-optics approximation
  propagating massless spin 0, 1, and 2 waves, including the special
  case of Einstein's general relativity.
\end{abstract}
%

{\hypersetup{linkcolor=black}
\tableofcontents}
\section{Introduction}\label{Sec.1}

Lately we followed in \cite{Hehl:2016glb} the program of Kottler of
1922 {\it to remove the metric tensor of spacetime,} the gravitational
potential within general relativity theory (GR), from the fundamental
laws of classical electromagnetism and gravity as far as possible. In
particular, we applied this to the theory of gravity
\cite{Itin:2016nxk} in that we started from a translational gauge
theory of gravity, also known as teleparallel theory of gravity
(TG)\footnote{For the application of teleparallelism in continuum
  mechanics, see Delphenich \cite{Delphenich:2013}.}. We assume that
our readers are familiar with \cite{Itin:2016nxk}. The TG approach is
reviewed in Blagojevi\'c et al.\ \cite{Blagojevic:2013xpa}, see also
Maluf \cite{Maluf:2013gaa}, and Aldrovandi \& Pereira \cite{Aldrovandi:2013}.
For a somewhat related tetrad approach to gravity, in which affine symmetry is
exploited, one should compare S{\l}awianowski et al.\ \cite{Slawianowski:2018}.

The spacetime geometry of TG is represented by a four-dimensional (4d)
manifold equipped with a coframe 1-form $\vt^\a=e_i{}^\a \,dx^i$ and
with a linear connection 1-form
$\Gamma_\a{}^\b= \Gamma_{i\hspace{1pt}\a}{}^\b\,dx^i$, the curvature
2-form of which, $R_\a{}^\b=\frac 12 R_{ij\a}{}^\b dx^i \wedge dx^j$,
vanishes globally: $R_\a{}^\b\equiv 0$. In the notation we follow
basically the book \cite{Birkbook}: $\a$ and $\b$ are co- and
contravariant frame indices, with
$\a,\b,... =\hat{0},\hat{1},\hat{2},\hat{3}$, whereas
$i,j,... =0,1,2,3$ are coordinate indices.

Because of the vanishing curvature, we can pick suitable frames such
that $\Gamma_\a{}^\b$ vanishes globally:
\begin{equation}\label{telegauge}
\Gamma_\a{}^\b\stareq 0\qquad\text{(everywhere in spacetime)}\,.
\end{equation}
{{In this `teleparallel gauge,' the covariant  exterior  derivative
taken with respect to the  connection $\Gamma_\a{}^\b$ is reduced
to the ordinary  exterior derivative.}} This will simplify our
formalism. However,
we will drop the star $^\ast$ over the equality sign in future since
\eqref{telegauge} is assumed to be valid throughout our paper.

The essence of the premetric approach can be formulated as follows.
This universal field-theoretic scheme is bas\-ed on conservation laws
which hold true for the two types of variables: extensive fields
(``how much?'') and intensive (``how strong?'') ones. These variables
satisfy the {\it fundamental equ\-aions} which are metric-free,
whereas the metric comes in only via the {\it linking equations} which
establish consitutive relations between the extensive and intensive
variables.

Premetric electrodynamics \cite{Birkbook} is based on the conservation
laws of electric charge and magnetic flux which give rise to the
fundamental equations $dH = J$ and $dF = 0$. Here $H$ is the
electromagnetic excitation 2-form (extensive variable) and $F$ the
electromagnetic field strength 2-form (intensive variable).  By
introducing the constitutive relation $H = \kappa[F]$, one obtains a
predictive physical theory.

The premetric gravity framework \cite{Itin:2016nxk} can be constructed
along the same lines by replacing the electric charge with a
``gravitational charge'' $\rightarrow$ mass $\rightarrow$
energy-momentum. In this introductory section, we provide a short
overview of the premetric teleparallel approach.

\subsection{Field equations}

The field equations of TG will be our starting point. The
inhomogeneous gravitational field equation of TG reads
\cite[Table I]{Itin:2016nxk}
\begin{equation}\label{gr-kot2}
  \boxed{dH_\a
  ={}^{(\vartheta)}\!\Sigma_\a+{}^{\text{(m)}}\!\Sigma_\a}\quad
\qquad(4\times 6\text{ components})\,.
\end{equation}
It relates the twisted gravitational excitation 2-form
\begin{equation}\label{gr-kot2a}
{H}_\a=\frac 12 H_{ij\a}dx^i\wedge dx^j=\frac
12{H}_{\b\g}{}_\a\,\vt^{\b\g} =\frac 12
\check{H}^{\b\g}{}_\a\,\epsilon_{\b\g}
\end{equation}
to its source, the sum of the twisted energy-momentum 3-forms of
gravity $^{(\vartheta)}\!\Sigma_\a$ and of matter $^{\text{(m)}}\!\Sigma_\a$,
respectively. Here $\vt^{\b\g}:=\vt^\b\wedge \vt^\g$ and
$\epsilon_{\b\g}:=\frac{1}{2} \epsilon_{\b\g\d\ve}\,\vt^\d\wedge
\vt^\ve$, see \cite[p.39]{Birkbook}.

As was already pointed out in \cite[p.52]{Hehl:1979}, ``...for
consistency, we cannot allow spinning matter (other than as test
particles) in such a T$_4$...'' that is, in a 4d Weitzenb\"ock
spacetime. In other words, the field equation \eqref{gr-kot2} is valid
only for hydrodynamic and for electromagnetic matter. A careful proof
of this stipulation was given by Obukhov \& Pereira\footnote{Within
  the framework of TG as a translational gauge theory, the coupling to
  matter is achieved via the minimal coupling procedure, strictly in
  the sense of a bona fide gauge theory. Dispensing with the minimal
  coupling principle (as Maluf \cite{Maluf:2003fs} does, e.g.) is
  against the spirit of gauge theory.}  \cite{Obukhov:2004hv}. For
matter {\it with spin,} the Dirac field, for example, the Lorentz
group should also be gauged, which removes the teleparallelism
constraint $R_\a{}^\b=0$.  Then one arrives at a Poincar\'e gauge
theory operating in a Riemann-Cartan spacetime with torsion and with
Cartan curvature $R_\a{}^\b\ne 0$.

The homogeneous field equation of TG reads
 \begin{equation}\label{gr-kot3}
 \boxed{d{ F}^\a=0}\quad\qquad(4\times 6\text{ components})\,,
\end{equation}
with the untwisted vector-valued 2-form $F^\a:=d\vt^\a$, the torsion of
spacetime, which has the expansion
 \begin{equation}\label{gr-kot5}
F^\a=\frac 12 F_{ij}{}^\a dx^i\wedge dx^j=\frac 12
F_{\b\g}{}^\a\vartheta^{\b\g}\,.
\end{equation}
Note that (\ref{gr-kot3}) is the first Bianchi identity of a linearly
connected spacetime with vanishing curvature, see
\cite[Eq.(5.41)]{Schouten:1954} or \cite[Eq.(C.1.67)]{Birkbook}.

The analogy to the Maxwell equations of electrodynamics should be
apparent, see \cite{Birkbook}. Eq.(\ref{gr-kot2}) represents four
inhomogeneous Maxwell type equations and Eq.(\ref{gr-kot3}) four
homogeneous Maxwell type equations. Since TG is a translational gauge
theory, it has the four one-form potentials $\vt^\a=e_i{}^\a\,dx^i$,
where four is the number of generators of the translation group.

\subsection{Local and linear constitutive law}

In order to complete the two field equations of TG to a predictive
system of equations, one has to adopt a constitutive law between
excitation $H$ and field strength $F$. The simplest assumption is that
the functional $H=\kappa[F]$ is local and linear,
\begin{equation}\label{const1}
  \boxed{H_\a={\boldsymbol
      \kappa}_{\a\b}\boldsymbol{[}F^\b\boldsymbol{]}\,.}
\end{equation}
To deduce the corresponding component representation, we
remember that the functional ${\boldsymbol\kappa}_{\a\b}$ acts on
2-forms and creates as response other 2-forms. Since any 2-form can be
decomposed with respect to the 2-form basis $\vt^{\nu\rho}$, it is
sufficient to study the behavior of $\vt^{\nu\rho}$ under the
application of ${\boldsymbol\kappa}_{\a\b}$. Because of the assumed
locality and linearity, we have
\begin{equation}\label{kappa_dec}
  \boldsymbol{\kappa}_{\a\b}\boldsymbol{[}\vt^{\nu\rho}
\boldsymbol{]}=
  \frac 12\,\kappa_{\lambda\mu\a}{}^{\nu\rho}{}\!_{\b}\,\vt^{\lambda\mu}\,.
\end{equation}
Let us come back to \eqref{const1}. We decompose $H_\a$ and $F^\b$
and, by using \eqref{kappa_dec}, we find:
\begin{eqnarray}\label{const1a}
  H_\a&=&\frac 12
H_{\lambda\mu\a}\vt^{\lambda\mu}=\boldsymbol{\kappa}_{\a\b}\boldsymbol{[}
          F^\b\boldsymbol{]}=\boldsymbol{\kappa}_{\a\b}\boldsymbol{\left[}\frac
          12F_{\nu\rho}{}^{\b}\vt^{\nu\rho} \boldsymbol{\right]}\cr
  &=&\frac 12F_{\nu\rho}{}^{\b}\boldsymbol{\kappa}_{\a\b}\boldsymbol{[}
      \vt^{\nu\rho} \boldsymbol{]}=\frac
      14\kappa_{\lambda\mu\a}{}^{\nu\rho}{}\!_{\b}
      \, F_{\nu\rho}{}^\b\vt^{\lambda\mu}\,.
\end{eqnarray}
By renaming some indices, we eventually {{derive}} the final formula
\begin{equation}\label{const2}
  H_{\b\g\a}=\frac{1}{2}\,\kappa_{\b\g\a}{}^{\nu\rho}{}\,_{\mu}
  F_{\nu\rho}{}^{\mu}\,.
\end{equation}
see \cite[Eq.(47)]{Itin:2016nxk}.

If we use Schouten's \cite{Schouten:1954} notation
$(\a\b):= \frac 12 \{\a\b+\b\a\}$ and, moreover,
$[\a\b]:= \frac 12 \{\a\b-\b\a\}$, we have here the antisymmetries
$H_{(\b\g)\a}=0$ and $F_{(\nu\rho)}{}^\mu =0$. Thus, the constitutive
tensor obeys the identities
\begin{equation}\label{const_symm}
  \kappa_{(\b\g)\a}{}^{\nu\rho}{}_{\mu}=0 \quad\text{and}\quad
 \kappa_{\b\g\a}{}^{(\nu\rho)}{}_{\mu}=0\,.
\end{equation}
Accordingly, $\kappa_{\b\g\a}{} ^{\nu\rho}{}_{\mu}$ has
$(6\times 4)^2=576$ independent components. In the corresponding
electrodynamics case, the constitutive law reads
$H_{\b\g}=\frac 12\kappa_{\b\g}{}^{\nu\rho}F_{\nu\rho}$. Thus, by
contrast, we have only $6^2=36$ independent components. As we will see
further {{down}}, if we study only reversible processes, then this number
is appreciably downsized in both cases.

A concise Hamiltonian formulation of teleparallel gravity was
given by Ferraro and Guzm\'an \cite{Ferraro:2016wht}. Recently Hoh\-mann
et al. \cite{Hohmann:2017duq} studied teleparallel gravity, but
instead of taking local and linear constitutive equations, they turned
to local and {\it non}linear ones in order to incorporate
$f(T)$-theories into the general TG formalism. These investigations
\cite{Hohmann:2017duq} are very helpful since they bring order into
the plethora of $f(T)$-theories and make them more transparent. Such
nonlinear models provide an interesting development of gravitational
theory based on an analogy with Born-Infeld-Pleba\'nski
electrodynamics. We believe, though, that there is, at the present
time, no real need to push nonlinear constitutive laws, since gravity
is nonlinear anyway, due to its self-interaction---and this in spite
of a {\it linear} constitutive law, which guarantees the
quasi-linearity of the emerging field equation.

Kosteleck\'y \& Mewes \cite{Kostelecky:2017zob} investigated Lorentz
and diffeomorphism violations in {\it linearized} gravity. In this
context, they introduced tensors which are of a similar type as our
constitutive tensor $\chi$. Our group-theoretical treatment of $\chi$
in Sec.3 is reminiscent of their method. However, in our article the
full nonlinearity of gravity is treated in a premetric framework.

We would like to stress the following fact about teleparallelism
theories: We have always a frame $e_\a=e^i{}_\a\partial_i$ and a
coframe $\vt^\b=e_j{}^\b dx^j$ with us, that is, we can always {{change}}
from holonomic to anholonomic indices by using $e^i{}_\a$ and
$e_j{}^\b$, respectively---and vice versa. This implies that all the
indices in premetric TG are fundamentally equal, in particular, all
those {{occurring}} in $\kappa_{\b\g\a}{} ^{\nu\rho}{}_{\mu}$ in
(\ref{const2}). The `group' indices $\a$ and $\mu$ are of the same
quality as the `form' indices $\b,\g,\nu,\rho$. Contractions with all
indices are always allowed. Of course, at the premetric stage, raising
and lowering of indices is only possible with the totally
antisymmetric Levi-Civita symbols $\eps_{\a\b\g\d}=\pm 1,0$
and $\eps^{\mu\nu\rho\sigma}=\mp 1,0$, since no metric is available so
far in our premetric framework.

If we follow the pattern of electrodynamics as formulated in {\it
  tensor calculus,} see Post \cite{Post:1962}, then it is obvious that
we should introduce the excitation with the components
$\check{H}^{\b\g}{}_\a=\frac 12 \epsilon^{\b\g\mu\nu}H_{\mu\nu\a}$, as
already defined in Eq.(\ref{gr-kot2a}). Thus,
\begin{equation}\label{gr-kot9}
  \check{H}^{\b\g}{}_{\a}=\frac 12 \chi^{\b\g}{}_{\a}{}^{\nu\rho}{}_\mu
  \,F_{\nu\rho}{}^\mu\,,
\end{equation}
with
\begin{equation}\label{gr-kot9a}
  \chi^{\b\g}{}_{\a}{}^{\nu\rho}{}_\mu:=\frac 12
\epsilon^{\b\g\d\varepsilon}\,\kappa_{\d\varepsilon\a}{}^{\nu\rho}{}_{\mu}\,.
\end{equation}
The constitutive tensor density\footnote{This tensor density emerged
  already in \cite{Hehl:1980}, see also
  \cite{Aldrovandi:2013,Boehmer:2014jsa}.}
$\chi^{\b\g}{}_{\a}{}^{\nu\rho}{}_\mu$ is equivalent to
$ \kappa_{\b\g\a}{} ^{\nu\rho}{}_{\mu}$, in particular,
 \begin{equation}\label{gr-kot10}
\chi^{(\b\g)}{}_{\a}{}^{\nu\rho}{}_\mu  =0 \quad\text{and}\quad
\chi^{\b\g}{}_{\a}{}^{(\rho\nu)}{}_\mu =0\,.
\end{equation}

But there is still a third useful version of the constitutive tensor
available. For the purpose of the irreducible decomposition under the
$\text{GL}(4,\mathbb{R})$, it is optimal to have the indices of the
constitutive tensor all exclusively either in lower or in upper
position. Since so far we have no metric at our disposal, we can only
move the antisymmetric sets of indices.  If we have a look at
(\ref{const2}), we can rewrite it with
$\check{F}^{\nu\rho\mu}=\frac 12\epsilon^{\nu\rho\nu'\!\rho'\!}
F_{\nu'\!\rho'} {}^\mu$ as
\begin{equation}\label{const3}
 H_{\b\g\a}=\frac 12 \boldsymbol
  {\check}{\chi}_{\b\g\a\nu\rho\mu}\,\boldsymbol{\check}{F}^{\nu\rho\mu}\,,
\end{equation}
that is,
\begin{equation}\label{chi_check}
 \boldsymbol
  {\check}{\chi}_{\b\g\a\nu\rho\mu}:=\frac 12
\epsilon_{\nu\rho\nu'\!\rho'}\,
  \kappa_{\b\g\a}{}^{\nu'\!\rho'\!}{}_{\mu}=\frac 14
\epsilon_{\b\g\b'\!\g'\!}\,
  \epsilon_{\nu\rho\nu'\!\rho'\!}\,
  \chi^{\b'\!\g'\!}{}_{\a}{}^{\nu'\!\rho'\!}{}_\mu\,.
\end{equation}
We have the symmetries
\begin{equation}\label{const3a}
  \boldsymbol {\check}{\chi}_{(\b\g)\a\nu\rho\mu}=0\quad\text{and}\quad
  \boldsymbol {\check}{\chi}_{\b\g\a(\nu\rho)\mu}=0\,.
\end{equation}

\subsection{Reversibility}
\begin{footnotesize}
   {\it A process which can in no way be completely reversed is termed
 {{ {\emph {irreversible,}}}} all other processes {{{\emph
{reversible.}}}} That a
    process may be irreversible, it is not sufficient that it cannot
    be directly reversed. This is the case with many mechanical
    processes which are not irreversible... ...The full requirement is,
    that it be impossible, even with the assistance of all agents in
    nature, to restore everywhere the exact initial state when the
    process has once taken place... the generation of heat by
    friction, the expansion of a gas without the performance of
    external work and the absorption of external heat, the conduction
    of heat, etc., are irreversible processes...}
 \hspace{62pt} Max Planck \cite[p.84]{Planck}\end{footnotesize}\bigskip
\footnote{Similarly, we have: {\it Suppose that when a system under
    consideration changes from a state, $\a$, to another state, $\a'$,
    the environment changes from $\b$ to $\b'$. If in some way it is
    possible to return the system from $\a'$ to $\a$ and at the same
    time to return the environment from $\b'$ to $\b$, the process
    $(\a,\b)\rightarrow(\a',\b')$ is said to be reversible.}
 \hfill Ryogo~Kubo \cite[p.61]{Kubo}}

  The merit of formulating a field theory only in terms of its field
  equations---here Eqs.\eqref{gr-kot2} and \eqref{gr-kot3}---and an
  associated constitutive law---here Eq.\eqref{const1}---is that it
  covers both processes, irreversible and reversible ones.  At first,
  like in the conventional treatment of general relativity
  \cite{Meaning}, we turn our attention to reversible processes. From
  their definition it is clear that, for instance, periodic processes
  and those the equations of motion of which are formulated in a time
  symmetric way, are reversible. Time symmetry means that we can
  substitute in the field equations $t$ by $-t$ without changing them.

  Thus dissipation is not allowed in reversible processes and we can
  define for each such process an energy function and, by a Legendre
  transformation, a Lagrangian. Accordingly, reversible processes can
  always be formulated by means of an action principle. For TG, the
  twisted Lagrange 4-form reads
\begin{equation}\label{Lagr}
  ^{(\vt)}\!\!\Lambda =-\frac 12 F^\a\wedge H_\a=-\frac 12
 \, F^\a \wedge
\boldsymbol{\kappa}_{\a\b}\boldsymbol{[}F^\b\boldsymbol{]}\,.
\end{equation}
We substitute \eqref{gr-kot5} and \eqref{const1a} into
  \eqref{Lagr}
  and find:
\begin{eqnarray}\label{Lagr0}
  ^{(\vt)}\!\!\Lambda
  &=&-\frac 12\left(\frac 12  F_{\eta\zeta}{}^\a\vt^{\eta\zeta}
      \right)\wedge\left(\frac 14
\kappa_{\lambda\mu\a}{}^{\nu\rho}{}_{\b}F_{\nu\rho}{}^\b\vt^{\lambda\mu}
\right)\nonumber\\
  &=&-\frac{1}{16}\,\kappa_{\lambda\mu\a}{}^{\nu\rho}{}\!_\b
F_{\eta\zeta}{}^\a
      F_{\nu\rho}{}^\b\,\vt^{\eta\zeta\lambda\mu}\,.
\end{eqnarray}
With the definition \eqref{gr-kot9a} and the volume 4-form
vol, Eq.(\ref{Lagr0}) can be rewritten as
\begin{equation}\label{Lagr1}
 ^{(\vt)}\!\! \Lambda =-\frac 18 \chi^{\b\g}{}_{\alpha}{}^{\nu\rho}{}_{\mu}\,
  F_{\b\g}{}^{\a} F_{\nu\rho}{}^{\mu}\,\text{vol}
\end{equation}
 or, by renaming the indices of the $F$'s, equivalently as
\begin{equation}\label{Lagr1a}
 ^{(\vt)}\!\! \Lambda =-\frac 18 \chi^{\nu\rho}{}_{\mu}{}^{\b\g}{}_{\alpha}\,
  F_{\b\g}{}^{\a} F_{\nu\rho}{}^{\mu}\,\text{vol}\,.
\end{equation}
Consequently, only those components of the constitutive tensor enter
the Lagrangian that satisfy the relations
\begin{equation}\label{rev-cond}
\chi^{\b\g}{}_{\alpha}{}^{\nu\rho}{}_{\mu}=
\chi^{\nu\rho}{}_{\mu}{}^{\b\g}{}_{\alpha}\,.
\end{equation}
If we assume that the model is completely specified by the Lagrangian
(\ref{Lagr}), we restrict our considerations to reversible
processes. Then the relations \eqref{rev-cond} are necessary
conditions, which correspond to the symmetry of a $24\times 24$
matrix. Thus, the set of the 576 independent components of
$\chi^{\b\g}{}_{\alpha}{}^{\nu\rho}{}_{\mu}$ reduces to only 300 ones.

It is possible to rewrite the constitutive law in a more {{ compact}}
way. We introduce the 6-dimensional co-basis
$\vt^{\a\b}$ $\rightarrow\,\vt^I$ in the space of 2-forms, with the
collective indices $I,J,\dots=(01,02,03;23,31,12)=(1,2,\dots,6)$, see
\cite[p.40]{Birkbook}. Then excitation and field strength decompose as
$ H_\a=H_{I\a}\vt^I$ and $F^\b =F_{J}{}^{\b} \vt^J$, respectively, and the
constitutive law reads,
\begin{equation}\label{6dDecomp1}
  H_{I\a}=\kappa_{I\a}{}^{J}{}_{\!\b}F_{J}{}^{\b}\,,\quad
  \check{H}^{I}{}_{\a}=\frac
  12\chi^{I}{}_{\a}{}^{J}{}_{\!\b}\,F_{J}{}^\b\,,
\end{equation}
with the 300 independent components 
\begin{equation}\label{6dDecomp1a}
  {{ \chi^{I}{}_{\a}{}^{J}{}_{\!\b}=\chi^{J}{}_{\b}{}^{I}{}_{\!\a}}}\,.
\end{equation}
Thus, for the Lagrangian we find
\begin{equation}
  ^{(\vt)}\!\!  \Lambda =-\frac 12 \chi^{I}{}_{\a}{}^{J}{}_{\b}\,
  F_{I}{}^{\a}F_{J}{}^{\b}\text{vol}\,.
\end{equation}

This closes our short introduction to TG. Like all classical theories,
whose form is established, we can put TG into a Tonti type diagram
\cite{Tonti:2013} in order to clearly display its structure. This will
be done for the first time in Sec.\ref{Sec.2}. Then we turn to a
closer examination of the constitutive tensor of TG. In
Sec.\ref{Sec.3}, we decompose it into smaller pieces, in particular
into the irreducible pieces with respect to the linear group
$\text{GL}(4,\mathbb{R})$. In Sec.\ref{Sec.4}, metric dependent
constitutive tensors will be addressed, in particular those which
relate TG to general relativity.  In Sec.\ref{Sec.5}, we will study
the propagation of gravitational waves in TG within the geometric
optics approximation. We will follow the procedure that we developed
for electrodynamics \cite{Birkbook,Itin:2009aa,Baekler:2014kha}. Since
in TG we have four generators of the gauge group, things become a bit
more complicated than in electrodynamics. In Sec.\ref{Sec.6}, we will
specialize these considerations on gravitational waves to metric
dependent models.

\section{Tonti diagram of the premetric teleparallel
  theory of gravity}\label{Sec.2}

\begin{figure}[!htb]
\centering
\includegraphics[width=\textwidth]{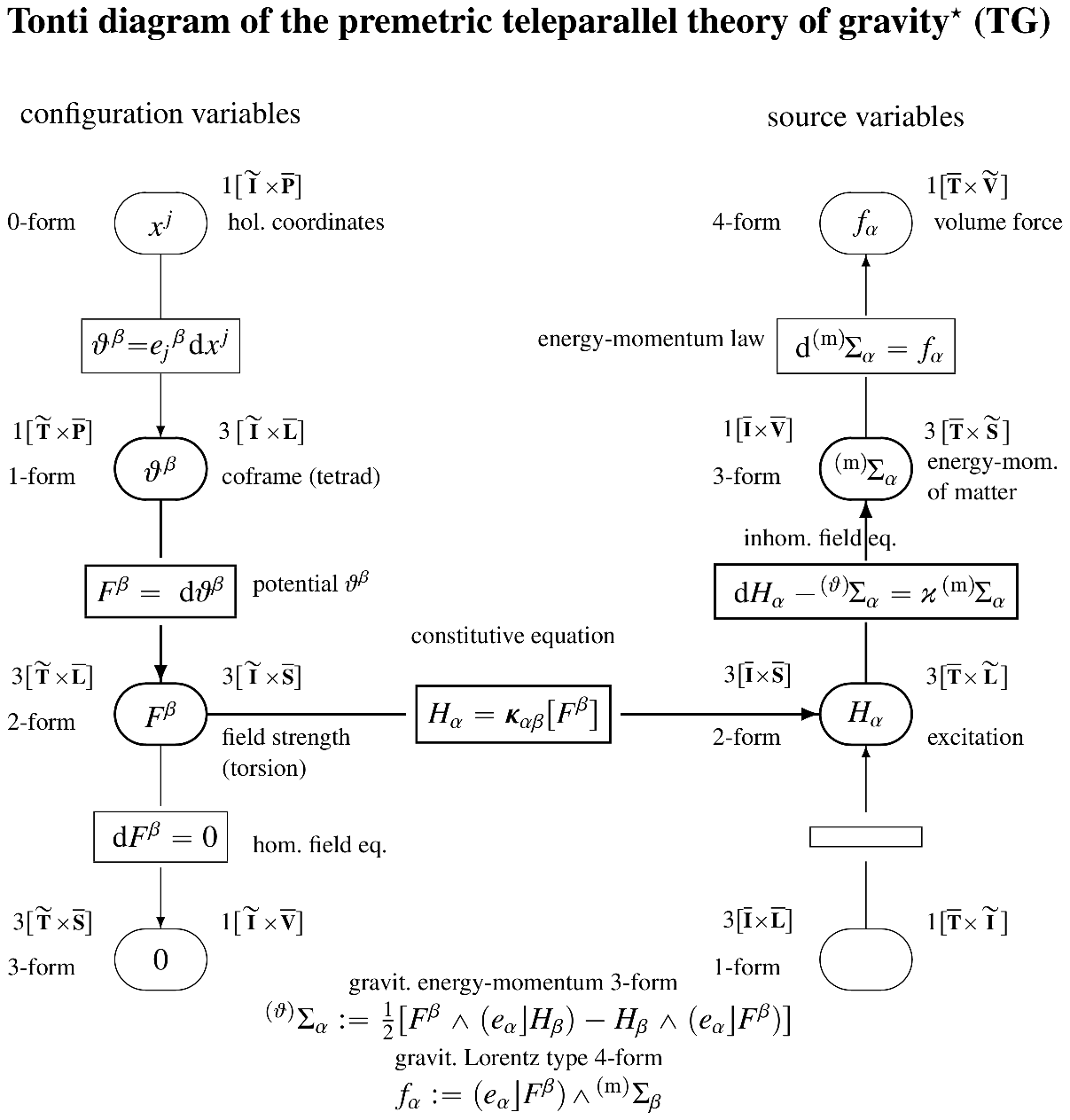}
\caption{$\bullet$ Patterned after E.~Tonti: {\it The Mathematical Structure of
  Classical and Relativistic Physics,}  A general classification
  diagram (Birkh\"auser-Springer, New York, 2013) pages 402 and 315.\\
$\bullet$ Notation: Y.~Itin, F.~W.~Hehl, Yu.~N.~Obukhov, Phys. Rev.
D {\bf 95}, 084020 (2017), arXiv:1611.05759. We denoted {\it here} the
torsion 2-form with $F^\beta$ in order to underline its function as a
field strength. Usually, however, we use for torsion $T^\beta$.\\
$\bullet$ We chose everywhere the `teleparallel gauge' such that the
connection 1-form vanishes globally:
$\Gamma^{\alpha\beta}(x)\stackrel{*}{=}0$.\\$\bullet$ For the reversible case, we
have the gravitational Lagrangian as
$^{(\vartheta)}\!\Lambda = -\,\frac{1}{2} F^{\alpha} \wedge H_{\alpha}$
and $\Lambda={}^{(\vartheta)}\!\Lambda+{}^{({\rm m})}\!\Lambda$. The
gravitational constant is denoted by $\varkappa$.\\
\noindent $^\star$Also known as translation gauge theory of gravity.}
\label{Fig1}
\end{figure}

Over the past decades, Tonti \cite{Tonti:2013} developed a general
classification program for classical and relativistic theories in
physics, such as, e.g., for particle dynamics, electromagnetism, the
mechanics of deformable media, fluid mechanics, thermodynamics, and
gravitation. Here we will display in Figure \ref{Fig1} for the first
time an appropriate and consistent diagram of the teleparallel theory
of gravity (TG).

If a theory is well-understood, its configuration and its source
variables can be clearly identified and their interrelationships
displayed in the form of a Tonti diagram. Such a diagram defines what
one may call the skeleton of a theory. In Tonti's book
\cite{Tonti:2013}, for all classical theories, including the
relativistic ones, a corresponding framework was
established---and this step by step, based on an operational
definition of the quantities involved.

Tonti \cite[p.402]{Tonti:2013} has also displayed a diagram for
relativistic gravitation. In Tonti's own words, it was an ``attempt''
of a diagram based on an ansatz for a tetrad theory of gravity by
Kreisel \& Treder, see \cite[pp.60--67, pp.71--91]{Treder}. Due to our
enhanced knowledge of TG, see \cite[Chapters 5 and
6]{Blagojevic:2013xpa}, we can now improve on Tonti's attempt, see
\cite{DPG:2017} and our Figure \ref{Fig1}. The notation in Figure
\ref{Fig1} is based on our recent paper \cite{Itin:2016nxk} and on the
present one.

Let us have a look at our new diagram. The configuration variables of
TG are the coordinates $x^i$ of the 4-dimensional spacetime (four
0-forms) and the coframes $\vt^\a$ (four 1-forms). By
differentiation, we find the torsion $F^\b$ (four 2-forms) and by
further differentiation the homogeneous field equation of gravity
$dF^\b=0$ (four 3-forms), the right hand side of which vanishes.

The round boxes on the left column depict geometrical objects and the
square boxes interrelate these geometrical objects. The formula
$dF^\b=0$, for example, corresponds to the first Bianchi identity of a
teleparallel spacetime. In a Riemann-Cartan space, we have
$DF^\b=\vt^\a\wedge R_\a{}^\b$. In teleparallelism, the curvature
vanishes, $R_\a{}^\b=0$, and, in the teleparallel gauge, $dF^\b=0$.

For global variables, Tonti distinguishes spatial domains, such as
{\it volumes} $\bf V$, {\it surfaces} $\bf S$, {\it lines} $\bf L$,
and {\it points} $\bf P$, with respect to time he introduces {\it
  instants} $\bf I$ and {\it intervals} $\bf T$. Furthermore, for the
respective domains, he has {\it inner} (interior)
$^{\overline{\hspace{5pt}}}$ and outer (exterior)
$\widetilde{\hspace{5pt}}$ orientation.

Hence $[\widetilde{\bf I}\times\overline{\bf S}]$, for example, refers
to a time domain $\widetilde{\bf I}$ with exterior orientation and a
space domain $\overline{\bf S}$ with interior orientation. And this
domain $[\widetilde{\bf I}\times\overline{\bf S}]$ supports the
torsion two-form. The holonomic coordinates $x^j$, to take another
example, depend on an instant of time $\bf I$ and a spatial point
$\bf P$.  i.e.  $[{\bf I}\times{\bf P}]$.  Then we have to add the
orientation. All exterior forms on the left columns are forms without
twist, see \cite{Birkbook}, those on the right columns all carry
twist. This can be read off from the corresponding orientations. This
comes about as follows:

For configuration variables the associated space elements are endowed
always with an interior orientation, whereas in the case of source
variables it is the exterior orientation which plays a role. This
behavior is found by phenomenologically examining the different
theories. According to Tonti, the underlying theoretical reason for
this correspondence is not clear, but phenomenology does not allow any
other attributions. The configuration variables are related to the
theory of chains of algebraic topology, whereas the source variables
are associated to co-chains. For more details, we refer to the exhaustive
monograph of Tonti \cite{Tonti:2013}.

Since the gravitational field itself carries energy-momen\-tum, it is
also the source of a new gravitational field, which likewise carries
energy-momentum, etc.. Thus, like general relativity, TG is an
intrinsically nonlinear theory, even it carries a linear constitutive
law. Within the field equation of TG,
$d H_{\alpha}-{}^{(\vartheta)\!}  \Sigma_{\alpha}={}^{({\rm m})\!}
\Sigma_{\alpha}\,,$ the gravitational energy-momen\-tum 3-form
\begin{equation}\label{gravcharge}
  ^{(\vartheta)}\Sigma_\alpha:=\frac 12[F^\beta\wedge(e_\alpha\rfloor
  H_\beta)-H_\beta\wedge(e_\alpha\rfloor F^\beta)]
\end{equation}
shows up explicitly and is a manifestation of this nonlinearity. By
differentiation of the field equation, we find
\begin{equation}\label{ddh}
ddH_\a=0=d\left(^{(\vartheta)\!}  \Sigma_{\alpha}+{}^{({\rm m})\!}
\Sigma_{\alpha}\right)\,.
\end{equation}
Thus,
$d{}^{({\rm m})\!}  \Sigma_{\alpha}=-d ^{(\vartheta)\!}
\Sigma_{\alpha}$
is {\it nonvanishing} in general, which clearly shows up in our Tonti
diagram.

The Tonti diagram displayed in Fig. \ref{Fig1} was constructed for the
{\it premetric} version of teleparallelism, no metric is involved at
all. We developed the corresponding formalism in our previous paper
in \cite[Sec.II]{Itin:2016nxk}. Above, in Sec.1.1, we discussed
already that teleparallel gravity is only equivalent to GR as long as
the matter which is involved does {\it not} carry intrinsic spin. If this is
the case, the energy-momentum of matter is described by the 3-form
$^{({\rm m})}\sigma_\a$, with
$\vt_{[\a}\!\wedge\!{}^{({\rm m})}\!\sigma_{\b]}=0$, i.e., the
corresponding energy-momentum tensor is symmetric. Since
$\vt_\a:=g_{\a\g}\vt^\g$, we need a metric for the specification of
such an energy-momentum $^{({\rm m})}\sigma_\a$. And the metric
$g_{\a\b}$ induces an associated Levi-Civita connection 1-form
$\widetilde{\Gamma}_\a{}^\b$.

The energy-momentum law, in the teleparallel gauge, reads
$d\,^{({\rm m})}\!\Sigma_\a=(e_\a \rfloor F^\b)$ $\wedge{}^{({\rm
    m})}\!\Sigma_\b$.
If we dispense with the teleparallel gauge for the moment, the
derivative $d\,^{({\rm m})}\!\Sigma_\a$ can be substituted by
$D\,^{({\rm m})}\!\Sigma_\a$, with $D$ as the covariant exterior
derivative operator with respect to the teleparallel connection
$\Gamma_\a{}^\b$. Having now a metric available, we can assume that
the teleparallel connection is metric compatible, that is,
$Dg_{\a\b}=0$. For the symmetric energy-momentum, we have then the
energy-momentum law
\begin{equation}\label{sym_em}
  D\,^{({\rm m})}\!\sigma_\a=(e_\a
  \rfloor F^\b)\wedge^{({\rm m})}\!\sigma_\b\,.
\end{equation}

Due to a lemma of Meyer \cite{Meyer:1983}, the right side of
\eqref{sym_em} can be absorbed by the left side. This can be
demonstrated by expanding the covariant derivative:
\begin{equation}\label{sym_em1}
d\,^{({\rm m})}\!\sigma_\a-(\Gamma_\a{}^\b+e_\a\rfloor
F^\b)\wedge^{({\rm m})}\! \sigma_\b=0\,.
\end{equation}
Let us now introduce a tilde to denote the Riemannian objects and
operators: $\widetilde{\Gamma}_\a{}^\b$ is the Christoffel connection,
for example, and $\widetilde{D}$ the Riemannian covariant derivative.
Since $\Gamma_\a{}^\b=\widetilde{\Gamma}_\a{}^\b-K_\a{}^\b$, with the
contortion 1-form $K_\a{}^\b$, we can rewrite \eqref{sym_em1} as
\begin{equation}\label{sym_em2}
\widetilde{D}\,^{({\rm m})}\!\sigma_\a-(-K_\a{}^\b+e_\a\rfloor
F^\b)\wedge^{({\rm m})}\! \sigma_\b=0\,.
\end{equation}
The contortion, is related to the torsion via
$F^\b=K^\b{}_\g\wedge\vt^\g$. If we substitute this into
\eqref{sym_em2}, it can be recast into
\begin{equation}\label{sym_em3}
\widetilde{D}\,^{({\rm m})}\!\sigma_\a+\left[K_\a{}^\b-(e_\a\rfloor
K^\b{}_\g)\vt^\g+K^\b{}_\g\d_\a^\g\right]\wedge^{({\rm m})}\! \sigma_\b=0\,.
\end{equation}
Because of the antisymmetry of the contortion, $K_{(\a\b)}=0$,
Eq.\eqref{sym_em3} simplifies to
\begin{equation}\label{sym_em4}
\widetilde{D}\,^{({\rm m})}\!\sigma_\a-(e_\a\rfloor
K^{\b\g})\vt_{[\g}\wedge^{({\rm m})}\! \sigma_{\b]}=0\,.
\end{equation}
Now we recall that $^{({\rm m})}\!\sigma_\a$ is symmetric,
$\vt_{[\g}\!\wedge\!{}^{({\rm m})}\!\sigma_{\b]}=0$. Consequently,
\begin{equation}\label{em-GR}
\boxed{\widetilde{D}\,^{({\rm m})}\!\sigma_\a=0\,.}
\end{equation}
This is the energy-momentum law of GR. Accordingly, our entry
$d\,^{({\rm m})}\!\Sigma_\a=(e_\a \rfloor$ $F^\b)\wedge{}^{({\rm
    m})}\!\Sigma_\b$ in the Tonti diagram becomes, provided a
symmetric energy-momentum tensor is prescribed,
$\widetilde{D}\,^{({\rm m})}\!\sigma_\a=0$, well in accordance with
the vanishing divergences in analogous Tonti diagrams.

\section{Decompositions of the constitutive tensor}\label{Sec.3}

In this section we characterize the constitutive tensor due to its
symmetry transformations relative to the general linear group
$\text{GL}(4,\mathbb{R})$ and the permutation (symmetry) group.  We
refer to Hamermesh \cite{Hamermesh:1962} and to Barut \& Raczka
\cite{Barut:1978} as background information for group theory.

\subsection{Two forms of the constitutive tensor}

According to \eqref{gr-kot9}, the constitutive tensor density
$\chi{}^{\alpha\beta}{}_\mu{}^{\gamma\delta}{}_\nu$ is a
order-\!\!\tvect{4}{2} tensor density that is skew-symmetric in the two
pairs of upper indices, see \eqref{gr-kot10}:
\begin{align}
\label{eq:chi-symmetries}
  \chi{}^{(\alpha\beta)}{}_\mu{}^{\gamma\delta}{}_\nu =0\,,\quad 
  \chi{}^{\alpha\beta}{}_\mu{}^{(\gamma\delta)}{}_\nu=0 .
\end{align}
In $n$ dimensions, we have
$ n\tvect{n}{2}n\tvect{n}{2} = \frac 14 n^4 (n-1)^2$ independent
components. For $n=4$, it gives 576 relevant components. Due to this
large number, it is instructive to consider decomposition of the
constitutive tensor into smaller pieces. We apply Young's
decomposition technique that yields irreducible decomposition under
the group $\text{GL}(4,\mathbb{R})$. For our conventions and the
details of the Young decomposition, see
\ref{app:conventions-decomposition}.  For the tensor
$\chi{}^{\alpha\beta}{}_\mu{}^{\gamma\delta}{}_\nu$, separate
$S_4$-permutations of four upper indices and $S_2$-permutations of two
lower indices are available, where $S_n$ denotes the $n$-dimensional
permutation group. Correspondingly, we are dealing with the Cartesian
product group $S_4\times S_2$.

Although the metric tensor is not available in our construction, we
can, as we have shown in \eqref{chi_check}, lower two pairs of
skew-symmetric indices by the Levi-Civita symbol.  Thus, we have a
order-\tvect{0}{6} tensor
\begin{align}
\label{eq:chi-check-symmetries}
  \check{\chi}{}_{\alpha\beta\mu\gamma\delta\nu} := 
  \frac 14 \epsilon{}_{\alpha\beta\omega\lambda}
  \epsilon{}_{\gamma\delta\rho\sigma} 
  \chi{}^{\omega\lambda}{}_\mu{}^{\rho\sigma}{}_\nu , 
\end{align}
with the identities
\begin{align}
\label{eq:chi-check-symmetries*}
  \check{\chi}{}_{(\alpha\beta)\mu\gamma\delta\nu} = 0\,,\quad
  \check{\chi}{}_{\alpha\beta\mu(\gamma\delta)\nu}=0 .
\end{align}
This tensor is naturally decomposed under the permutation group $S_6$.
Both types of decompositions are invariant under the action of the
basis transformation group $\text{GL}(4,\mathbb{R})$.

\subsection{$S_4\times S_2$ decomposition of
  $\chi{}^{\alpha\beta}{}_\mu{}^{\gamma\delta}{}_\nu$}

Treating the covariant and contravariant indices as belonging to two
separate tensor spaces, the irreducible decomposition of
$\chi{}^{\alpha\beta}{}_\mu{}^{\gamma\delta}{}_\nu$ is defined as a
product of the irreducible decompositions. The corresponding Young
diagrams are expanded as
\begin{align}
\begin{split}
\label{eq:chi-gl4r-gl4r-decomp}
&\Yvcentermath1 {\yng(1,1) \otimes \yng(1) \otimes \yng(1,1) \otimes
  \yng(1)}= \Yvcentermath1 \left( \,\yng(1,1) \otimes \yng(1,1) \,
\right) \otimes \yng(1) \otimes \yng(1)\\ \\
& =\Yvcentermath1 {\left( \, \yng(2,2) \oplus \yng(2,1,1) \oplus
    \yng(1,1,1,1) \, \right) ~\otimes~ \left( \, \yng(2) \oplus
    \yng(1,1) \, \right). }
\end{split}
\end{align}
Here, the three first diagrams relate to the upper indices. This
decomposition repeats the known irreducible decomposition of the
electromagnetic constitutive tensor. The remaining two diagrams
represent the symmetric and antisymmetric parts of the second order
tensor. Collecting all possible combinations, we obtain the
decomposition of the tensor
$\chi{}^{\alpha\beta}{}_\mu{}^{\gamma\delta}{}_\nu$ into 6 independent
pieces:
\begin{align}
\begin{split}
  \Yvcentermath1 \left( \, \yng(2,2)\otimes\yng(2) \, \right) ~\oplus~
  \left( \, \yng(2,2)\otimes\yng(1,1) \, \right) ~\oplus~ \left( \,
    \yng(2,1,1)\otimes\yng(1,1) \, \right) \\~\oplus~ \left( \,
    \Yvcentermath1 \yng(2,1,1)\otimes\yng(2) \, \right)
  \Yvcentermath1 
  { ~\oplus~ \left( \, \yng(1,1,1,1)\otimes\yng(1,1) \, \right)
    ~\oplus~ \left( \, \yng(1,1,1,1)\otimes\yng(2) \, \right)}.
\end{split}
\end{align}
In tensor notation, the above decomposition reads 
\begin{align}\label{DeComp}
  \chi{}^{\alpha\beta}{}_\mu{}^{\gamma\delta}{}_\nu 
  &= \sum\limits_{I=1}^6 {}^{[I]}
    \mathbb{P}^{\alpha\beta}_{\omega\lambda}{}^\kappa_\mu{}^{\gamma\delta}_{\rho\sigma}
    {}^\epsilon_\nu \, \chi{}^{\omega\lambda}{}_\kappa{}^{\rho\sigma}{}_\epsilon
    = \sum\limits_{I=1}^6 \!\!{}^{[I]}\! \chi{}^{\alpha\beta}{}_\mu{}^{\gamma\delta}{}_\nu\,. 
\end{align}
Here we used the set of six projection operators ${}^{[I]}\mathbb{P}$.
We chose the labeling of $I=1,\dots,6$ such that it corresponds to the
same sequence of Young diagrams as depicted in the second equality of
Eq.\eqref{DeComp}. 
Explicitly, the $S_4\times S_2$-irreducible pieces of the constitutive
tensor are expressed as
\begin{align}
\label{eq:s4s2-1}
{}^{[1]}\chi{}^{\alpha\beta}{}_\mu{}^{\gamma\delta}{}_\nu & := \chi{}^{\alpha\beta}{}_{(\mu}{}^{\gamma\delta}{}_{\nu)}
- {}^{[3]}\chi{}^{\alpha\beta}{}_\mu{}^{\gamma\delta}{}_\nu
- {}^{[5]}\chi{}^{\alpha\beta}{}_\mu{}^{\gamma\delta}{}_\nu, \\
  \label{eq:s4s2-2}
  {}^{[2]}\chi{}^{\alpha\beta}{}_\mu{}^{\gamma\delta}{}_\nu & := \chi{}^{\alpha\beta}{}_{[\mu}{}^{\gamma\delta}{}_{\nu]}
                                                              - {}^{[4]}\chi{}^{\alpha\beta}{}_\mu{}^{\gamma\delta}{}_\nu
                                                              - {}^{[6]}\chi{}^{\alpha\beta}{}_\mu{}^{\gamma\delta}{}_\nu, \\
  \label{eq:s4s2-3}
  {}^{[3]}\chi{}^{\alpha\beta}{}_\mu{}^{\gamma\delta}{}_\nu & := \frac12\left( \chi{}^{\alpha\beta}{}_{(\mu}{}^{\gamma\delta}{}_{\nu)} - \chi{}^{\gamma\delta}{}_{(\mu}{}^{\alpha\beta}{}_{\nu)} \right) , \\
\label{eq:s4s2-4}
{}^{[4]}\chi{}^{\alpha\beta}{}_\mu{}^{\gamma\delta}{}_\nu & := \frac12\left( \chi{}^{\alpha\beta}{}_{[\mu}{}^{\gamma\delta}{}_{\nu]} - \chi{}^{\gamma\delta}{}_{[\mu}{}^{\alpha\beta}{}_{\nu]} \right), \\
\label{eq:s4s2-5}
{}^{[5]}\chi{}^{\alpha\beta}{}_\mu{}^{\gamma\delta}{}_\nu & := \chi{}^{[\alpha\beta}{}_{(\mu}{}^{\gamma\delta]}{}_{\nu)}, \\
\label{eq:s4s2-6}
{}^{[6]}\chi{}^{\alpha\beta}{}_\mu{}^{\gamma\delta}{}_\nu & := \chi{}^{[\alpha\beta}{}_{[\mu}{}^{\gamma\delta]}{}_{\nu]} .
\end{align}
In terms of independent components in $n=4$ dimensions, this
decomposition corresponds to
\begin{align}
576 &= 200 ~\oplus~ 120 ~\oplus~ 150 ~\oplus~ 90 ~\oplus~ 10 ~\oplus~ 6 .
\end{align}

It is straightforward to show
that the above operators
${}^{[I]}\mathbb{P}={}^{[I]}\mathbb{P}
{}^{\alpha\beta\kappa\gamma\delta\epsilon}_{\omega\lambda\mu\rho\sigma\nu}$
are orthogonal projectors:
\begin{align} {}^{[I]}\mathbb{P} \circ {}^{[J]}\mathbb{P} =
  {}^{[I]}\mathbb{P} \delta{}^{IJ}, \quad \delta{}^{IJ}
  = \begin{cases} 1 & \text{ for } I=J \\ 0 & \text{ for } I \not=
    J \end{cases} ~.
\end{align}

In (\ref{eq:s4s2-1}) to (\ref{eq:s4s2-6}), we first decompose the
gravitational constitutive tensor into three independent pieces
relative to its four upper indices. This $S_4$ decomposition is done
completely similar to electrodynamics. Then we extract relative to the
pair of the lower indices the symmetric and antisymmetric parts. Since
the principal and axion part of the electromagnetic constitutive
tensor are reversible, the corresponding parts of the gravitational
constitutive tensor are reversible when symmetrized in the lower
indices. The electrodynamics the skewon part is irreversible. Thus its
gravitational analog is reversible when antisymmetrized in the lower
indices. The remaining three parts are irreversible (change their sign
under permutation of triads of indices).

In order to demonstrate this behavior explicitly, we decompose the
constitutive tensor into its reversible and irreversible parts,
\begin{align}
  \chi{}^{\alpha\beta}{}_\mu{}^{\gamma\delta}{}_\nu = 
  \overset{+}{\chi}{}^{\alpha\beta}{}_\mu{}^{\gamma\delta}{}_\nu
  + \overset{-}{\chi}{}^{\alpha\beta}{}_\mu{}^{\gamma\delta}{}_\nu \,,
\end{align}
with 
\begin{align}
  \overset{+}{\chi}{}^{\alpha\beta}{}_\mu{}^{\gamma\delta}{}_\nu 
  &:= \frac12( \chi{}^{\alpha\beta}{}_\mu{}^{\gamma\delta}{}_\nu 
    + \chi{}^{\gamma\delta}{}_\nu{}^{\alpha\beta}{}_\mu )=
    \overset{+}{\chi}{}^{\gamma\delta}{}_\nu{}^{\alpha\beta}{}_\mu\,, \\
  \overset{-}{\chi}{}^{\alpha\beta}{}_\mu{}^{\gamma\delta}{}_\nu 
  &:= \frac12( \chi{}^{\alpha\beta}{}_\mu{}^{\gamma\delta}{}_\nu 
    - \chi{}^{\gamma\delta}{}_\nu{}^{\alpha\beta}{}_\mu)
=-\overset{-}{\chi}{}^{\gamma\delta}{}_\nu{}^{\alpha\beta}{}_\mu .
\end{align}
The first part
$\overset{+}{\chi}{}^{\alpha\beta}{}_\mu{}^{\gamma\delta}{}_\nu$
describes {\it reversible} processes, it can be derived from a
Lagrang\-ian. In contrast, the second part
$\overset{-}{\chi}{}^{\alpha\beta}{}_\mu{}^{\gamma\delta}{}_\nu$
corresponds to irreversibility, and it does not enter the
Lagrangian. Following the nomenclature in electrodynamics, we will
call $\overset{-}{\chi}{}^{\alpha\beta}{}_\mu{}^{\gamma\delta}{}_\nu$
the {\it skewon} part.

In the next step, we can decompose each of these parts into the
symmetric/antisymmetric piece in the lower indices:
\begin{align}\label{+-decomposed}
 \overset{\pm}{\chi}{}^{\alpha\beta}{}_\mu{}^{\gamma\delta}{}_\nu =
 \overset{\pm}{\chi}{}^{\alpha\beta}{}_{(\mu}{}^{\gamma\delta}{}_{\nu)} +
 \overset{\pm}{\chi}{}^{\alpha\beta}{}_{[\mu}{}^{\gamma\delta}{}_{\nu]}\,.
\end{align}
Two immediate consequences are
\begin{align}
\overset{+}{\chi}{}^{[\alpha\beta}{}_{[\mu}{}^{\gamma\delta]}{}_{\nu]} \equiv 0,\qquad
\overset{-}{\chi}{}^{[\alpha\beta}{}_{(\mu}{}^{\gamma\delta]}{}_{\nu)} \equiv 0.
\end{align}
Thus, axion pieces can be extracted only from
$\overset{+}{\chi}{}^{\alpha\beta}{}_{(\mu}{}^{\gamma\delta}{}_{\nu)}$
and from
$\overset{-}{\chi}{}^{\alpha\beta}{}_{[\mu}{}^{\gamma\delta}{}_{\nu]}$.

After these preliminaries, we can recast the irreducible decompositions
(\ref{eq:s4s2-1}) to (\ref{eq:s4s2-6}) into a more transparent form,
\begin{align}
  {}^{[1]}\chi{}^{\alpha\beta}{}_\mu{}^{\gamma\delta}{}_\nu & 
:= \overset{+}{\chi}{}^{\alpha\beta}{}_{(\mu}{}^{\gamma\delta}{}_{\nu)} 
- \overset{+}{\chi}{}^{[\alpha\beta}{}_{(\mu}{}^{\gamma\delta]}{}_{\nu)} , \\
  {}^{[2]}\chi{}^{\alpha\beta}{}_\mu{}^{\gamma\delta}{}_\nu & 
:=\overset{-}{\chi}{}^{\alpha\beta}{}_{[\mu}{}^{\gamma\delta}{}_{\nu]} 
- \overset{-}{\chi}{}^{[\alpha\beta}{}_{[\mu}{}^{\gamma\delta]}{}_{\nu]} , \\
  {}^{[3]}\chi{}^{\alpha\beta}{}_\mu{}^{\gamma\delta}{}_\nu & 
:= \overset{-}{\chi}{}^{\alpha\beta}{}_{(\mu}{}^{\gamma\delta}{}_{\nu)} , \\
  {}^{[4]}\chi{}^{\alpha\beta}{}_\mu{}^{\gamma\delta}{}_\nu & 
:= \overset{+}{\chi}{}^{\alpha\beta}{}_{[\mu}{}^{\gamma\delta}{}_{\nu]} , \\
  {}^{[5]}\chi{}^{\alpha\beta}{}_\mu{}^{\gamma\delta}{}_\nu & 
:=  \overset{+}{\chi}{}^{[\alpha\beta}{}_{(\mu}{}^{\gamma\delta]}{}_{\nu)}, \\
  {}^{[6]}\chi{}^{\alpha\beta}{}_\mu{}^{\gamma\delta}{}_\nu & 
:=  \overset{-}{\chi}{}^{[\alpha\beta}{}_{[\mu}{}^{\gamma\delta]}{}_{\nu]}.
\end{align}
In other words, we have a decomposition of the reversible and skewon
parts into, respectively,
\begin{align}
  \overset{+}{\chi}{}^{\alpha\beta}{}_\mu{}^{\gamma\delta}{}_\nu 
&= {}^{[1]}\chi{}^{\alpha\beta}{}_\mu{}^{\gamma\delta}{}_\nu 
+ {}^{[4]}\chi{}^{\alpha\beta}{}_\mu{}^{\gamma\delta}{}_\nu 
+ {}^{[5]}\chi{}^{\alpha\beta}{}_\mu{}^{\gamma\delta}{}_\nu , \\
  \overset{-}{\chi}{}^{\alpha\beta}{}_\mu{}^{\gamma\delta}{}_\nu 
&= {}^{[2]}\chi{}^{\alpha\beta}{}_\mu{}^{\gamma\delta}{}_\nu 
+ {}^{[3]}\chi{}^{\alpha\beta}{}_\mu{}^{\gamma\delta}{}_\nu 
+ {}^{[6]}\chi{}^{\alpha\beta}{}_\mu{}^{\gamma\delta}{}_\nu .
\end{align}
We may call ${}^{[1]}\chi$ a reversible symmetric principal part
(prin\-cipal-1), ${}^{[4]}\chi$ a reversible antisymmetric principal
part (prin\-cipal-2), ${}^{[5]}\chi$ a reversible axion (axion-1);
likewise we may call ${}^{[2]}\chi$ a skewon antisymmetric principal
part (skewon-1), ${}^{[3]}\chi$ a skewon symmetric principal part
(skewon-2), and ${}^{[6]}\chi$ a skewon axion (axion-2). This is
summarized in Table~\ref{tab.1}. 

{
\begin{table}[!htb]
    \centering
    \begin{tabular}{lllll}
      Irr.~Parts & Nomenclature & Lagr. & En-mom. & Disp. \\
      \hline ${}^{[1]}\chi^{\a\b}{}_\g{}^{\mu\nu}{}_\rho$ & principal-1 & yes & yes & yes \\
      ${}^{[2]}\chi^{\a\b}{}_\g{}^{\mu\nu}{}_\rho$ & skewon-1 & no & yes & yes \\
      ${}^{[3]}\chi^{\a\b}{}_\g{}^{\mu\nu}{}_\rho$ &  skewon-2  & no & yes & yes \\
      ${}^{[4]}\chi^{\a\b}{}_\g{}^{\mu\nu}{}_\rho$ &  principal-2 & yes & yes & yes \\
      ${}^{[5]}\chi^{\a\b}{}_\g{}^{\mu\nu}{}_\rho$ & axion-1 & yes & no & no \\
      ${}^{[6]}\chi^{\a\b}{}_\g{}^{\mu\nu}{}_\rho$ & axion-2 & no & yes & no
    \end{tabular}
\caption{Physical identifications of the irreducible parts of the
  constitutive tensor $\chi^{\a\b}{}_\g{}^{\mu\nu}{}_\rho$.}\label{tab.1}
\end{table}
For completeness, we present here the results about the contributions
of the irreducible pieces into the dispersion relation for the
gravitational waves. These facts will be derived in
Sec.\ref{Sec.6}.

\subsection{$S_6$ decomposition of $\check{\chi}{}_{\alpha\beta\mu\gamma\delta\nu}$}

Now we consider the decomposition of the constitutive tensor under the
permutation of its six lower indices. We follow Hamermesh's
\cite{Hamermesh:1962} Eqs.\ (7-159) to (7-162). Then, for $n=4$
dimensions, the symmetries \eqref{eq:chi-check-symmetries*} imply the
Young decomposition of $\check{\chi}{}_{\alpha\beta\mu\gamma\delta\nu}$:
\begin{align}
\begin{split}
  &{\Yvcentermath1 \yng(1,1) \otimes \yng(1) \otimes \yng(1,1) \otimes
    \yng(1) } 
  ={\Yvcentermath1 {\yng(4,2) ~\oplus~ \yng(4,1,1)} ~\oplus~ \yng(3,3)} \\
  &{\Yvcentermath1 ~\oplus~ 4~\yng(3,2,1) }
{\Yvcentermath1 
\oplus~ 2~\yng(2,2,2) 
\oplus~ 3~\yng(3,1,1,1)
\oplus~ 4~\yng(2,2,1,1)}\;.
\end{split}
\end{align}
Here we omitted vanishing diagrams in $n=4$ (i.e.\ diagrams that
contain antisymmetrization with respect to more than four indices). In
terms of independent components, we have
\begin{align}
  576 = 126 ~\oplus~ 70 ~\oplus~ 50 ~\oplus~ 4\times 64 ~\oplus
  ~ 2\times 10 \nonumber\\
  ~\oplus~ 3\times 10 ~\oplus~ 4\times 6 . \label{eq:chi-dimensions}
\end{align}
In tensorial language, this corresponds to
\begin{align}
  \check{\chi}{}_{\alpha\beta\mu\gamma\delta\nu} 
  &= \sum\limits_{I=1}^7
    {}^{(I)}\mathbb{P}{}_{\alpha\beta\mu\gamma\delta\nu}^{\omega\lambda\kappa\rho\sigma\epsilon}
    \, \check{\chi}{}_{\omega\lambda\kappa\rho\sigma\epsilon} 
    =\sum\limits_{I=1}^7 {}^{(I)}\check{\chi}{}{}_{\alpha\beta\mu\gamma\delta\nu} ,
\end{align}
where again the expressions
${}^{(I)}\mathbb{P}{}_{\alpha\beta\mu\gamma\delta\nu}^{\omega\lambda\kappa\rho\sigma\epsilon}$
denote orthogonal projectors,
\begin{align} {}^{(I)}\mathbb{P} \circ {}^{(J)}\mathbb{P} =
  {}^{(I)}\mathbb{P}\delta{}^{IJ} , \quad \delta{}^{IJ}
  = \begin{cases} 1 & \text{ for } I=J \\ 0 & \text{ for } I \not=
    J \end{cases} ~.
\end{align}
The explicit expressions of the seven terms
${}^{(I)}\check{\chi}{}_{\alpha\beta\mu\gamma\delta\nu}$ are quite
involved and we do not display them here; they can be found in
\ref{app:conventions-decomposition}.

\subsection{Relation between the decompositions}
In order to relate the two inequivalent decompositions discussed
above, it is useful to define
\begin{align}
  {}^{\{I\}}\chi{}^{\alpha\beta}{}_\mu{}^{\gamma\delta}{}_\nu
  &:= \frac 14 \epsilon{}^{\alpha\beta\omega\lambda}
  \epsilon{}^{\gamma\delta\rho\sigma}\, {}^{(I)}\!
  \check{\chi}{}_{\omega\lambda\mu\rho\sigma\nu} .
\end{align}
By using
Eqs.~\eqref{eq:cc-decomposition-1} to \eqref{eq:cc-decomposition-7}, one
may verify that the tensors
${}^{\{I\}}\chi{}^{\alpha\beta}{}_\mu{}^{\gamma\delta}{}_\nu$ satisfy
the relations
\begin{alignat}{4}
  \label{eq:aux-cc-1}
  {}^{\{I\}}\chi{}^{[\alpha\beta}{}_\mu{}^{\gamma\delta]}{}_\nu 
&= 0 & \text{ for } I &= 1,2,3,4,5 , \\
  \label{eq:aux-cc-2}
  {}^{\{I\}}\chi{}^{\alpha\beta}{}_{[\mu}{}^{\gamma\delta}{}_{\nu]}
  &= 0 & \text{ for } I &= 1,2 , \\
  \label{eq:aux-cc-3}
  {}^{\{I\}}\chi{}^{\alpha\beta}{}_{(\mu}{}^{\gamma\delta}{}_{\nu)}
  &= 0 & \text{ for } I &= 3 ,
\end{alignat}
as well as
\begin{alignat}{4}
  \label{eq:aux-cc-4}
  {}^{\{I\}}\chi{}^{\alpha\beta}{}_\mu{}^{\gamma\delta}{}_\nu
  &= +\, {}^{\{I\}}
  \chi{}^{\gamma\delta}{}_\mu{}^{\alpha\beta}{}_\nu
  & \quad \text{ for } I &= 1,3 , \\
  \label{eq:aux-cc-5}
  {}^{\{I\}}\chi{}^{\alpha\beta}{}_\mu{}^{\gamma\delta}{}_\nu
  &= -\,
  {}^{\{I\}}\chi{}^{\gamma\delta}{}_\mu{}^{\alpha\beta}{}_\nu
  & \text{ for } I &= 2 , \\
  \label{eq:aux-cc-6}
  {}^{\{I\}}\chi{}^{\alpha\beta}{}_\mu{}^{\gamma\delta}{}_\nu
  &= +\,
  {}^{\{I\}}\chi{}^{\gamma\delta}{}_\nu{}^{\alpha\beta}{}_\mu
  & \text{ for } I &= 1,5 , \\
  \label{eq:aux-cc-7}
  {}^{\{I\}}\chi{}^{\alpha\beta}{}_\mu{}^{\gamma\delta}{}_\nu
  &= -\,
  {}^{\{I\}}\chi{}^{\gamma\delta}{}_\nu{}^{\alpha\beta}{}_\mu
  & \text{ for } I &= 2,3 .
\end{alignat}
Eventually, we find
\begin{align}
  \label{eq:aux-cc-8}
  {}^{\{6\}}\chi{}^{\alpha\beta}{}_{[\mu}{}^{\gamma\delta}{}_{\nu]} 
  + {}^{\{6\}}\chi{}^{\gamma\delta}{}_{[\mu}{}^{\alpha\beta}{}_{\nu]} 
  &= 0 , \\
  \label{eq:aux-cc-9}
  {}^{\{6\}}\chi{}^{\alpha\beta}{}_{(\mu}{}^{\gamma\delta}{}_{\nu)}
  +{}^{\{6\}}\chi{}^{\gamma\delta}{}_{(\mu}{}^{\alpha\beta}{}_{\nu)} 
&= {}^{\{6\}}\chi{}^{[\alpha\beta}{}_{(\mu}{}^{\gamma\delta]}{}_{\nu)} , \\
  \label{eq:aux-cc-10}
  {}^{\{7\}}\chi{}^{\alpha\beta}{}_{(\mu}{}^{\gamma\delta}{}_{\nu)}
  +
  {}^{\{7\}}\chi{}^{\gamma\delta}{}_{(\mu}{}^{\alpha\beta}{}_{\nu)} 
&= 0 . 
\end{align}
We can apply directly the $S_4\times S_2$ decomposition to these
order-\tvect{4}{2} tensors and compare the
results to the tensors \\
${}^{[I]}\chi{}^{\alpha\beta}{}_\mu{}^{\gamma\delta}{}_\nu$. By
extensive use of computer algebra, we find the following relations:
\begin{align}
  {}^{[1]}\chi{}^{\alpha\beta}{}_\mu{}^{\gamma\delta}{}_\nu
  &= {}^{\{1\}}\chi{}^{\alpha\beta}{}_{(\mu}{}^{\gamma\delta}{}_{\nu)} 
   +{}^{\{5\}}\chi{}^{\alpha\beta}{}_{(\mu}{}^{\gamma\delta}{}_{\nu)} 
    \label{eq:relation-c-cc-1} \\
    &\hspace{12pt}+ \frac12\left[ {}^{\{4\}}\chi{}^{\alpha\beta}{}_{(\mu}{}^{\gamma\delta}{}_{\nu)}
    + {}^{\{4\}}\chi{}^{\gamma\delta}{}_{(\mu}{}^{\alpha\beta}{}_{\nu)}  \right] , \nonumber \\
  {}^{[2]}\chi{}^{\alpha\beta}{}_\mu{}^{\gamma\delta}{}_\nu
  &= {}^{\{3\}}\chi{}^{\alpha\beta}{}_{[\mu}{}^{\gamma\delta}{}_{\nu]} - {}^{\{7\}}
  \chi{}^{[\alpha\beta}{}_{[\mu}{}^{\gamma\delta]}{}_{\nu]} \label{eq:relation-c-cc-2} \\
  &\hspace{12pt} + \frac12 \sum\limits_{I=4,7} \left[ {}^{\{I\}}\chi{}^{\alpha\beta}
  {}_{[\mu}{}^{\gamma\delta}{}_{\nu]} + {}^{\{I\}}\chi{}^{\gamma\delta}{}_{[\mu}{}^{\alpha\beta}{}_{\nu]} \right], \nonumber \\
  {}^{[3]}\chi{}^{\alpha\beta}{}_\mu{}^{\gamma\delta}{}_\nu
  &= {}^{\{2\}}\chi{}^{\alpha\beta}{}_{(\mu}{}^{\gamma\delta}{}_{\nu)} \label{eq:relation-c-cc-3} \\
  &\hspace{12pt}+ \frac 12 \sum\limits_{I=4,6,7} \left[ {}^{\{I\}}\chi{}^{\alpha\beta}
  {}_{(\mu}{}^{\gamma\delta}{}_{\nu)} - {}^{\{I\}}\chi{}^{\gamma\delta}{}_{(\mu}{}^{\alpha\beta}{}_{\nu)} \right] ,\nonumber\\
  {}^{[4]}\chi{}^{\alpha\beta}{}_\mu{}^{\gamma\delta}{}_\nu &= {}^{\{5\}}\chi{}^{\alpha\beta}
  {}_{[\mu}{}^{\gamma\delta}{}_{\nu]} \label{eq:relation-c-cc-4} \\
  &\hspace{12pt}+ \frac 12 \sum\limits_{I=4,6,7} \left[ {}^{\{I\}}\chi{}^{\alpha\beta}
  {}_{[\mu}{}^{\gamma\delta}{}_{\nu]} - {}^{\{I\}}\chi{}^{\gamma\delta}{}_{[\mu}{}^{\alpha\beta}{}_{\nu]} \right] ,\nonumber\\
  {}^{[5]}\chi{}^{\alpha\beta}{}_\mu{}^{\gamma\delta}{}_\nu &= {}^{\{6\}}
  \chi{}^{[\alpha\beta}{}_{(\mu}{}^{\gamma\delta]}{}_{\nu)} , \label{eq:relation-c-cc-5} \\
  {}^{[6]}\chi{}^{\alpha\beta}{}_\mu{}^{\gamma\delta}{}_\nu &={}^{\{7\}}
  \chi{}^{[\alpha\beta}{}_{[\mu}{}^{\gamma\delta]}{}_{\nu]} . \label{eq:relation-c-cc-6}
\end{align}

\subsection{Reversibility}
For reversible processes, as shown in \eqref{rev-cond}, the
constitutive tensor satisfies
\begin{align}
  \chi{}^{\alpha\beta}{}_\mu{}^{\gamma\delta}{}_\nu = 
  \chi{}^{\gamma\delta}{}_\nu{}^{\alpha\beta}{}_\mu\, .
\end{align}
As we demonstrated above, the reversible and irreversible parts of the
constitutive tensor are presented, respectively, as
\begin{align}
  \overset{+}{\chi}{}^{\alpha\beta}{}_\mu{}^{\gamma\delta}{}_\nu 
  &= \frac12\left( \chi{}^{\alpha\beta}{}_\mu{}^{\gamma\delta}{}_\nu 
    + \chi{}^{\gamma\delta}{}_\nu{}^{\alpha\beta}{}_\mu \right) \\
  &= {}^{[1]}\chi{}^{\alpha\beta}{}_\mu{}^{\gamma\delta}{}_\nu 
    + {}^{[4]}\chi{}^{\alpha\beta}{}_\mu{}^{\gamma\delta}{}_\nu 
    + {}^{[5]}\chi{}^{\alpha\beta}{}_\mu{}^{\gamma\delta}{}_\nu , \nonumber \\
  \overset{-}{\chi}{}^{\alpha\beta}{}_\mu{}^{\gamma\delta}{}_\nu 
  &= \frac12\left( \chi{}^{\alpha\beta}{}_\mu{}^{\gamma\delta}{}_\nu 
    - \chi{}^{\gamma\delta}{}_\nu{}^{\alpha\beta}{}_\mu \right) \nonumber \\
  &= {}^{[2]}\chi{}^{\alpha\beta}{}_\mu{}^{\gamma\delta}{}_\nu 
    + {}^{[3]}\chi{}^{\alpha\beta}{}_\mu{}^{\gamma\delta}{}_\nu 
    + {}^{[6]}\chi{}^{\alpha\beta}{}_\mu{}^{\gamma\delta}{}_\nu .
\end{align}
This corresponds to $576 = 300 \oplus 276$, that is, the reversible
constitutive tensor comprises 300 independent components. Employing
relations Eqs.~\eqref{eq:relation-c-cc-1}--\eqref{eq:relation-c-cc-6},
one finds
\begin{align}
  \overset{+}{\chi}{}^{\alpha\beta}{}_\mu{}^{\gamma\delta}{}_\nu 
  &= {}^{\{1\}}\chi{}^{\alpha\beta}{}_{(\mu}{}^{\gamma\delta}{}_{\nu)} 
    + {}^{\{5\}}\chi{}^{\alpha\beta}{}_\mu{}^{\gamma\delta}{}_\nu \\
  &\hspace{12pt} + \frac12 \sum\limits_{I=4,6,7}
    \left[ {}^{\{I\}}\chi{}^{\alpha\beta}{}_\mu{}^{\gamma\delta}{}_\nu 
    + {}^{\{I\}}\chi{}^{\gamma\delta}{}_\nu{}^{\alpha\beta}{}_\mu 
    \right] , \nonumber \\
  \overset{-}{\chi}{}^{\alpha\beta}{}_\mu{}^{\gamma\delta}{}_\nu 
  &={}^{\{2\}}\chi{}^{\alpha\beta}{}_{(\mu}{}^{\gamma\delta}{}_{\nu)} 
    + {}^{\{3\}}\chi{}^{\alpha\beta}{}_{[\mu}{}^{\gamma\delta}{}_{\nu]} \\
  &\hspace{12pt} + \frac12 \sum\limits_{I=4,6,7}\left[ {}^{\{I\}}\chi
    {}^{\alpha\beta}{}_\mu{}^{\gamma\delta}{}_\nu 
    -{}^{\{I\}}\chi{}^{\gamma\delta}{}_\nu{}^{\alpha\beta}{}_\mu 
    \right] . \nonumber
\end{align}
Let us emphasize that Eqs.~\eqref{eq:aux-cc-2} and
\eqref{eq:aux-cc-3} imply
\begin{align}
 {}^{\{1\}}\chi{}^{\alpha\beta}{}_{[\mu}{}^{\gamma\delta}{}_{\nu]}
  = 0\,, \>\;
  {}^{\{2\}}\chi{}^{\alpha\beta}{}_{[\mu}{}^{\gamma\delta}{}_{\nu]}
  = 0\,, \>\;
  {}^{\{3\}}\chi{}^{\alpha\beta}{}_{(\mu}{}^{\gamma\delta}{}_{\nu)}
  = 0\, .
\end{align}
In summary, we see that the reversible piece of the constitutive
tensor $\chi{}^{\alpha\beta}{}_\mu{}^{\gamma\delta}{}_\nu$ is
comprised of the order-\tvect{0}{6} irreducible pieces $I=1,5$ as well
as the appropriately (anti-)\allowbreak symmetrized parts of
$I=4,6,7$.

\subsection{Lagrangian}

Let us first consider how these independent pieces contribute to the
gravitational Lagrangian
$^{(\vartheta)\!} \Lambda=-\,\frac{1}{2} F^{\alpha} \wedge
H_{\alpha}$. In components, we have
\begin{equation}\label{str-dec6}
  ^{(\vartheta)\!}\!\Lambda=-\,{\frac 14} F_{\b\g}{}^\a\,
  \check{H}^{\b\g}{}_\a{\rm vol}\,.
\end{equation}
Substitution of the constitutive law (\ref{gr-kot9}) yields
\begin{equation}\label{str-dec7}
  ^{(\vartheta)\!}\!\Lambda=-\,{\frac 18}
  \chi^{\b\g}{}_\a{}^{\nu\rho}{}_\mu \,F_{\b\g}{}^\a 
  F_{\nu\rho}{}^\mu\,{\rm vol}\,.
\end{equation}
Consequently only those parts of the constitutive tensor contribute to
the Lagrangian that satisfy the ``pair commutation'' symmetries
\begin{equation}\label{str-dec8}
 \chi^{\b\g}{}_\a{}^{\nu\rho}{}_\mu=\chi^{\nu\rho}{}_\mu{}^{\b\g}{}_\a\,.
\end{equation}
Accordingly, only the following terms are left over in the Lagrangian:
\begin{equation}\label{str-dec9}
  ^{(\vartheta)\!}\!\Lambda=-\,{\frac 18} \sum\limits_{I=1,4,5}
  {}^{[I]}\chi^{\b\g}{}_\a{}^{\nu\rho}{}_\mu
F_{\b\g}{}^\a \, F_{\nu\rho}{}^\mu\,{\rm vol}\,.
\end{equation}

\subsection{Parametrization of irreducible parts}

In premetric electrodynamics, there exists a convenient
para\-metrization of the axion and skewon parts by a pseudo-scalar and a
traceless second order tensor, see Secs.D.1.4 and D.1.5 of the book
\cite{Birkbook}.

One may wish to develop a similar formalism for the gravitational case
under consideration.  By analogy with electrodynamics, let us
introduce the following new objects:
\begin{eqnarray}
{\!\not\!S}_{\c\r\s}{}^{\nu} &:=& {\frac 14}\epsilon_{\a\b\mu\s}\,
{}^{[3]}\chi{}^{\a\b}{}_\c{}^{\mu\nu}{}_\r,\label{S3}\\
{\!\not\!Q}_{\c\r\s}{}^{\nu} &:=& {\frac 14}\epsilon_{\a\b\mu\s}\,
{}^{[4]}\chi{}^{\a\b}{}_\c{}^{\mu\nu}{}_\r.\label{Q4}
\end{eqnarray}
By construction (hint: use the definitions of ${}^{[3]}\chi$ and
${}^{[4]}\chi$), these tensors obey
\begin{eqnarray}
{\!\not\!S}_{[\a\b]\mu}{}^{\nu} &=& 0\,,\qquad
{\!\not\!S}_{\a\b\nu}{}^{\nu} = 0\,;\label{Spro}\\
{\!\not\!Q}_{(\a\b)\mu}{}^{\nu} &=& 0\,,\qquad
{\!\not\!Q}_{\a\b\nu}{}^{\nu} = 0.\label{Qpro}
\end{eqnarray}
Accordingly, the number of independent components for\;\;
${\!\not\!S}_{\a\b\mu}{}^{\nu}$ is $10\times 4\times 4 - 10 = 150$
and for ${\!\not\!Q}_{\a\b\mu}{}^{\nu}$ $6\times 4\times 4 - 6 = 90$,
respectively. It is straightforward to see that
\begin{eqnarray}
{}^{[3]}\chi{}^{\a\b}{}_\c{}^{\mu\nu}{}_\r &=& \epsilon^{\a\b\lambda[\mu}\,{\!\not\!S}_{\c\r\lambda}{}^{\nu]}
- \epsilon^{\mu\nu\lambda[\a}\,{\!\not\!S}_{\c\r\lambda}{}^{\b]},\label{chi3S}\\
{}^{[4]}\chi{}^{\a\b}{}_\c{}^{\mu\nu}{}_\r &=& \epsilon^{\a\b\lambda[\mu}\,{\!\not\!Q}_{\c\r\lambda}{}^{\nu]}
- \epsilon^{\mu\nu\lambda[\a}\,{\!\not\!Q}_{\c\r\lambda}{}^{\b]}.\label{chi4Q}
\end{eqnarray}
Thereby, we found a complete parametrization of ${}^{[3]}\chi$ and
${}^{[4]}\chi$. This construction obviously generalizes the
representation of a skewon in electrodynamics in terms of the
traceless second order tensor \cite{Birkbook}. In the context of the
gravitational theory, however, only (\ref{chi3S}) refers to the
skewon, whereas (\ref{chi4Q}) parametrizes the {\it reversible}
irreducible part.

Now, let us turn to the irreducible
parts ${}^{[3]}\chi$ and ${}^{[4]}\chi$ for which we introduce
\begin{eqnarray}
  {}^{[5]}A_{\c\r} &:=& {\frac 1{4!}}\epsilon_{\a\b\mu\nu}\,{}^{[5]}
  \chi{}^{\a\b}{}_\c{}^{\mu\nu}{}_\r= {}^{[5]}A_{\r\c}\,,\label{A5}\\
  {}^{[6]}A_{\c\r} &:=& {\frac
 1{4!}}\epsilon_{\a\b\mu\nu}\,{}^{[6]}\chi{}^{\a\b}{}_\c{}^{\mu\nu}{}_\r
=-{}^{[6]}A_{\r\c}\,.\label{A6}
\end{eqnarray}
The reciprocal relations read
\begin{eqnarray}
  {}^{[5]}\chi{}^{\a\b}{}_\c{}^{\mu\nu}{}_\r &=
  & {}^{[5]}A_{\c\r}\,\epsilon^{\a\b\mu\nu},\label{chi5A}\\
  {}^{[6]}\chi{}^{\a\b}{}_\c{}^{\mu\nu}{}_\r &=
  & {}^{[6]}A_{\c\r}\,\epsilon^{\a\b\mu\nu}.\label{chi6A}
\end{eqnarray}
As a result, we find for the axion type part of the constitutive
tensor
\begin{equation} {}^{\rm (ax)}\chi{}^{\a\b}{}_\c{}^{\mu\nu}{}_\r =
  {\mathcal A}_{\c\r}\,\epsilon^{\a\b\mu\nu},\;\, {\mathcal A}_{\a\b}
  := {}^{[5]}A_{\a\b} + {}^{[6]}A_{\a\b}.\label{AX}
\end{equation}

Summarizing, we demonstrated that the four irreducible parts
${}^{[3]}\chi$, ${}^{[4]}\chi$, ${}^{[5]}\chi$, and ${}^{[6]}\chi$ of
the constitutive tensor can be conveniently parametrized, see
Eqs.\eqref{chi3S}, \eqref{chi4Q}, \eqref{chi5A}, and
\eqref{chi6A}. They are expressed in terms of the tensors (\ref{S3}),
(\ref{Q4}), (\ref{A5}), and (\ref{A6}) with the correct number of
independent components: for ${\!\not\!S}_{\a\b\mu}{}^{\nu}$ 150, for
${\!\not\!Q}_{\a\b\mu}{}^{\nu}$ 90, for ${}^{[5]}A_{\a\b}$ 10, and for
${}^{[6]}A_{\a\b}$ 6. The two remaining pieces ${}^{[1]}\chi$ and
${}^{[2]}\chi$ do not admit simple parametrizations, though.

\subsection{Energy-momentum current}

The energy-momentum current of the coframe field is defined as
\cite{Itin:2016nxk}
\begin{equation}\label{str-dec10}
  ^{(\vartheta)\!}\Sigma_\a =\frac 12 \left[ F^\b\wedge(e_\a\rfloor{
      H}_\b)-H_\b\wedge(e_\a    \rfloor F^\b)\right]\,.
\end{equation}
The constitutive law for the pure axion piece (\ref{AX}) is expressed as
\begin{equation}\label{str-dec11}
 {}^{\rm (ax)}H_\a={\cal A}_{\a\b}F^\b\,.
\end{equation}
In this case, the energy-momentum current takes the form
\begin{equation}\label{str-dec12}
 {}^{\rm (ax)}\Sigma_\a={\cal A}_{[\b\g]} F^\b\wedge(e_\a\rfloor { F}^\g)\,.
\end{equation}
Consequently, the symmetric combination of the axionic part
${}^{[5]}\chi^{\b\g}{}_\a{}^{\nu\rho}{}_\mu$ does {\it not} contribute
to the energy-momentum current.

\subsection{Contracting $\chi{}^{\a\b}{}_\c{}^{\mu\nu}{}_\r$ twice:
  four second order tensors} 

In electrodynamics, we can extract from the constitutive tensor
$\chi^{\a\b\c\d}$ a pseudo-scalar density, the axion field, by
contracting it with the Levi-Civita symbol:
$ {}^{\rm
  (ax)}\!\a:=\frac{1}{4!}\varepsilon_{\a\b\c\d}\;$ $\chi^{\a\b\c\d}.$
In gravity, we can only reduce the constitutive tensor
$\chi{}^{\a\b}{}_\c{}^{\mu\nu}{}_\r$ of type $\tvect{4}{2}$ to
four different 2nd order tensors:
\begin{eqnarray}
m^{\a\b}&:=& \chi^{\a\mu}{}_{\mu}{}^{\b\nu}{}_{\nu}  \,,\label{trm}   \\
n^{\a\b}&:=& \chi^{\a\mu}{}_{\nu}{}^{\b\nu}{}_{\mu} \,,\label{trn}   \\
k^{\a\b}&:=& \chi^{\mu\nu}{}_{\mu}{}^{\a\b}{}_{\nu}  \,,\label{trk}   \\
l^{\a\b}&:=& \chi^{\a\b}{}_{\mu}{}^{\mu\nu}{}_{\nu}  \,.\label{trl}
\end{eqnarray}
The construction of scalars is impossible at this premetric stage.

The tensors $k^{\a\b} = - k^{\b\a}$ and $l^{\a\b} = - l^{\b\a}$ are
antisymmetric, that is, $k^{(\a\b)}=0\,,\, l^{(\a\b)}=0\,$, whereas
$m^{\a\b}$ and $n^{\a\b}$ are asymmetric.  However, in the reversible
case, $m^{\a\b}$ and $n^{\a\b}$ turn out to be symmetric, whereas the
two other antisymmetric tensors turn out to be transposed to each
other, $k^{\a\b} = l^{\b\a}$. This can be recognized more clearly by
collecting all the traces into the specific constitutive tensor
\begin{eqnarray}
  \chi^{\a\b}{}_{\c}{}^{\mu\nu}{}_{\r} 
  &=&\hspace{13pt} {\mathcal L}^{\a\b}\,\d^{[\mu}_{\;\c}\d^{\nu]}_{\;\r} +
      {\mathcal K}^{\mu\nu}\,\d^{[\a}_{\;\r}\d^{\b]}_\c  \nonumber\\
  && +\,\d^{[\a}_{\;\c} {\mathcal M}^{\b][\mu}\d^{\nu]}_{\;\r} + \d^{[\a}_{\;\r} 
     {\mathcal N}^{\b][\mu}\d^{\nu]}_{\;\c}\,.\label{Tr0}
\end{eqnarray}
This tensor is determined by four tensors of 2nd order: two of them
are antisymmetric, ${\mathcal L}^{\a\b} = -\,{\mathcal L}^{\b\a}$ and
${\mathcal K}^{\a\b} = -\,{\mathcal K}^{\b\a}$, that is,
${\mathcal L}^{(\a\b)} =0\,,\,{\mathcal K}^{(\a\b)} =0$, and the two
other ones are general 2nd order tensors ${\mathcal M}^{\a\b}$ and
${\mathcal N}^{\a\b}$.

It is straightforward to establish one-to-one correspondences between
these objects and the traces (\ref{trm}) to (\ref{trl}):
\begin{eqnarray}
m^{\a\b}&=& {\frac 32}\left({\mathcal L}^{\a\b} - {\mathcal K}^{\a\b}\right)\nonumber\\
&& - \,{\frac 14}\left(9{\mathcal M}^{\a\b} + 2{\mathcal N}^{\a\b} + {\mathcal N}^{\b\a}\right)\,,\label{trm2}\\
n^{\a\b}&=& {\frac 32}\left({\mathcal K}^{\a\b} - {\mathcal L}^{\a\b}\right)\nonumber\\
&& - \,{\frac 14}\left(9{\mathcal N}^{\a\b} + 2{\mathcal M}^{\a\b} + {\mathcal M}^{\b\a}\right)\,,\label{trn2}\\
k^{\a\b}&=& {\mathcal L}^{\a\b} - 6{\mathcal K}^{\a\b} 
- \,{\frac 32}{\mathcal M}^{[\a\b]} + {\frac 32}{\mathcal N}^{[\a\b]}\,,\label{trk2}\\
l^{\a\b}&=& 6{\mathcal L}^{\a\b} - {\mathcal K}^{\a\b} 
- \,{\frac 32}{\mathcal M}^{[\a\b]} + {\frac 32}{\mathcal N}^{[\a\b]}\,.\label{trl2}
\end{eqnarray}
The inverse relations read
\begin{eqnarray}
{\mathcal L}^{\a\b} &=& {\frac 1{10}}\left(3l^{\a\b} + k^{\a\b} - 3m^{[\a\b]} + 3n^{[\a\b]}\right),\label{trL}\\
{\mathcal K}^{\a\b} &=& {\frac 1{10}}\left(-l^{\a\b} - 3k^{\a\b} + 3m^{[\a\b]} - 3n^{[\a\b]}\right),\label{trK}\\
{\mathcal M}^{\a\b} &=& {\frac 1{10}}\left(3l^{\a\b} + 3k^{\a\b} - 9m^{[\a\b]} + 5n^{[\a\b]}\right)\nonumber\\
&& -\,{\frac 16}\left(3m^{(\a\b)} - n^{(\a\b)}\right),\label{trM}\\
{\mathcal N}^{\a\b} &=& {\frac 1{10}}\left(- 3l^{\a\b} - 3k^{\a\b} + 5m^{[\a\b]} - 9n^{[\a\b]}\right)\nonumber\\
&& +\,{\frac 16}\left(m^{(\a\b)} - 3n^{(\a\b)}\right).\label{trN}
\end{eqnarray}

Decomposing the specific constitutive tensor (\ref{Tr0}) into
reversible and irreversible (skewon) parts yields
\begin{eqnarray}
  \overset{\pm}{\chi}{}^{\a\b}{}_{\c}{}^{\mu\nu}{}_{\r} 
  &=& \overset{\pm}{\mathcal L}{}^{\a\b}
      \,\d^{[\mu}_{\;\c}\d^{\nu]}_{\;\r} \pm 
      \overset{\pm}{\mathcal L}{}^{\mu\nu}\,\d^{[\a}_{\;\r}\d^{\b]}_{\;\c}\nonumber\\
  && +\,\d^{[\a}_{\;\c}\overset{\pm}{\mathcal M}{}^{\b][\mu}\d^{\nu]}_{\;\r}
     + \d^{[\a}_{\;\r}\overset{\pm}{\mathcal N}{}^{\b][\mu}\d^{\nu]}_{\;\c}\,.\label{Tr1}
\end{eqnarray}
Here we introduced 
\begin{eqnarray}
  \overset{\pm}{\mathcal L}{}^{\a\b} &=& {\frac 12}\left({\mathcal L}^{\a\b} \pm {\mathcal K}^{\a\b}\right),\label{LL}\\
\overset{+}{\mathcal M}{}^{\a\b} &=& {\mathcal M}^{(\a\b)},\quad
\overset{-}{\mathcal M}{}^{\a\b} = {\mathcal M}^{[\a\b]},\label{MM}\\
\overset{+}{\mathcal N}{}^{\a\b} &=& {\mathcal N}^{(\a\b)},\quad
\overset{-}{\mathcal N}{}^{\a\b} = {\mathcal N}^{[\a\b]}.\label{NN}
\end{eqnarray}
This confirms that the irreversible, the skewon part vanishes,
$\overset{-}{\chi}{}^{\a\b}{}_{\c}{}^{\mu\nu}{}_{\r} = 0$, provided
$\overset{-}{\mathcal L}{}^{\a\b}$ = 0,
$\overset{-}{\mathcal M}{}^{\a\b} = 0$, and
$\overset{-}{\mathcal N}{}^{\a\b} = 0$, i.e., when $m^{\a\b}$ and
$n^{\a\b}$ are symmetric and $k^{\a\b} = l^{\b\a}$.

In the general case, the specific constitutive trace tensor (\ref{Tr0}) is
characterized by 5 nontrivial irreducible parts:
\begin{align}
  {}^{[1]}\chi{}^{\a\b}{}_\c{}^{\mu\nu}{}_\r =
  &\, \d^{[\a}_{(\c}\overset{s}{\mathcal R}{}^{\b][\mu}\d^{\nu]}_{\r)},\label{1T} \\
  {}^{[2]}\chi{}^{\a\b}{}_{\c}{}^{\mu\nu}{}_{\r} =
  &\, \overset{-}{\mathcal L}{}^{\a\b}
    \,\d^{[\mu}_{\;\c}\d^{\nu]}_{\;\r}
    - \overset{-}{\mathcal L}{}^{\mu\nu}\,\d^{[\a}_{\;\r}\d^{\b]}_{\;\c}
    \nonumber\\
  & + \,\d^{[\a}_{[\c}\overset{a}{\mathcal S}{}^{\b][\mu}\d^{\nu]}_{\r]} - {\mathcal A}{}^{[\a\b}\d^{\mu}_\c\d^{\nu]}_\r,\label{2T} \\
  {}^{[3]}\chi{}^{\a\b}{}_\c{}^{\mu\nu}{}_\r =&\, \d^{[\a}_{(\c}\overset{s}{\mathcal S}{}^{\b][\mu}\d^{\nu]}_{\r)},\label{3T} \\
  {}^{[4]}\chi{}^{\a\b}{}_\c{}^{\mu\nu}{}_\r =
  &\, \overset{+}{\mathcal L}{}^{\a\b}
    \,\d^{[\mu}_{\c}\d^{\nu]}_\r + \overset{+}{\mathcal L}{}^{\mu\nu}\,\d^{[\a}_\r\d^{\b]}_\c \nonumber\\
  & + \,\d^{[\a}_{[\c}\overset{a}{\mathcal R}{}^{\b][\mu}\d^{\nu]}_{\r]},\label{4T} \\
  {}^{[5]}\chi{}^{\a\b}{}_\c{}^{\mu\nu}{}_\r =&\, 0,\label{5T} \\
  {}^{[6]}\chi{}^{\a\b}{}_\c{}^{\mu\nu}{}_\r =
  &\, {\mathcal A}{}^{[\a\b}\d^{\mu}_\c\d^{\nu]}_\r.\label{6T}
\end{align}
Here we denoted
\begin{eqnarray}
{\mathcal A}{}^{\a\b} &:=& 2\overset{-}{\mathcal L}{}^{\a\b} + \overset{-}{\mathcal M}{}^{\a\b}
- \overset{-}{\mathcal N}{}^{\a\b},\label{AA}\\
\overset{s}{\mathcal R}{}^{\a\b} &:=& \overset{+}{\mathcal M}{}^{\a\b} 
+ \overset{+}{\mathcal N}{}^{\a\b},\label{RS}\\
\overset{a}{\mathcal R}{}^{\a\b} &:=& \overset{+}{\mathcal M}{}^{\a\b} 
- \overset{+}{\mathcal N}{}^{\a\b},\label{RA}\\
\overset{s}{\mathcal S}{}^{\a\b} &:=& \overset{-}{\mathcal M}{}^{\a\b} 
+ \overset{-}{\mathcal N}{}^{\a\b},\label{SS}\\
\overset{a}{\mathcal S}{}^{\a\b} &:=& \overset{-}{\mathcal M}{}^{\a\b} 
- \overset{-}{\mathcal N}{}^{\a\b}.\label{SA}
\end{eqnarray}
The first object (\ref{AA}) represents a skewonic axion, whereas the
four other objects (\ref{RS}) to (\ref{SA}) describe contributions to
the principal reversible and irreversible (skewon) parts.  In the
reversible case, only the first irreducible piece of Eq.(\ref{1T})
survives and the fourth one of Eq.(\ref{4T}).

\section{Metric-dependent constitutive tensor}\label{Sec.4}

In general relativity as well as in teleparallelism, the existence of
a metric tensor $g$ is conventionally assumed. We choose the metric
signature $(+,-,-,-)$.  The question is then how the constitutive
tensor density $\chi^{\a\b}{}_{\m}{}^{\g\d}{}_{\n}$ can be express\-ed
in terms of the metric, namely in terms of its covariant and/or
contravariant components $g_{\a\b}$ and $g^{\c\d}$,
respectively. Since $\chi$ is a 6th order tensor of type\tvect{4}{2},
it seems reasonable to start with an ansatz of a purely contravariant
dimensionless 6th order tensor $K^{\a\b\m\g\d\n}$. Then the
contravariant metric components $g^{\g\d}$ are the only metric
components that have to be taken into account.  In general, one can
come up with a polynomial ansatz of arbitrary order in the metric
tensor. However, since the resulting constitutive tensor is of 6th
order, all indices of the metric factors except six should be
necessarily contracted. Recalling that a contraction of the covariant
and contravariant metric yields a Kronecker delta, we then find that a
polynomial of an arbitrary order is automatically reduced to a general
polynomial expression in the metric which is just {\it cubic} in
$g^{\a\b}$.

\subsection{A metric-dependent ansatz with parity conserving terms}

If we look only for parity conserving (even) pieces in $\chi$, the
totally antisymmetric unit tensor is not allowed in the polynomial
expression. Thus, for the 6th order tensor $K^{\a\b\m\g\d\n}$, the most
general cubic expression in the metric, omitting terms explicitly
symmetric in the index pairs $\a\b$ and $\g\d$, reads
\begin{align}\label{K_ansatz}
  K^{\a\b\m\g\d\n} &=\!\a_1g^{\a\m}g^{\b\g}g^{\d\n}+\a_2g^{\a\m}g^{\b\d}g^{\g\n}
  +\a_3 g^{\a\g}g^{\b\m}g^{\d\n}\nonumber\\
  &\,+\a_4g^{\a\g}g^{\b\d}g^{\m\n}+
  \a_5g^{\a\g}g^{\b\n}g^{\m\d}+\a_6g^{\a\d}g^{\b\m}g^{\g\n}\nonumber\\
 &\,+\a_7g^{\a\d}g^{\b\g}g^{\m\n}+
    \a_8g^{\a\d}g^{\b\n}g^{\g\m}+\a_9g^{\a\n}g^{\b\g}g^{\m\d}\nonumber\\
  &\,+\a_{10}g^{\a\n}g^{\b\d}g^{\m\g}\,,
\end{align}
where $\a_1,\dots,\a_{10}$ are constants. 

In order to tailor the ansatz \eqref{K_ansatz} for our constitutive
tensor $\chi^{\a\b}{}_{\m}{}^{\g\d}{}_{\n}$, we lower the indices $\m$
and $\n$ and, at the same time, we reorder terms with only one metric
tensor such that the metric tensor is sandwiched in between two
Kronecker deltas:
\begin{align}\label{K_ansatz1}
 K^{\a\b}{}_{\m}{}^{\g\d}{}_{\n}
  &=\,\,\a_1\d^{\a}_{\m}g^{\b\g}\d^{\d}_{\n}\hspace{4pt}+\a_2\d^{\a}_{\m}g^{\b\d}\d^{\g}_{\n}
    +\a_3\d^{\b}_{\m}g^{\a\g}\d^{\d}_{\n}\nonumber\\
  &\,+\a_4g^{\a\g}g^{\b\d}g_{\m\n}+
    \a_5\d^{\b}_{\n}g^{\a\g}\d_{\m}^{\d}+\a_6\d^{\b}_{\m}g^{\a\d}\d^{\g}_{\n}\nonumber\\
 &\,+\a_7g^{\a\d}g^{\b\g}g_{\m\n}+
    \a_8\d^{\b}_{\n}g^{\a\d}\d^{\g}_{\m}+\a_9\d^{\a}_{\n}g^{\b\g}\d_{\m}^{\d}\nonumber\\
  &\,+\a_{10}\d^{\a}_{\n}g^{\b\d}\d_{\m}^{\g}\,.
\end{align}

Since $\chi^{\a\b}{}_{\m}{}^{\g\d}{}_{\n}$ is antisymmetric in the
index pairs $\a\b$ and $\g\d$, we have to antisymmetrize $K$
correspondingly by subsequently putting brackets around the indices
$\a\b$ and $\g\d$, respectively. Also recall that the metric is
symmetric, hence its index ordering is optional:
\begin{align}\label{K_ansatz2}
  & K^{[\a\b]}{}_{\m}{}^{[\g\d]}{}_{\n}
    =\nonumber\\
  &\hspace{11pt}\a_1\d^{[\a}_{\,\,\m}g^{\b][\g}\d^{\d]}_{\n}\hspace{2pt}
    +\,\,\a_2\d^{[\a}_{\,\,\m}g^{\b][\d}\d^{\g]}_{\n}
    +\a_3\d^{[\b}_{\,\,\m}g^{\a][\g}\d^{\d]}_{\n}\nonumber\\
  &\,+\a_4g^{\g[\a}g^{\b]\d}g_{\m\n}+
    \a_5\d^{[\b}_{\,\,\n}g^{\a][\g}\d_{\m}^{\d]}
    +\a_6\d^{[\b}_{\,\,\m}g^{\a][\d}\d^{\g]}_{\n}\nonumber\\
  &\,+\a_7g^{\d[\a}g^{\b]\g}g_{\m\n}+
    \a_8\d^{[\b}_{\,\,\n}g^{\a][\d}\d^{\g]}_{\m}
    +\a_9\d^{[\a}_{\,\,\n}g^{\b][\g}\d_{\m}^{\d]}\nonumber\\
  &\,+\a_{10}\d^{[\a}_{\,\,\n}g^{\b][\d}\d_{\m}^{\g]}\,.
\end{align}
Note that the $\a_4$ and the $\a_7$ terms are already antisymmetric in
$\c\d$. Thus, there is no need to put brackets $[\;]$ around $\g\d$ or
$\d\g$, respectively. We can now collect all terms by recalling the
antisymmetry in both index pairs $\a\b$ and $\g\d$. We find, without
any further intermediary step, the compact equation
\begin{align}\label{K_ansatz3}
 K^{[\a\b]}{}_{\m}{}^{[\g\d]}{}_{\n}=
&\hspace{10pt}(\hspace{8pt}\a_4-\a_7)\,g^{\g[\a}g^{\b]\d}g_{\m\n}\nonumber\\
&+(\hspace{8pt}\a_1-\a_2-\a_3+\a_6)\,
\d_{\,\,\m}^{[\a}g^{\b][\g}\d^{\d]}_{\,\n}\nonumber\\
&+(-\a_5+\a_8+\a_9-\a_{10})\,\d^{[\a}_{\,\,\n}g^{\b][\g}\d_{\,\m}^{\d]}\,.
\end{align}

We know that $\chi$ is a density, like $\sqrt{-g}$, with
$g:= \det g_{\mu\nu}$.  Accordingly, we can identify the parity even
part of the constitutive tensor density as follows,
\begin{equation}\label{identify}
  ^{\text{even}\!}\chi^{\a\b}{}_{\m}{}^{\g\d}{}_{\n}(g)
  =\frac{\sqrt{-g}}{\varkappa}\, K^{[\a\b]}{}_{\m}{}^{[\g\d]}{}_{\n}\,,
\end{equation}
with $\varkappa$ as the gravitational constant. In our formalism up to
now, including the Tonti diagram, $\varkappa$ was put to $1$ for
simplicity. Here we do display it for clarity. Thus, we find, with the
three constants,
\begin{equation}
\left.\begin{aligned}
\b_1 &:= \a_4-\a_7\\
\b_2 &:= \a_1-\a_2-\a_3+\a_6 \\
\b_3 &:= -\,\a_5+\a_8+\a_9-\a_{10},\end{aligned}\right\}\!
\end{equation}
the expression
\begin{align}
  {}^{\text{even}}\chi^{\a\b}{}_{\m}{}^{\g\d}{}_{\n}(g)
  =&\frac{\sqrt{-g}}{\varkappa}\left(\b_1\,g^{\g[\a}g^{\b]\d}g_{\m\n}\right.
     \nonumber\\
&  \hspace{-5pt}\left. +\,\b_2\,\d^{[\a}_{\,\,\m}g^{\b][\g}\d^{\d]}_{\n} 
+ \b_3\,\d^{[\a}_{\,\,\n}
     g^{\b][\g}\d_{\m}^{\d]}\right)\!.\label{even_endresult}
\end{align}
This proves our ansatz with three independent constants in
\cite[Eq.(80)]{Itin:2016nxk}.\footnote{In \cite{Itin:2016nxk}, there
  was a typo in Eq.(80): the last plus sign $+$ in this formula should
  have been a minus sign $-$. Moreover, for simplification, we changed
  now our conventions with respect to the $\b$'s slightly:
  $4\b^{\text{old}}_1 = \b_1$, $8\b^{\text{old}}_2 = -\b_2$,
  $8\b^{\text{old}}_3 = -\b_3$.}  It is, indeed, the most general
expression which, up to a factor of $\sqrt{-g}$, turns out to be a
{\it cubic} polynomial in the metric.

\subsection{Parity violating terms}

Already for quite some time, see Pellegrini \& Plebanski
\cite{Pellegrini} and M\"uller-Hoissen \& Nitsch \cite{Muller:1983},
also parity violating (odd) constitutive tensors for torsion square
Lagrangians have been considered in the literature; for recent
reviews, see
\cite{Baekler:2011,Karananas:2014pxa,Blagojevic:2018dpz,Obukhov:2018}. There
it has been shown that {\it two} parity odd terms occur, which are
both linear in the totally antisymmetric Levi-Civita symbol
$\eps^{\a\b\c\d}$, which is, as we may recall, a tensor density:
\begin{equation}\label{met1}
  {}^{\rm odd} \chi^{\a\b}{}_{\m}{}^{\g\d}{}_{\n}\,(g) =
{\frac{1}{\varkappa}}\left\{
  {\beta}_4\,\eps^{\a\b\g\d}\,g_{\mu\n} + {\beta}_5\,
  \eps^{\a\b[\g}{}_{[\m}\,\delta^{\d]}_{\,\nu]}\right\}.
\end{equation}
One can also take into account such additional parity violating terms
in the framework of the Poincar\'e gauge theory of gravity. In that
more general theory, it leads to interesting new cosmological models
\cite{Baekler:2010,Ho:2015ulu} and to new gravitational wave solutions
\cite{Blagojevic:2017ssv,Obukhov:2006gea}.

We would like to understand the generality of the result in
\eqref{met1} in a similar way as we did it for the parity even
case. For parity reasons, the Levi-Civita symbol has to be linearly or
in odd powers in the corresponding ansatz.  Since again, as in the
even part of $\chi^{\a\b}{}_{\mu}{}^{\g\d}{}_\nu$, we start with an
tensor expression of type $\tvect{6}{0}$, we need additionally a
tensor of type $\tvect{2}{0}$. Clearly the contravariant components of
the metric $g^{\a\b}$ qualify for such a purpose. Thus, the simplest
possible ansatz, see also \eqref{met1}, is to start with a 6th order
tensor density as follows:
\begin{align}\label{eq5}
{M}^{\a\b\m\g\d\n}&=\m_1g^{\a\m}\ve^{\b\g\d\n}+\m_2g^{\a\g}\ve^{\b\m\d\n}+
\m_3g^{\a\d}\ve^{\b\m\g\n}\nonumber\\
&+\m_4g^{\a\n}\ve^{\b\m\g\d}+\m_5g^{\b\m}\ve^{\a\g\d\n}+
\m_6g^{\b\g}\ve^{\a\m\d\n}\nonumber\\
&+\m_7g^{\b\d}\ve^{\a\m\g\n} +\m_8g^{\b\n}\ve^{\a\m\g\d}+
\m_9g^{\m\g}\ve^{\a\b\d\n}\nonumber\\
&+\m_{10}g^{\m\d}\ve^{\a\b\g\n}+
\m_{11}g^{\m\n}\ve^{\a\b\g\d}+\m_{12}g^{\g\n}\ve^{\a\b\m\d}\nonumber\\
&+\m_{13}g^{\d\n}\ve^{\a\b\m\g}\,.
\end{align}
The $\m_1,\m_2,\dots, \m_{13}$ are arbitrary constants. We now lower
again the indices $\m$ and $\n$ and antisymmetrize at the same time
with respect to the index pairs $\a\b$ and $\c\d$, respectively:
\begin{align}
  &{M}^{[\a\b]}{}_{\m}{}^{[\g\d]}{}_{\n} =
  \m_{11}g_{\m\n}\ve^{\a\b\g\d} \nonumber\\
  &+ (\m_1-\m_5)\,\d_{\m}^{[\a}\ve^{\b]\g\d}{}_{\n}+
  (\m_4-\m_{8})\,\d_{\n}^{[\a}\ve^{\b]\g\d}{}_{\m}\nonumber\\
  &+(\m_9-\m_{10})\,\d_{\m}^{[\g}\ve^{\d]\a\b}{}_{\n}-
  (\m_{12}-\m_{13})\,\d_{\n}^{[\g}\ve^{\d]\a\b}{}_{\m}\nonumber\\
  & + {\frac 12}(-\m_2+\m_3+\m_6-\m_7)\,
  \left(g^{\g[\a}\ve^{\b]\d}{}_{\m\n}
    -g^{\d[\a}\ve^{\b]\g}{}_{\m\n}\right)\,.\label{MdoubleAnti_2}
\end{align}

Having established the required antisymmetries, we can now identify
the odd piece of the constitutive tensor as
\begin{align}
^{\text{odd}}\chi^{\a\b}{}_{\m}{}^{\g\d}{}_{\n}\,(g)=\frac{1}{\varkappa}\,
    {M}^{[\a\b]}{}_{\m}{}^{[\g\d]}{}_{\n}\,.
\label{MdoubleAnti_3}
\end{align}

Here we have to recall an algebraic trick that we had already applied
in discussing the electromagnetic 4th order constitutive tensor, see
\cite[Eq.(D.1.65)]{Birkbook}. Since in four dimensions the indices
have always four values, any expression antisymmetrized over five
indices vanishes identically. Thus, for an arbitrary tensor
$T^{\a\b\g\d\varepsilon\cdots} _{\;\cdots}$, we have
$T^{[\a\b\g\d\varepsilon]\cdots}_{\;\cdots}\equiv 0$.  Accordingly, we
have, for example, $T^{[\rho}\varepsilon^{\a\b\g\d]}=0$ or, if we
evaluate the brackets,
\begin{align}
  T^\r\ve^{\a\b\g\d}\equiv T^\a\ve^{\r\b\g\d}+T^\b\ve^{\a\r\g\d}
  +T^\g\ve^{\a\b\r\d}+T^\d\ve^{\a\b\g\r}\,.
\end{align}
Looking now at \eqref{MdoubleAnti_2}, we have typically
$\d^{[\a}_{\;\mu}\ve^{\b]\g\d\nu}$. With $T^\a{}_\mu=\d^\a_\mu$, we
find
\begin{align}\label{identity_1}
\d^{[\a}_{\;\mu}\ve^{\b]\g\d\nu}\equiv\d^{[\b}_{\;\mu}\ve^{\a]\g\d\nu}+\d^{\g}_\mu\ve^{\b\a\d\nu}
+\d^{\d}_\mu\ve^{\b\g\a\nu}+\d^{\nu}_\mu\ve^{\b\g\d\a}\,.
\end{align}
By such methods, we can derive several identities for the Levi-Civita
symbol in 4 dimensions:
\begin{align}
  \varepsilon^{\a\b[\g}{}_{(\m}\,\delta^{\d]}_{\n)} 
  + \varepsilon^{\g\d[\a}{}_{(\m}\,\delta^{\b]}_{\n)} 
  &\equiv {\frac 12}\,\varepsilon^{\a\b\g\d}\,g_{\m\n},\label{id1}\\
  \varepsilon^{\a\b[\g}{}_{[\m}\,\delta^{\d]}_{\n]} 
  + \varepsilon^{\g\d[\a}{}_{[\m}\,\delta^{\b]}_{\n]} 
  &\equiv 0,\label{id2}\\
  g^{\g[\a}\varepsilon^{\b]\d}{}_{\m\n} -
  g^{\d[\a}\varepsilon^{\b]\g}{}_{\m\n} 
  &\equiv 2\,\varepsilon^{\a\b[\g}{}_{[\m}\,\delta^{\d]}_{\n]}\,.\label{id3}
\end{align}

Let us come back to \eqref{MdoubleAnti_3} with
\eqref{MdoubleAnti_2}. We decompose the $\d\ve$-terms in symmetric and
antisymmetric pieces with respect to the lower indices:
\begin{align}
  \d_{\;\mu}^{[\a}\ve^{\b]\g\d}{}_{\n} &=\d_{(\mu}^{[\a}\ve^{\b]\g\d}{}_{\n)} 
  + \d_{[\mu}^{[\a}\ve^{\b]\g\d}{}_{\n]},\label{t1*}\\
  \d_{\;\mu}^{[\g}\ve^{\d]\a\b}{}_{\n} &=\d_{(\mu}^{[\g}\ve^{\d]\a\b}{}_{\n)} 
  + \d_{[\mu}^{[\g}\ve^{\d]\a\b}{}_{\n]}\,.\label{t1}
\end{align}
As a result, the $\d\ve$-terms in {\it the second and third lines of}
(\ref{MdoubleAnti_2}) can be rearranged as follows:
\begin{align}
  &\left(\m_1 - \m_5 + \m_4 - \m_8\right)\d_{(\mu}^{[\a}\ve^{\b]\g\d}{}_{\n)}
 \nonumber\\ &+ \left(\m_9 - \m_{10} - \m_{12} + \m_{13}\right)
\d_{(\mu}^{[\g}\ve^{\d]\a\b}{}_{\n)}\nonumber\\
  &+ \left(\m_1 - \m_5 - \m_4 + \m_8\right)\d_{[\mu}^{[\a}\ve^{\b]\g\d}{}_{\n]}
\nonumber\\  &+ \left(\m_9 - \m_{10} + \m_{12} -
  \m_{13}\right)\d_{[\mu}^{[\g}\ve^{\d]\a\b}{}_{\n]}.\label{t2}
\end{align}

Next, we write 
\begin{align}
  \d_{(\mu}^{[\a}\ve^{\b]\g\d}{}_{\n)} &= 
\frac{1}{2}\left(\d_{(\mu}^{[\a}\ve^{\b]\g\d}{}_{\n)} + 
  \d_{(\mu}^{[\g}\ve^{\d]\a\b}{}_{\n)}\right)\nonumber\\
&\hspace{3pt}+ \frac{1}{2}\left(\d_{(\mu}^{[\a}\ve^{\b]\g\d}{}_{\n)}-
  \d_{(\mu}^{[\g}\ve^{\d]\a\b}{}_{\n)}\right),\label{t3}\\
\d_{(\mu}^{[\g}\ve^{\d]\a\b}{}_{\n)} &=
\frac{1}{2}\left(\d_{(\mu}^{[\g}\ve^{\d]\a\b}{}_{\n)} +
  \d_{(\mu}^{[\a}\ve^{\b]\g\d}{}_{\n)}\right)\nonumber\\ 
&\hspace{3pt}+ {\frac 12}\left(\d_{(\mu}^{[\g}\ve^{\d]\a\b}{}_{\n)} -
  \d_{(\mu}^{[\a}\ve^{\b]\g\d}{}_{\n)}\right).\label{t4}
\end{align}
Consequently, the expression (\ref{t2}) transforms into
\begin{align}
&   \left(\m_1 - \m_5 - \m_4 + \m_8 - \m_9 + \m_{10} - \m_{12} 
+ \m_{13}\right)\d_{[\mu}^{[\a}\ve^{\b]\g\d}{}_{\n]}\label{t5}\nonumber\\
&  +  {\frac 12}\left(\m_1 - \m_5 + \m_4 - \m_8 - \m_9 + \m_{10} + \m_{12} 
- \m_{13}\right)\times
 \nonumber\\
&\hspace{40pt}\left(\d_{(\mu}^{[\a}\ve^{\b]\g\d}{}_{\n)} 
- \d_{(\mu}^{[\g}\ve^{\d]\a\b}{}_{\n)}\right)\nonumber\\
 & +{\frac 12} \left(\m_1 - \m_5 + \m_4 - \m_8 + \m_9 - \m_{10} 
- \m_{12} + \m_{13}\right)\times\nonumber\\
& \hspace{40pt}\left(\d_{(\mu}^{[\a}\ve^{\b]\g\d}{}_{\n)} 
+ \d_{(\mu}^{[\g}\ve^{\d]\a\b}{}_{\n)}\right)\,.
\end{align}
where we used the identity (\ref{id2}). We recall that \eqref{t5}
belongs to the second and third line of \eqref{MdoubleAnti_2}.

In the last step, it remains to use (\ref{id1}) in the last line of
(\ref{t5}) and to substitute (\ref{id3}) into the last line of
(\ref{MdoubleAnti_2}). Then, with the constants
\begin{align}
  & \b_4:=\,{\frac 14}\left(\mu_1 - \mu_5 + \mu_4 - \mu_8 + \mu_9 
    - \mu_{10} - \mu_{12} + \mu_{13} +
    4\m_{11}\right)\,, \nonumber\\ 
  &\b_5:=\mu_1 - \mu_5 - \mu_4 + \mu_8 - \mu_9 + \mu_{10}
\nonumber\\
  &\hspace{20pt} +
    \mu_{12}     - \mu_{13} + \m_2-\m_6-\m_3+\m_7\,, \nonumber\\ 
  &\b_6:= {\frac 12}\left(\mu_1 - \mu_5 + \mu_4 - \mu_8
    - \mu_9 + \mu_{10} + \mu_{12} - \mu_{13}\right)\,, \nonumber
\end{align}
the result eventually reads
\begin{align}\label{odd_endresult}
  {}^{\text{odd}}\chi^{\a\b}{}_{\m}{}^{\g\d}{}_{\nu}(g) =& 
  \frac{\a}{\varkappa}\left[ {\beta}_4\,\ve^{\a\b\g\d}\,g_{\m\n} + {\beta}_5\,
    \ve^{\a\b[\g}{}_{[\m}\,\delta^{\d]}_{\n]}\right.\nonumber\\
  &\left. + \b_6\left(\delta_{(\m}^{[\a}\ve^{\b]\g\d}{}_{\n)} 
    - \delta_{(\m}^{[\g}\ve^{\d]\a\b}{}_{\n)}\right)\right].
\end{align}
Here $\a$ is dimensionless pseudoscalar. It is necessary in order to
take into account the transformation law of this part under improper
reflections.

Therefore, the most general parity-odd part of the constitutive tensor
in the end boils down to just the three independent terms with $\b_4$,
$\b_5$, and $\b_6$. The $\b_4$-term, being totally antisymmetric in
$\a\b\!\g\d$, represents an $\b_4$-axion, whereas the $\b_5$-term
describes a $\b_5$-axion. They both represent reversible processes. In
contrast, the $\b_6$-term corresponds to {\em irreversi\-bil\-ity.} When
contracted by $F_{\a\b}{}^{\m}$ and $F_{\g\d}{}^{\n}$, this term drops
out from the general teleparallel Lagrangian. Moreover, one can reduce
the number of independent terms in the Lagrang\-ian by making use of the
Nieh-Yan topological invariant
\cite{Nieh:1981ww,Nieh:2007zz,Obukhov:1997}.

\subsection{The general case and its irreducible decomposition}

The general constitutive tensor
\begin{equation}\label{chigen}
  \chi^{\a\b}{}_{\m}{}^{\g\d}{}_\nu(g) =   {}^{\rm
even}\chi^{\a\b}{}_{\m}{}^{\g\d}{}_\nu(g)
  + {}^{\rm odd}\chi^{\a\b}{}_{\m}{}^{\g\d}{}_\nu(g)
\end{equation}
encompasses the parity-even piece of Eq.\eqref{even_endresult} and
the parity-odd piece of Eq.\eqref{odd_endresult},

Computing the irreducible parts, we find
\begin{equation}\label{met6}
  {}^{[2]}\chi^{\a\b}{}_\m{}^{\g\d}{}_\n(g)= 
  {}^{[6]}\chi^{\a\b}{}_\m{}^{\g\d}{}_\n(g)=0\,,
\end{equation}
whereas the nontrivial pieces read
\begin{align} {}^{[1]}\chi^{\a\b}{}_\m{}^{\g\d}{}_\n(g) 
&=\nonumber\\ &\hspace{-40pt}  \frac{\sqrt{-g}}{\varkappa}
\Big[ \b_1\,g^{\g[\a}g^{\b]\d}g_{\m\n} + (\b_2 +
            \b_3)\,\d_{(\m}^{[\a}g^{\b][\g}\d^{\d]}_{\n)} \Big],
\label{chim1}\\
  {}^{[3]}\chi^{\a\b}{}_\m{}^{\g\d}{}_\n(g) &=\frac{\b_6}{\varkappa}
\left(\delta_{(\m}^{[\a}\ve^{\b]\g\d}{}_{\n)} 
- \delta_{(\m}^{[\g}\ve^{\d]\a\b}{}_{\n)}\right),\label{chim3}\\
  {}^{[4]}\chi^{\a\b}{}_\m{}^{\g\d}{}_\n(g) &= \nonumber\\
&\hspace{-40pt} \frac{1}{\varkappa}\Big[(\b_2 -
  \b_3)\,\sqrt{-g}\d_{[\m}^{[\a}g^{\b][\g}\d^{\d]}_{\n]} 
+ {\beta}_5\,\ve ^{\a\b[\g}{}_{[\m}\,\delta^{\d]}_{\n]}\Big],\label{chim4}\\
  {}^{[5]}\chi^{\a\b}{}_\m{}^{\g\d}{}_\n(g) &= \frac{{\beta}_4}{\varkappa}\,
\ve^{\a\b\g\d}\,g_{\m\n}.\label{chim5}
\end{align}
Summarizing, a general metric-dependent constitutive tensor
encompasses two principal ${}^{[1]}\chi$ and ${}^{[4]}\chi$ parts, one
axion ${}^{[5]}\chi$ part, and one skewon ${}^{[3]}\chi$ part.

Calculating the traces \eqref{trm}--\eqref{trl} of the constitutive tensor
(\ref{chigen}), we obtain
\begin{equation}\label{met3}
m^{\a\b}=\frac{ 3\sqrt{-g}}{4\varkappa}(2\b_1-3\b_2-\b_3)g^{\a\b}\,,
\end{equation}
and
\begin{equation}\label{met4}
n^{\a\b}=\frac{ 3\sqrt{-g}}{4\varkappa}(2\b_1-\b_2-3\b_3)g^{\a\b}\,.
\end{equation}
The antisymmetric tensors vanish, $k^{\a\b}=l^{\a\b}=0$. The parity
odd terms drops out altogether in the trace building.

\subsection{Specific Lagrangians. }

\subsubsection{GR$_{||}$: the teleparallel equivalent of GR}

We found in the framework of the {\it teleparallel equivalent
  GR$_{||}$ of general relativity} (GR) for the $\b$'s in
\cite[Eq.(88)]{Itin:2016nxk}\footnote{There occurred an computing
  error. The correct values in the old conventions are
  $\beta^\text{old}_1 =-\frac 14\,,\;\beta^\text{old}_2 =\frac 12\,,\;
  \beta^\text{old}_3 = -\frac 14$.},
\begin{equation}\label{met_GR}
\left.
\begin{aligned}
\b_1 &= -\,1\,,& \b_2&= -\,4\,,& \b_3 &= 2 \\
\b_4 &= 0\,,& \b_5 &= 0\,,& \b_6 &= 0
\end{aligned}\quad\right\}\quad\text{GR$_{||}\,.$}
\end{equation}
Accordingly, the three $\b$'s are related to GR in a quite
definite way, and we get a feeling for their interpretation.

The GR$_{||}$ Lagrangian is distinguished from the other
tele\-parallelism Lagrangians as being {\it locally} Lorentz
invariant, see, e.g., Cho \cite{Cho:1975dh} and M\"uller-Hoissen \&
Nitsch \cite{Muller:1983}. Any other additional term of even parity in
the gravitational Lagrangian removes this local invariance.

Ferrarro and Guzm\'an \cite{Ferraro:2016wht} called the constitutive
tensor $ \chi^{\a\b}{}_{\m}{}^{\g\d}{}_\nu(g)$ of \eqref{chigen}
together with the parameters of \eqref{met_GR} the {\it supermetric}. In
fact, up to a conventional factor of $-2$, our result agrees with
their Eq.(17) in \cite{Ferraro:2016wht}.

\subsubsection{Von der Heyde Lagrangian}

Earlier, different parity even pieces of torsion square Lagrangians
were compared by Muench et al.\ \cite{Muench:1998}, see also the
literature cited therein. A particular role played in these
considerations the torsion square piece of a Lagrangian of {\it von
  der Heyde} \cite{vdH:1976}. Its constitutive tensor \cite{Hehl:1979} reads
\begin{align}\label{chi_vdH1}
  ^{\text{vdH}}\chi^{\a\b}{}_{\m}{}^{\c\d}{}_\n= -\,\frac{2\sqrt{-g}}{\varkappa}
  \left(g^{\g[\a}g^{\b]\d}g_{\m\n}
  + 2\delta^{[\a}_{\,\,\m} g^{\b][\g}\delta_{\n}^{\d]}\right)\!\!.
\end{align}
For the corresponding constitutive law, we have
\begin{align}
\check{H}^{\a\b}{}_{\m} &= {{\frac
{1}{2}}}\,^{\text{vdH}}\chi^{\a\b}{}_{\m}{}^{\g\d}{}_\n
\,F_{\g\d}{}^\n \nonumber\\
&= -\,\frac{\sqrt{-g}}{\varkappa}\left(F^{\a\b}{}_{\m}
+ 2\delta^{[\a}_{\,\,\mu} F^{\b]}{}_{\!\g}{}^\g\right).\label{gr-kot9'}
\end{align}
This law, in the teleparallel case, leads already to the correct
Newtonian approximation. The constitutive tensor \eqref{chi_vdH1}
carries the following $\b$-values:
\begin{align}\label{met_vdH}
\left.\begin{aligned}
\b_1 &= -\,2\,,& \b_2&= -\,4\,,& \b_3 &=0\\
\b_4 &= 0\,,& \b_5 &= 0\,,& \b_6 &= 0
\end{aligned}\quad\right\}\quad\text{vdH}
\end{align}
 As we saw, the teleparallel equivalent GR$_{||}$
has a slightly different constitutive law (\ref{met_GR}). 

\subsubsection{Torsion-square Lagrangian in nonlocal gravity (NLG)}

In a nonlocal extension of GR---{\it Nonlocal Gravity (NLG)} ---the
starting point \cite{Hehl:2008eu,Hehl:2009es} was the parameter set of
GR$_{||}$, namely \eqref{met_GR}. Later, Mashhoon found it necessary,
to enlarge it by adding a supplementary piece proportional to the
axial\footnote{In exterior calculus, we have
  $^{(3)}F^\a=\frac 13 e^\a\rfloor(F^\b\wedge\vartheta_\b)$.} torsion
$^{(3)}F_{\a\b}{}^\mu$, see \cite[Eq.(6.109)]{Mashhoon:2017}. This
yields, with an unknown dimensionless parameter $\check{p}$,
\begin{align}\label{met_Mashhoon}
  \left.\begin{aligned}
      \b_1 &= -\,1,& \b_2&= -\,4,& \b_3 &= 2,\\
      \b_4 &= -\,\frac{\check{p}}{6}\,,& \b_5 &= -\,4\,
      \frac{\check{p}}{6}\,,& \b_6 &=  2\,\frac{\check{p}}{6}.
\end{aligned}\quad\right\}\quad\text{Mashh.\hspace{-2pt}}
\end{align}    
It is amazing that
$(\b_4\,,\,\b_5\,,\,\b_6)=\frac{\check{p}}{6}
(\b_1\,,\,\b_2\,,\,\b_3)$.
Probably this means that Mashhoon's additional parity odd term is
locally Lorentz invariant. Incidentally, there is a conventional
factor between Mashhoon's $\check{p}$ and ours.  

According to Mashhoon,\footnote{Private communication, August 2018.}
``...the mere fact of postulating a {\sl nonlocal} constitutive relation
violates local Lorentz invariance.'' Still, the version of NLG of 2009
is based on the reversible and locally Lorentz invariant parameter set
\eqref{met_GR} and Mashhoon's version corresponds to an
irreversible and possibly also locally Lorentz invariant parameter
set, see the 2nd line of \eqref{met_Mashhoon}.

\section{Propagation of gravitational waves}\label{Sec.5}

In this section, we discuss the wave propagation in our premetric
teleparallel theory (TG) in linear approximation.

\subsection{Geometric optics approximation}

We start with the source-free field equations 
\begin{equation}\label{wf0}
 dH_\a=0\qquad {\rm and} \qquad d F^\a =0\,.
\end{equation}
Notice that the source term $\Sigma$ in (\ref{gr-kot2}) contains two
independent expressions: the matter energy-momentum current
${}^{({\rm m})}\Sigma$ and the gravitational energy-momentum current
${}^{(\vt)}\!\Sigma$. In mat\-ter-free regions, we have
${}^{({\rm m})}\!\Sigma=0$. Similar to standard GR, we will assume, in
addition, ${}^{(\vt)}\!\Sigma=0$.  This requirement is applicable for
small waves and means linearization of the field equations.

In the geometric optics approximation (and, equivalently, in
Hadamard's approach, see \cite{Birkbook}), one derives the linear
system
\begin{equation}\label{wf1}
  \check{H}^{\mu\nu}{}_{\a}\,q_\nu = 0\qquad {\rm and} \qquad
  \epsilon^{\mu\nu\rho\sigma}\,F_{\rho\sigma}{}^\a\,q_\nu = 0\,.
\end{equation}
Here the components $q_\nu$ of the wave covector are determined by the
differential of the phase $\varphi$ function of the wave:
$d\varphi = q_\nu\vartheta^\nu$.

The second equation of (\ref{wf1}) has the solution
\begin{equation}
  F_{\rho\sigma}{}^\a = A_\rho{}^\a\,q_\sigma - A_\sigma{}^\a\,q_\rho\,,\label{wf3}
\end{equation}
with an arbitrary tensor $A_\b{}^\a$. It is a gravitational analog of
the electromagnetic potential. We observe the {\it gauge invariance}
of this gravitational potential. An expression of the form
$A_\b{}^\a=q_\b C^\a$ does not contribute to the field strength
$F_{\rho\sigma}{}^\a$.  In other words, the model is invariant under
the transformations
\begin{equation}
A_\b{}^\a\to A_\b{}^\a + q_\b C^\a\,,\label{wf4}
\end{equation}
with an arbitrary vector $C^\a$. 

We substitute (\ref{wf3}) into (\ref{wf1}) and use the constitutive
relation
\begin{equation}\label{const}
  \check{H}^{\mu\nu}{}_{\a} = {\frac 12}\,\chi^{\mu\nu}{}_{\a}{}^{\rho\sigma}{}_\b\,
  F_{\rho\sigma}{}^\b\,.
\end{equation}
This yields the {\it characteristic equation}
\begin{equation}
M^\mu{}_\alpha{}^\nu{}_\beta\,A_\nu{}^\b = 0\,,\label{Ma0}
\end{equation}
with fourth order {\it characteristic  tensor}
\begin{equation}\label{Mabmn}
  M^\mu{}_{\alpha}{}^\nu{}_\beta := 
  \chi^{\mu\rho}{}\!_{\a}{}^{\nu\sigma}{}\!_\b\,q_\rho\,q_\sigma.
\end{equation}
Inserting here the decomposition (\ref{DeComp}) of the constitutive
tensor, we observe that the axion-type parts
${}^{[5]}\chi^{\mu\rho}{}_{\a}{}^{\nu\sigma}{}_\b$ and
${}^{[6]}\chi^{\mu\rho}{}_{\a}{}^{\nu\sigma}{}_\b$ do not enter the
tensor $M^\mu{}_{\alpha}{}^\nu{}_\beta$. Consequently, these parts do
not contribute to the wave propagation as it was already outlined in
Table 1.

In the reversible case, the characteristic tensor
$M^\mu{}_\alpha{}^\nu{}_\beta$ satisfies the symmetry relations
\begin{equation}\label{M-sym}
  M^\mu{}_\alpha{}^\nu{}_\beta = M^\nu{}_\beta{}^\mu{}_\alpha\,.
\end{equation}

A generalized Fresnel equation can be derived as the condition for the
solvability of the equation (\ref{Ma0}) along the lines of the
algebraic computations of Itin \cite{Itin:2014gba}.

\subsection{Dispersion relation---general facts}

Observe that the characteristic equation
\begin{equation}\label{eq1}
E^\mu{}_\alpha =0\,,\quad\text{with}\quad E^\mu{}_\alpha
:=M^\mu{}_\alpha{}^\nu{}_\beta\,A_\nu{}^\b\,,
\end{equation}
represents 16 equations for the 16 variables $A_\nu{}^\b$.

For a compact representation of this system, we denote a pair of upper
and lower indices by a multi-index
\begin{equation}\label{spec0}
\{{}^\mu{}_\alpha{}\}=I\qquad I=1,\cdots 16\,.
\end{equation}
In this notation, the system (\ref{eq1}) reads
\begin{equation}\label{eq1a}
  E^I=M^{IJ}\,A_J = 0\,.
\end{equation}

The gauge transformation (\ref{wf4}) can be rewritten as
\begin{equation}\label{eq1b}
A_J \to A_J+Q_J\,,\qquad {\rm where}\qquad Q_J=q_\b C^\a\,.
\end{equation}
The 4 linearly independent vectors $C^\a$ imply that there are
likewise 4 linearly independent vectors $Q_J$. The identities
\begin{equation}\label{spec1}
M^\mu{}_\alpha{}^\nu{}_\beta \,q_\mu=M^\nu{}_\alpha{}^\mu{}_\beta \,q_\mu=0
\end{equation}
translate into
\begin{equation}\label{eq1c}
M^{IJ}\,Q_J =M^{JI}\,Q_J= 0\,.
\end{equation}
These equations can be understood as 4 linear relations between the
rows (and the columns) of the matrix $M^{IJ}$.  Consequently, the
matrix $M^{IJ}$ has rank $12=16-4$. The condition for the existence of
non-trivial solutions of Eq.(\ref{eq1a}) reads now as
\begin{equation}\label{eq11c} {}^{(4)}\!\!\text{Adj}(M)= 0\,.
\end{equation}
We used here the fourth order adjoint of the matrix $M^{IJ}$. It can
be constructed by evaluating the determinants of the matrices that are
left in the $16\times 16$ matrix $M^{IJ}$ by removing four rows and
four columns.  Consequently ${}^{(4)}\!\!\text{Adj}(M)$ is a set of
12th order polynomials of the entries $M^{IJ}$. Accordingly, it is a
set of 12th order polynomials of the variables $q_\a$.  The 4th
order adjoint can be written as a tensor with eight free indices:
\begin{equation}\label{eq1d}
{}^{(4)}\!\!\text{Adj}(M)_{I_1\cdots I_4\,J_1\cdots J_4}= 0\,.
\end{equation}
In electromagnetism with only one gauge invariance constraint, there
appears the 1st order adjoint matrix $\text{Adj}(M)_{\a\b}$. This
matrix is expressed as a scalar function multiplied by a tensor
product of $q_\a$, namely $\text{Adj}(M)_{\a\b}=\lambda(q)q_\a q_\b$.
Thus, the electromagnetic dispersion relation takes the form
$\lambda(q)=0$.

Similarly, in the case of the gravitational equation (\ref{eq1d}), we
have
\begin{equation}\label{eq1e} {}^{(4)}\!\!\text{Adj}(M)_{I_1\cdots
    I_4\,J_1\cdots J_4}= \Lambda(Q) Q_{I_1}\cdots Q_{I_4}Q_{J_1}\cdots
  Q_{J_4}=0\,.
\end{equation}
Thus, the gravitational dispersion relation takes the form 
\begin{equation}\label{eq1f}
\Lambda(q)=0. 
\end{equation}
Here $\Lambda(q)$ is a homogeneous polynomial of the order $16=24-8$
in the variable $q_\a$.  We recall that in the electromagnetic case,
the corresponding form is of 4th order. This fact yields birefringence
in the wave propagation. In 3-dimensional elasticity theory, the
dispersion relation is of 6th order, with 3 different waves in
general.  In our generalized gravitational model, there are 8
different waves in general.

In the simplest case, Eq.(\ref{eq1f}) reads
$\left(g^{\a\b} q_\a q_\b\right)\!^8=0$. We discuss the metrical case
in Sec.\ref{Sec.6}.

\subsection{Dispersion relation decomposed}

In order to clarify the nature of the equation (\ref{eq1}), we
decompose the solution under the $\text{GL}(4,\mathbb{R})$ irreducibly
into the scalar $A\d^\b_\a$ and the traceless part $\ANT_\a{}^\b$,
\begin{equation}\label{spec2}
  A_\a{}^\b=\ANT_\a{}^\b+A\d^\b_\a\,,\;\,\text{with}\;\, A:=\frac 14
  A_\c{}^\c\;\,\text{and}\;\, \ANT_\a{}^\a=0\,.
\end{equation}
Consequently, Eq.(\ref{eq1}) decomposes into
\begin{equation}\label{spec3}
  E^\mu{}_\a = M^\mu{}_\alpha{}^\nu{}_\beta\ANT_\nu{}^\b 
  +  M^\mu{}_\alpha{}^\beta{}_\beta\,A = 0\,.
\end{equation}

A similar decomposition can be performed with $E^\mu{}_\a$. Thus, we
find the $1+15$ equations
\begin{equation}\label{spec2a}
E^\a{}_\a=0\, \qquad{\rm and}\qquad E^\mu{}_\a-\frac 14 E^\g{}_\g\,\d^\mu_\a =0\,
\end{equation}
or, explicitly, one equation
\begin{equation}\label{spec2b}
M^\a{}_\alpha{}^\nu{}_\beta\ANT_\nu{}^\b + M^\a{}_\alpha{}^\beta{}_\beta\,A = 0\,
\end{equation}
and 15 equations
\begin{align}
  &\left(M^\mu{}_\alpha{}^\nu{}_\beta-\frac 14\d^\mu_\a
    M^\g{}_\g{}^\nu{}_\beta\right)
    \ANT_\nu{}^\b \nonumber\\ 
  &\hspace{40pt}+ \left(M^\mu{}_\alpha{}^\beta{}_\beta
    -\frac 14\d^\mu_\a  M^\g{}_\g{}^\beta{}_\beta\right)A= 0\,.\label{spec2c}
\end{align} 
Substituting (\ref{spec2b}) into (\ref{spec2c}), we derive the
algebraic system for the traceless variable:
\begin{equation}
N^\mu{}_\alpha{}^\nu{}_\beta\!\ANT_\nu{}^\b = 0,\label{NA}
\end{equation}
where we introduced 
\begin{equation}
  N^\mu{}_\alpha{}^\nu{}_\beta := M^\mu{}_\alpha{}^\nu{}_\beta 
  M^\rho{}_\rho{}^\sigma{}_\sigma
  - M^\mu{}_\alpha{}^\rho{}_\rho M^\sigma{}_\sigma{}^\nu{}_\beta.\label{N}
\end{equation}
The latter tensor is apparently traceless for both pairs of indices:
\begin{equation}
N^\mu{}_\alpha{}^\rho{}_\rho = 0,\qquad N^\sigma{}_\sigma{}^\nu{}_\beta = 0.\label{Nt}
\end{equation}
Accordingly, we find a system of 15 algebraic equations for 15
variables $\ANT_\nu{}^\mu$.  After one solves the homogeneous equation
(\ref{NA}), one immediately finds the scalar $A$ from (\ref{spec2b}).

\subsection{Scalar waves as a special case}

We consider now a special case of pure scalar waves. Let the field
$\ANT_\nu{}^\b$ be identically zero: $\ANT_\nu{}^\b=0$. Hence, we are
left with a system of two scalar equations
\begin{equation}
  \left(M^\mu{}_\alpha{}^\beta{}_\beta-\frac 14\d^\mu_\a
    M^\g{}_\g{}^\beta{}_\beta
  \right)A= 0\,,\quad  M^\a{}_\alpha{}^\beta{}_\beta\,A= 0\,.\label{spec4}
\end{equation}
The second equation has a non-trivial solution only if
$M^\alpha{}_\alpha{}^\beta{}_\beta$ $ = 0$.  Consequently we
have a dispersion relation
\begin{equation}
\chi^{\a\rho}{}_{\a}{}^{\b\sigma}{}_\b\,q_\rho\,q_\sigma=0\,.\label{spec5}
\end{equation}
Using the double-trace tensor (\ref{trm}), we can write it as
\begin{equation}
m^{\a\b}q_\a\,q_\b=0\,.\label{spec6}
\end{equation}
In the metric-dependent case (\ref{met3}), the tensor $m^{\a\b}$, up
to a factor, coincides with the Lorentz metric tensor $g^{\a\b}$. Thus
we recover the standard light cone of general relativity,
\begin{equation}
g^{\a\b}q_\a\,q_\b=0\,.\label{spec7}
\end{equation}

\subsection{A separable case}

We assume now that two terms in the left-hand side of Eqs.\
\eqref{spec2b} and \eqref{spec2c} vanish independently. In other words,
we assume that the system separates into two independent
subsystems. In \cite{Itin:2004ig}, a similar type of consideration
allowed to extract the teleparallel equivalent of GR from a set of
metric based models. Thus we have 15 equations for 15 variables
\begin{equation}
  \left(M^\mu{}_\alpha{}^\nu{}_\beta-\frac 14\d^\mu_\a 
    M^\g{}_\g{}^\nu{}_\beta\right)\ANT_\nu{}^\b=0\,.\label{spec8}
\end{equation}
 and 1 equation for 1 variable
\begin{equation}
M^\a{}_\alpha{}^\beta{}_\beta\,A= 0\,.\label{spec9}
\end{equation}

\subsection{Gauge conditions}

In electromagnetism, gauge invariance can be restricted by applying
a gauge fixing condition. The Lorenz gauge condition is the unique
diffeomorphism invariant expression. For wave solutions, it takes the
form $A_\a q^\a=0$. Note that this condition can be formulated only
on a metric manifold.

In teleparallel theory, we are able to formulate a similar condition
in a premetric form:
\begin{equation}
A_\nu{}^\b q_\b= 0\,.\label{gauge1}
\end{equation}
A more general gauge condition can be proposed in the form of
\begin{equation}
K_1A_\a{}^\b q_\b+K_2A_\b{}^\b q_\a= 0\,.\label{gauge2}
\end{equation}
Here $K_1$ and $K_2$ are arbitrary constants. The exceptional case
for $K_1=1$ and $K_2=-1$, namely
\begin{equation}
A_\a{}^\b q_\b-A_\b{}^\b q_\a= 0\,,\label{gauge3}
\end{equation}
is invariant under the gauge transformation
$A_\a{}^\b\to A_\a{}^\b+q_\a C^\b$ and, accordingly, does not
represent a gauge condition.

\section{Gravitational waves in a metric model}\label{Sec.6}

Let us now specialize to the metric-dependent constitutive tensor
(\ref{chigen}). The analysis of the wave propagation in the geometric
optics approximation is straightforward. Here we will confine our
attention to the {\it parity-even} case
(\ref{even_endresult}). Recently a related paper was published by
Hohmann et al.\ \cite{Hohmann:2018wxu}.

\subsection{Characteristic system}

We substitute the constitutive tensor (\ref{even_endresult}) with its
parity even pieces into (\ref{Mabmn}). As a result,
$M^{\b}{}_{\a}{}^{\nu}{}_\mu(g) =
\chi^{\b\g}{}_{\a}{}^{\nu\rho}{}_\mu(g)$ $\times$ $\!q_\g q_\rho$ takes the form
\begin{align}
  &M^{\b}{}_{\a}{}^{\nu}{}_\mu(g) =\frac{\sqrt{-g}}{4\varkappa}\Big\{
    {2\b_1}g_{\a\m} \left(g^{\b\n}q^2-q^{\n}q^{\b}\right)\nonumber\\
  & -\,{\b_2}\left[\left(q^2\delta^\b_\a - q^{\b}q_\a\right)\delta_\m^\n 
    - \left(q^{\n}\delta^\b_\a-g^{\b\n}q_\a\right)q_\m\right]\nonumber\\
  & -\,{\b_3}\left[\left(q^2\delta_\mu^\b -q^{\b}q_\mu\right)
    \delta^\nu_\a -\left(q^{\nu}\delta_\mu^\b - g^{\b\nu}q_\mu \right)
    q_\a\right]\Big\}\,.\label{Mq0}
\end{align}
We can immediately derive the contractions
\begin{equation}
  M^{\b}{}_{\a}{}^{\nu}{}_\nu = M^{\nu}{}_{\nu}{}^{\b}{}_\a 
  = \frac{\sqrt{-g}}{4\varkappa}\,
  (2\beta_1 - 3\beta_2 - \beta_3)\,\left(q^2\delta^\b_\a 
    - q^\b q_\a\right),\label{Mq1}
\end{equation}
where $q^2 := g^{\a\b}q_\a q_\b$. Consequently,
\begin{equation}
M^{\a}{}_{\a}{}^{\nu}{}_\nu = \frac{3\sqrt{-g}}{4\varkappa}\,
(2\beta_1 - 3\beta_2 - \beta_3)\,q^2\,.\label{Mq2}
\end{equation}
Furthermore, we verify that the tensor (\ref{Mq0}) has the following
properties:
\begin{align}\label{Mq3}
  M^{\b}{}_{\a}{}^{\nu}{}_\mu\,q_\nu 
  &= 0,\qquad M^{\b}{}_{\a}{}^{\nu}{}_\mu\,q_\beta = 0,\\
  M^{\b}{}_{\a}{}^{\nu}{}_\mu\,q^\mu 
  &= \frac{\sqrt{-g}}{4\varkappa}\,
      (2\beta_1 - \beta_2 - \beta_3)\,\left(q^2g^{\b\nu} 
      - q^\b q^\nu\right)q_\a,\label{Mq4}\\
  M^{\b}{}_{\a}{}^{\nu}{}_\mu\,q^\a 
  &= \frac{\sqrt{-g}}{4\varkappa}\,
      (2\beta_1 - \beta_2 - \beta_3)\,\left(q^2g^{\b\nu} 
      - q^\b q^\nu\right)q_\mu.\label{Mq5}
\end{align}
Interestingly, for the GR$_{||}$ case of (\ref{met_GR}), the
right-hand sides of (\ref{Mq4}) and (\ref{Mq5}) vanish, i.e., the
tensor $M^{\b}{}_{\a}{}^{\nu}{}_\mu$ turns out to be transversal to
the wave covector $q_\mu$ in all four indices.

Taking into account these properties, the wave propagation system
(\ref{eq1}) yields
\begin{equation}\label{Mq6}
  (2\beta_1 - \beta_2 - \beta_3)\,\left(q^2g^{\b\nu} 
    - q^\b q^\nu\right)q_\mu A_\nu{}^\mu =0.
\end{equation}
In the generic case, when $2\beta_1 - \beta_2 - \beta_3 \neq 0$, we
find the relation
\begin{equation}
  q^2\,A_\beta{}^\mu q_\mu = q_\beta\,A_\nu{}^\mu q_\mu q^\nu,\label{Mq7}
\end{equation}
which implies
\begin{equation}
q^2\ANT_\beta{}^\mu q_\mu = q_\beta\ANT_\nu{}^\mu q_\mu q^\nu.\label{Mq8}
\end{equation}
We will use this in subsequent derivations.

Let us now turn to the decomposed equations (\ref{spec2c}) and
(\ref{NA}), Using (\ref{Mq1}), (\ref{Mq2}), and $\ANT_\nu{}^\nu = 0$,
we can recast (\ref{spec2c}) into
\begin{equation}
3q^2A = \ANT_\nu{}^\mu q_\mu q^\nu.\label{Mq9}
\end{equation}
This is derived for the generic case by assuming
$2\beta_1 - 3\beta_2 - \beta_3 \neq 0$, otherwise (\ref{spec2c}) is
trivial.

Finally, we turn to the traceless part (\ref{NA}) of the wave
propagation equation. Substituting (\ref{Mq0})-(\ref{Mq2}) into
(\ref{N}), and then making use of (\ref{Mq8}), we can rewrite
(\ref{NA}) in the equivalent form
\begin{align}
  &\hspace{12pt} 6\beta_1q^2\!\left(q^2\!\!\ANT{\,}^\b{}_\a 
    - q^\b q_\gamma\!\!\ANT{\,}^\gamma{}_\a\right) \nonumber \\
  &-\,3\beta_3q^2\!\left(q^2\!\!\ANT_\a{}^\b 
    - q_\a q^\gamma\!\!\ANT_\gamma{}^\b\right) \label{NA1} \\
  &+ \,(2\beta_1 - \beta_3)\left(q^2\delta^\b_\a 
    - q^{\b}q_\a\right)\!\ANT_\nu{}^\mu q_\mu q^\nu = 0. \nonumber
\end{align}
Quite remarkably, the coupling constant $\beta_2$ does not contribute.

Decomposing (\ref{NA1}) into symmetric and antisymmetric parts, we
obtain the two equations
\begin{align}
  & (2\beta_1 - \beta_3)\Big\{3\left(q^2\delta^\mu_\a 
    - q^\mu q_\a\right)(q^2\delta^\nu_\b -
    q^\nu q_\b) \nonumber\\
  &\hspace{74pt}- (q^2g_{\a\b} - q_\a q_\b)
    q^\mu q^\nu\!\!\Big\}\ANT_{(\mu\nu)} = 0,\label{NA2}\\
  & (2\beta_1 + \beta_3)\left(q^2\delta^\mu_\a - q^\mu q_\a\right)
    (q^2\delta^\nu_\b - q^\nu q_\b)\!\!\ANT_{[\mu\nu]} = 0.\label{NA3}
\end{align}
Two important observations are in order. Firstly, we see that the
symmetric $\ANT_{(\mu\nu)}$ and skew-symmetric $\ANT_{[\mu\nu]}$
variables decouple from each other and are governed by two separate
dynamical equations. Secondly, the symmetric $\ANT_{(\mu\nu)}$ mode
remains the only one when coupling constants satisfy
$2\beta_1 = -\,\beta_3$ and, similarly, the antisymmetric
$\ANT_{[\mu\nu]}$ mode remains the only one when the coupling
constants satisfy $2\beta_1 = \beta_3$.

In the generic case (when $2\beta_1\neq -\,\beta_3$ and
$2\beta_1\neq \beta_3$), both modes are dynamical. In principle, one
can proceed by deriving the Fresnel equations for each mode in a way
similar to classical electrodynamics. However, one can choose an
alternative way and analyze the propagation equations by using the
method of spin-projection operators. The corresponding tools were
earlier developed in metric gravity \cite{Rivers:1964} and in its
teleparallel formulation \cite{Muller:1983}. Their application is
straightforward. Ultimately we conclude that the wave equations
describe the massless spin-0, spin-2 and spin-1 modes propagating
along the light cones $q^2 = 0$.

\subsection{Waves in GR$_{||}$, the teleparallel equivalent of GR}

In order to understand more clearly what happens in GR$_{||}$, it is
instructive to analyze the wave propagation equation from
scratch. Plugging (\ref{Mq0}) into (\ref{Mabmn}), we find for
\eqref{Ma0}, after some straightforward algebra and lowering the index
$\mu$, 
\begin{align}
  &(2\beta_1 - \beta_2 - \beta_3)\times\nonumber\\
  &\left\{4A(q^2g_{\a\b} - q_\a q_\b) 
    - g_{\a\b} q^\mu q^\nu A_{\mu\nu}
    + q_\a A_{\b\gamma}q^\gamma\right\}\nonumber\\
  &  -\,(2\beta_1 + \beta_3)\left\{q^2A_{[\a\b]} - q_\a q^\gamma
    A_{[\gamma\b]} 
    - q_\b q^\gamma A_{[\a\gamma]}\right\}\nonumber\\
  & +\,(2\beta_1 - \beta_3)\big\{q^2A_{(\a\b)} - q_\a q^\gamma
    A_{(\gamma\b)} 
    - q_\b q^\gamma A_{(\a\gamma)}\nonumber\\
  &-\,4A(q^2g_{\a\b} - q_\a q_\b) + g_{\a\b} q^\mu q^\nu
    A_{\mu\nu}\big\} = 0.
\label{MA0}
\end{align} 
Contracting with $g^{\a\b}$, we find
\begin{equation}
  (2\beta_1 - 3\beta_2 - \beta_3)\left(4Aq^2 
    - q^\mu q^\nu A_{\mu\nu}\right) = 0,\label{MA1}
\end{equation}
whereas contraction with $q^\a$ yields
\begin{equation}
  (2\beta_1 - \beta_2 - \beta_3)\left(- q_\b q^\mu q^\nu A_{\mu\nu}
    + q^2 A_{\b\gamma}q^\gamma\right) = 0.\label{MA2}
\end{equation}
In the generic case, we recover the earlier findings (\ref{Mq7})
and (\ref{Mq6}). Making use of this, we see that the first line in
(\ref{MA0}) vanishes, whereas the remaining equation reduces to the
system (\ref{NA2}) and (\ref{NA3}).

However, GR$_{||}$ represents a very special case when the coupling
constants (\ref{met_GR}) are such that both
$2\beta_1 - \beta_2 - \beta_3 = 0$ and $2\beta_1 + \beta_3 = 0$. As a
result, the first two lines of (\ref{MA0}) disappear, and the
propagation equation has only a symmetric traceless part. The bottom
line is that in GR$_{||}$ Eq.(\ref{MA0})
reduces to
\begin{equation}
q^2h_{\a\b} - q_\a q^\gamma h_{\gamma\b} - q_\b q^\gamma h_{\a\gamma} = 0,\label{MA3}
\end{equation}
after we introduce the familiar variable
\begin{equation}
h_{\a\b} := A_{(\a\b)} - 2Ag_{\a\b} = A_{(\a\b)} - {\frac 12}g_{\a\b}A_\gamma{}^\gamma.\label{hab}
\end{equation}
Accordingly, we indeed recover the result of GR with the massless
spin-2 graviton propagating along the light cone. In other words,
GR$_{||}$ is completely consistent with Einstein's theory with respect
to the propagation of gravitational waves.

\section{Conclusions and outlook}

Recently, in Ref.[29], we developed a novel framework for a premetric
teleparallel theory of gravity (TG). Here we continue this study on TG
by paying attention specifically to TG models with a general {\it
  local and linear} constitutive law.

In Sec.\ref{Sec.2}, we constructed the Tonti diagram of TG which
explicitly displays the generic structure of the theory.

Our main new results are presented in Sec.\ref{Sec.3} where the irreducible
decomposition of the premetric constitutive tensor is established in
full detail. This issue was not analysed in the earlier literature. Here
we considered two types of such decompositions with
respect to the permutation group, namely one related to
$S_4\times S_2$ and the other one with respect to $S_6$. The relations
between these two decompositions are explicitly derived and the
physical meaning of the different irreducible pieces clarified.

After establishing for the constitutive tensor these premetric
results, we turned in Sec.\ref{Sec.4} to a special case: The spacetime
continuum is supposed to carry a metric tensor. We constructed the
most general metric dependent constitutive tensor that is cubic in the
metric tensor. It includes both, parity even and parity odd parts.
Thereby extending earlier results, we for the first time obtained the
most general family of teleparallel consitutive tensors. In particular,
a parity odd Lagrangian was newly found, and its physical interpretation fixed. 

In Sec.\ref{Sec.5}, the propagation of gravitational waves are derived
for the premetric case in the geometrical optics approximation. Additional
propagating modes of spin 1 and 0 are extracted. Our results are generally
consistent with the recent findings reported in \cite{Hohmann:2018wxu}. 

Subsequently, in Sec.\ref{Sec.6}, the metric case is addressed. The GR
limit is naturally embedded in our formalism.

The results presented here can serve as a basis for the study of
expanded gravitational models including axion and skewon effects. In
particular, violation of Lorentz invariance can be meaningfully
addressed in our premetric set-ups.

\subsubsection*{Acknowledgments} We are grateful to Enzo Tonti
(Trieste) for help in creating our Tonti diagram. Moreover, we thank
Bahram Mashhoon (Tehran) for most helpful remarks in the context of
nonlocal gravity (NLG). Remarks of Mar\'{\i}a Jos\'e Guzm\'an (Buenos
Aires), Manuel Hohmann (Tartu), Jos\'e Maluf (Brasilia), and James
Nester (Chung-Li) are gratefully acknowledged. For Y.N.O.\ this work
was partially supported by the Russian Foundation for Basic Research
(Grant No. 16-02-00844-A). J.B.\ is grateful for a Vanier Canada
Graduate Scholarship administered by the Natural Sciences and
Engineering Research Council of Canada as well as for the Golden Bell
Jar Graduate Scholarship in Physics by the University of Alberta.

\appendix
\section{Young decomposition}
\label{app:conventions-decomposition}

Here we briefly explain how to determine the
$\text{GL}(4,\mathbb{R})$ decomposition of the order-\tvect{0}{6}
constitutive tensor
$\check{\chi}{}_{\alpha\beta\mu\gamma\delta\nu}$. The
decomposition rules for tensors of other valence, say order-2 or
order-4, can easily be derived. In four dimensions, the decomposition
of $\check{\chi}{}_{\alpha\beta\mu\gamma\delta\nu}$ with the symmetry
properties
\begin{align}
  \tag{\ref*{eq:chi-check-symmetries*}}
\check{\chi}{}_{(\alpha\beta)\mu\gamma\delta\nu} = 0\,,\quad
  \check{\chi}{}_{\alpha\beta\mu(\gamma\delta)\nu}=0 ,
\end{align}
consists of the following seven pieces:
\begin{align}
\begin{split}
{\Yvcentermath1
  \yng(4,2) ~,
\quad \yng(4,1,1) ~,
\quad \yng(3,3) ~,
\quad \yng(3,2,1) ~,}\\
{\Yvcentermath1
\quad \yng(2,2,2) ~,
\quad \yng(3,1,1,1) ~,
\quad \yng(2,2,1,1) ~. }
\end{split}
\end{align}
Let us label them from $\lambda_1$ to $\lambda_7$, respectively. Each diagram $\lambda_i$ ($i=1\dots 7$) has corresponding dimensions according to
\begin{align}
\begin{split}
\text{dim} ~ V{}^\lambda &= \prod\limits_{(\alpha,\beta) \in \lambda} \frac{n + \beta - \alpha}{\text{hook}(\alpha,\beta)} , \\
\text{dim} ~ S{}^\lambda_6 &= f{}^{\lambda} = \frac{6!}{\prod\limits_{x \in \lambda} \text{hook}(x)} . \label{eq:dimension}
\end{split}
\end{align}
We find
\begin{align}
\text{dim}\,V{}^{\lambda_1} &= 126 , &
\text{dim}\,V{}^{\lambda_2} &=  70 , &
\text{dim}\,V{}^{\lambda_3} &=  50 , \nonumber \\
\text{dim}\,V{}^{\lambda_4} &=  64 , &
\text{dim}\,V{}^{\lambda_5} &=  10 , &
\text{dim}\,V{}^{\lambda_6} &=  10 , \\
\text{dim}\,V{}^{\lambda_7} &=   6 , && \nonumber
\end{align}
as well as
\begin{align}
\text{dim}\,S{}^{\lambda_1}_6 &=  9 , &
\text{dim}\,S{}^{\lambda_2}_6 &= 10 , &
\text{dim}\,S{}^{\lambda_3}_6 &=  5 , \nonumber \\
\text{dim}\,S{}^{\lambda_4}_6 &= 16 , &
\text{dim}\,S{}^{\lambda_5}_6 &=  5 , &
\text{dim}\,S{}^{\lambda_6}_6 &= 10 , \label{eq:dimensions-dec-chi} \\
\text{dim}\,S{}^{\lambda_7}_6 &=  9 . && \nonumber
\end{align}
With the relevant Young diagrams now given, we can extract the projection operators in the standard fashion and arrive at the irreducible decomposition of $\chi{}_{\alpha\beta\gamma\mu\nu\rho}$.

For each $\lambda_i$ we may extract
projection operators in the standard way: we can fill
$f^{\lambda_i}$ copies of the $\lambda_i$ Young diagram with numbers
to create all allowed Young tableaux, and then create the
corresponding projection operator onto that subspace by summing over
the induced Young symmetrizers:
\begin{align}
  \mathbbm{P}{}^{\lambda_i}_p &:= \frac{f^{\lambda_i}}{p!} \sum \limits_{j=1}^{f^{\lambda_i}} \mathbbm{P}\left(Y^{\lambda_i}_j\right) , \\
  {}^{(I)}\check{\chi}{}_{\alpha\beta\mu\gamma\delta\nu} &:= \mathbbm{P}_p^{\lambda_I} \left( \check{\chi}{}_{\alpha\beta\mu\gamma\delta\nu} \right) ,
\end{align}
where $\mathbbm{P}\left(Y^{\lambda_i}_j\right)$ is the Young
symmetrizer corresponding to the $j$-th Young tableaux of $\lambda_i$,
and ${}^{(I)}\check{\chi}{}_{\alpha\beta\mu\gamma\delta\nu}$ denotes
the $I$-th irreducible piece of the constitutive tensor. The Young
tableaux with non-trivial symmetrizers are as follows:
\begin{align}
\lambda_1 &: \quad \Yvcentermath1 \young(1356,24) ~,~ \young(1346,25) ~, \\[5pt]
\lambda_2 &: \quad \Yvcentermath1 \young(1356,2,4) ~,~ \young(1346,2,5) ~, \\[5pt]
\lambda_3 &: \quad \Yvcentermath1 \young(135,246) ~,~ \young(134,256) ~, \\[5pt]
\lambda_4 &: \quad \Yvcentermath1 \young(134,26,5) ~,~ \young(135,24,6) ~,~ \young(135,26,4) ~,~ \young(134,25,6) ~,\\~
&\hspace{17pt}\Yvcentermath1\young(146,25,3) ~,~ \young(136,24,5) ~,~ \young(136,25,4) ~, \\[5pt]
\lambda_5 &: \quad \Yvcentermath1 \young(14,25,36) ~,~ \young(13,24,56) ~,~ \young(13,25,46) , \\[5pt]
\lambda_6 &: \quad \Yvcentermath1 \young(156,2,3,4) ~,~ \young(134,2,5,6) ~,~ \young(135,2,4,6) ~,~ \young(146,2,3,5) ~,~ \young(136,2,4,5) ~, \\[5pt]
\lambda_7 &: \quad \Yvcentermath1 \young(15,26,3,4) ~,~  \young(13,24,5,6) ~,~ \young(14,26,3,5) ~,~ \young(13,26,4,5) ~,~ \young(14,25,3,6) ~,~ \young(13,25,4,6) ~.
\end{align}
For example, $\mathbbm{P}\left( Y^{\lambda_4}_6 \right)$ can be computed as follows:
\begin{align}
\mathbbm{P}\left( Y^{\lambda_4}_6 \right) &= \mathbbm{P}\left( \tiny
\Yvcentermath1 \young(136,24,5) \right) = S_{136} \circ S{}_{24} \circ
A{}_{125} \circ A{}_{34} \\
&\hspace{-18pt} = 
\left[ 1 + (13) + (16) + (36) + (136) + (631) \right] \circ \left[ 1 +
(24) \right]
\nonumber \\ & \hspace{10pt}
\hspace{-24pt}\circ \left[ 1 - (12) - (15) - (25) + (125) + (521)
\right] \circ \left[ 1 - (34) \right], \nonumber
\end{align}
where $S_{136}$ denotes symmetrization between the 1st, 3rd, 6th
index; similarly, $A{}_{34}$ denotes antisymmetrization between the
3rd and 4th index, and so on. In the last lines we switched to the
more convenient cycle notation. Note that in our conventions the Young
symmetrizers are not normalized.
\begin{figure}[!htb]
  \centering
  \includegraphics[width=1.00\textwidth]{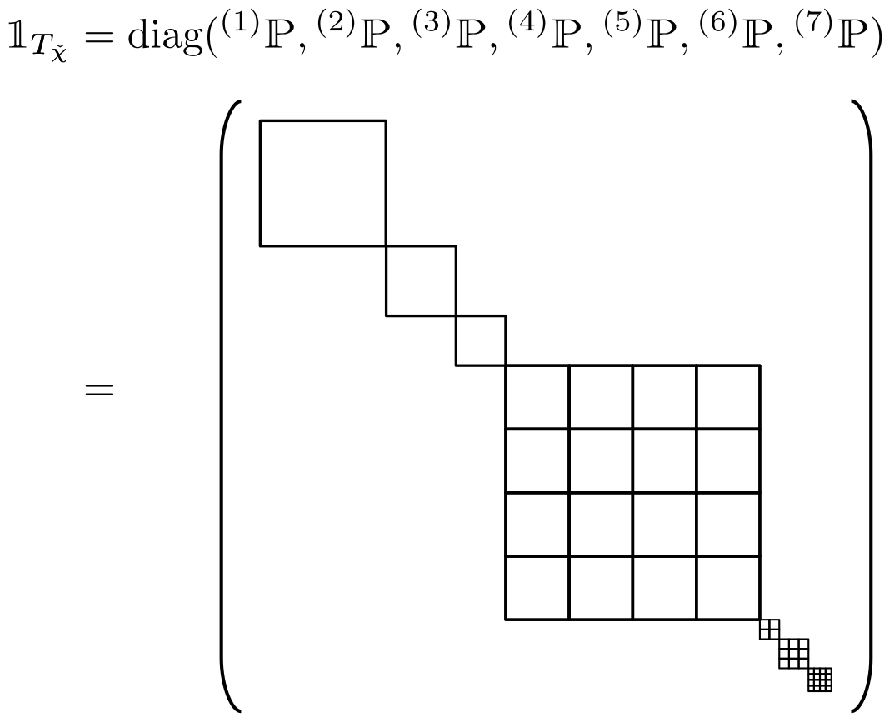}
  \caption{The resolution of the identity on the 576-dimensional
    vector space $T_{\check{\chi}}$ has a block-diagonal structure in
    terms of the canonical Young projection operators
    ${}^{\{I\}}\mathbbm{P} : T_{\check{\chi}} \rightarrow T^I_{\check{\chi}}$
    with $I=1,\dots,7$. We depict these operators by $1 \times 1$
    blocks, with the relative width of $f^{\lambda_I} \times d_I/576$.
    Provided $f^{\lambda_I} > 1$, these operators can be further
    decomposed into \emph{non-canonical} operators
    ${}^{(I,j)}\mathbbm{P} : T^I_{\check{\chi}} \rightarrow
    T^I_{\check{\chi}}$
    with $j=1,\dots,f^{\lambda_I}$. We depict this internal
    ambiguity by inserting $f^{\lambda_I} \times f^{\lambda_I}$
    smaller blocks.}
  \label{fig:chi-decomp-visualization}
\end{figure}
The exact expressions are lengthy, so we make some comments:

(i) Using the Young decomposition technique, the 576-dimensional
tensor space of $\check{\chi}{}_{\alpha\beta\gamma\mu\nu\rho}$, call
it $T{}_{\check{\chi}}$, can be decomposed into the direct sum of
lower-dimensional vector spaces $T{}^I_{\check{\chi}}$ with
$I=1,\dots,7$. Denoting the projection operators onto these
lower-dimensional subspaces via
${}^{(I)}\mathbbm{P} : T_{\check{\chi}} \rightarrow
T^I_{\check{\chi}}$,
the identity element on $T{}_{\check{\chi}}$ can be written as
\begin{align}
\mathbbm{1}_{T_{\check{\chi}}} = \bigoplus\limits_{I=1}^7 {}^{(I)} \mathbbm{P}  \qquad \Rightarrow \qquad {}^{(I)}\check{\chi} := {}^{(I)} \mathbbm{P}( \check{\chi} ) .
\end{align}

(ii) In a suitable matrix representation of $T_{\check{\chi}}$, call
it $\rho\,(T_{\check{\chi}}) = \mathbbm{R}^{576}$, the operators
${}^{(I)}\mathbbm{P}$ can be thought of as
$(f^{\lambda_I} d_I) \times (f^{\lambda_I} d_I)$ matrices, where
$f^{\lambda_I}$ denotes the degeneracy of $T^I_{\check{\chi}}$.

(iii) On $T{}^I_{\check{\chi}}$, various auxiliary projection
operators ${}^{(I,j)}\mathbbm{P}$ can be defined such that
\begin{align} {}^{(I)}\mathbbm{P} =
  \bigoplus\limits_{j=1}^{f^{\lambda_I}} {}^{(I,j)}\mathbbm{P} .
\end{align}
We emphasize that the operators ${}^{(I,j)}\mathbbm{P}$ are \emph{not
  canonical}, that is, they are only determined up to arbitrary linear
transformations
$G: \rho(T^I_{\check{\chi}}) \rightarrow
\rho(T^I_{\check{\chi}})$.
For a graphical representation of statements (i)--(iii), see
Fig.~\ref{fig:chi-decomp-visualization}.


We now use computer algebra \cite{Hearn,Boos:2018} to obtain explicit expressions which are a bit lengthy. To compress our results, we define a projector $\mathbf{C}$ (its index representation $C{}^{\rho\sigma\omega\lambda}_{\alpha\beta\gamma\delta}$) by
\begin{align}
\begin{split}
\label{symmetrizerC}
\mathbf{C} &:= \frac14 \left[1 - (12) - (45) + (12)(45) \right] \, , \\
C{}^{\rho\sigma\omega\lambda}_{\alpha\beta\gamma\delta} &:= \frac 14
\delta{}^{\rho\sigma}_{\alpha\beta}
\delta{}^{\omega\lambda}_{\gamma\delta} \, , \quad
\delta{}^{\rho\sigma}_{\alpha\beta} := \delta{}^\rho_\alpha
\delta{}^\sigma_\beta - \delta{}^\sigma_\alpha \delta{}^\rho_\beta \,
.
\end{split}
\end{align}
Observe that $\mathbf{C}^2 = \mathbf{C}$, indeed. This projector
antisymmetrizes in the two index pairs $\alpha\beta$ and
$\gamma\delta$ such that any constitutive tensor satisfying the
relation
$\check{\chi}_{\alpha\beta\mu\gamma\delta\nu} =
\check{\chi}_{[\alpha\beta]\mu[\gamma\delta]\nu}$
has eigenvalue 1 with respect to this operator:
\begin{align}
C{}^{\rho\sigma\omega\lambda}_{\alpha\beta\gamma\delta} \check{\chi}{}_{\rho\sigma\mu\omega\lambda\nu} = \check{\chi}{}_{\alpha\beta\mu\gamma\delta\nu} \, .
\end{align}
In general, $C{}^{\rho\sigma\omega\lambda}_{\alpha\beta\gamma\delta}$
projects a general order-6 tensor without any symmetries, call it
$T{}_{\alpha\beta\mu\gamma\delta\nu}$, onto a tensor that has the
symmetries of the constitutive tensor. In that sense we make use of
this projector to shorten our exact expressions for the irreducible
decomposition of the constitutive tensor by a factor of 4. We obtain:

  \begin{eqnarray}
  \label{eq:cc-decomposition-1}
    {}^{(1)}\check{\chi}{}_{\alpha\beta\mu\gamma\delta\nu} &=& \frac{1}{20} C{}^{\rho\sigma\omega\lambda}_{\alpha\beta\gamma\delta} \Big( -\check{\chi}{}_{\rho\sigma\omega\lambda\mu\nu} - \check{\chi}{}_{\rho\sigma\omega\lambda\nu\mu} + \check{\chi}{}_{\rho\sigma\mu\omega\lambda\nu}\\ \nonumber &&\hspace{59pt} + \check{\chi}{}_{\rho\sigma\mu\omega\nu\lambda} + \check{\chi}{}_{\rho\sigma\nu\omega\lambda\mu} + \check{\chi}{}_{\rho\sigma\nu\omega\mu\lambda}\\ \nonumber &&\hspace{59pt} - \check{\chi}{}_{\rho\omega\sigma\lambda\mu\nu} - \check{\chi}{}_{\rho\omega\sigma\lambda\nu\mu} - \check{\chi}{}_{\rho\omega\lambda\sigma\mu\nu}\\ \nonumber &&\hspace{59pt} - \check{\chi}{}_{\rho\omega\lambda\sigma\nu\mu} - \check{\chi}{}_{\rho\omega\mu\sigma\nu\lambda} - \check{\chi}{}_{\rho\omega\mu\lambda\nu\sigma}\\ \nonumber &&\hspace{59pt} - \check{\chi}{}_{\rho\omega\nu\sigma\mu\lambda} - \check{\chi}{}_{\rho\omega\nu\lambda\mu\sigma} + \check{\chi}{}_{\rho\mu\sigma\omega\lambda\nu}\\ \nonumber &&\hspace{59pt} + \check{\chi}{}_{\rho\mu\sigma\omega\nu\lambda} + \check{\chi}{}_{\rho\mu\omega\sigma\lambda\nu} - \check{\chi}{}_{\rho\mu\omega\lambda\nu\sigma}\\ \nonumber &&\hspace{59pt} - \check{\chi}{}_{\rho\mu\nu\sigma\omega\lambda} + \check{\chi}{}_{\rho\mu\nu\omega\lambda\sigma} + \check{\chi}{}_{\rho\nu\sigma\omega\lambda\mu} \\ \nonumber &&\hspace{59pt} + \check{\chi}{}_{\rho\nu\sigma\omega\mu\lambda} + \check{\chi}{}_{\rho\nu\omega\sigma\lambda\mu} - \check{\chi}{}_{\rho\nu\omega\lambda\mu\sigma}\\ \nonumber &&\hspace{59pt} - \check{\chi}{}_{\rho\nu\mu\sigma\omega\lambda} + \check{\chi}{}_{\rho\nu\mu\omega\lambda\sigma} + \check{\chi}{}_{\omega\lambda\rho\sigma\mu\nu}\\ \nonumber &&\hspace{59pt} + \check{\chi}{}_{\omega\lambda\rho\sigma\nu\mu} - \check{\chi}{}_{\omega\lambda\mu\rho\sigma\nu} - \check{\chi}{}_{\omega\lambda\mu\rho\nu\sigma}\\ \nonumber &&\hspace{59pt} - \check{\chi}{}_{\omega\lambda\nu\rho\sigma\mu} - \check{\chi}{}_{\omega\lambda\nu\rho\mu\sigma} + \check{\chi}{}_{\omega\mu\rho\sigma\lambda\nu}\\ \nonumber &&\hspace{59pt} + \check{\chi}{}_{\omega\mu\rho\sigma\nu\lambda} - \check{\chi}{}_{\omega\mu\lambda\rho\sigma\nu} - \check{\chi}{}_{\omega\mu\lambda\rho\nu\sigma} \\ \nonumber &&\hspace{59pt} - \check{\chi}{}_{\omega\nu\mu\rho\sigma\lambda} - \check{\chi}{}_{\omega\mu\nu\rho\sigma\lambda} - \check{\chi}{}_{\omega\mu\nu\rho\lambda\sigma} + \check{\chi}{}_{\omega\nu\rho\sigma\lambda\mu}\\ \nonumber &&\hspace{59pt} + \check{\chi}{}_{\omega\nu\rho\sigma\mu\lambda} - \check{\chi}{}_{\omega\nu\lambda\rho\sigma\mu} - \check{\chi}{}_{\omega\nu\lambda\rho\mu\sigma}  - \check{\chi}{}_{\omega\nu\mu\rho\lambda\sigma} \Big) ,
\end{eqnarray}
\begin{eqnarray}
  \label{eq:cc-decomposition-2}
{}^{(2)}\check{\chi}{}_{\alpha\beta\mu\gamma\delta\nu} &=& \frac{1}{18} C{}^{\rho\sigma\omega\lambda}_{\alpha\beta\gamma\delta} \Big( -\check{\chi}{}_{\rho\sigma\omega\lambda\mu\nu} - \check{\chi}{}_{\rho\sigma\omega\lambda\nu\mu} + \check{\chi}{}_{\rho\sigma\mu\omega\lambda\nu}\\ \nonumber &&\hspace{59pt} + \check{\chi}{}_{\rho\sigma\mu\omega\nu\lambda} + \check{\chi}{}_{\rho\sigma\nu\omega\lambda\mu} + \check{\chi}{}_{\rho\sigma\nu\omega\mu\lambda}\\ \nonumber &&\hspace{59pt} - \check{\chi}{}_{\rho\omega\sigma\lambda\mu\nu} - \check{\chi}{}_{\rho\omega\sigma\lambda\nu\mu} - \check{\chi}{}_{\rho\omega\lambda\sigma\mu\nu}\\ \nonumber &&\hspace{59pt} - \check{\chi}{}_{\rho\omega\lambda\sigma\nu\mu} - \check{\chi}{}_{\rho\omega\mu\sigma\nu\lambda} - \check{\chi}{}_{\rho\omega\mu\lambda\nu\sigma}\\ \nonumber &&\hspace{59pt} - \check{\chi}{}_{\rho\omega\nu\sigma\mu\lambda} - \check{\chi}{}_{\rho\omega\nu\lambda\mu\sigma} + \check{\chi}{}_{\rho\mu\sigma\omega\lambda\nu}\\ \nonumber &&\hspace{59pt} + \check{\chi}{}_{\rho\mu\sigma\omega\nu\lambda} + \check{\chi}{}_{\rho\mu\omega\sigma\lambda\nu} - \check{\chi}{}_{\rho\mu\omega\lambda\nu\sigma}\\ \nonumber &&\hspace{59pt} - \check{\chi}{}_{\rho\mu\nu\sigma\omega\lambda} + \check{\chi}{}_{\rho\mu\nu\omega\lambda\sigma} + \check{\chi}{}_{\rho\nu\sigma\omega\lambda\mu}\\ \nonumber &&\hspace{59pt} + \check{\chi}{}_{\rho\nu\sigma\omega\mu\lambda} + \check{\chi}{}_{\rho\nu\omega\sigma\lambda\mu} - \check{\chi}{}_{\rho\nu\omega\lambda\mu\sigma}\\ \nonumber &&\hspace{59pt} - \check{\chi}{}_{\rho\nu\mu\sigma\omega\lambda} + \check{\chi}{}_{\rho\nu\mu\omega\lambda\sigma} + \check{\chi}{}_{\omega\lambda\rho\sigma\mu\nu}\\ \nonumber &&\hspace{59pt} + \check{\chi}{}_{\omega\lambda\rho\sigma\nu\mu} - \check{\chi}{}_{\omega\lambda\mu\rho\sigma\nu} - \check{\chi}{}_{\omega\lambda\mu\rho\nu\sigma}\\ \nonumber &&\hspace{59pt} - \check{\chi}{}_{\omega\lambda\nu\rho\sigma\mu} - \check{\chi}{}_{\omega\lambda\nu\rho\mu\sigma} + \check{\chi}{}_{\omega\mu\rho\sigma\lambda\nu}\\ \nonumber &&\hspace{59pt} + \check{\chi}{}_{\omega\mu\rho\sigma\nu\lambda} - \check{\chi}{}_{\omega\mu\lambda\rho\sigma\nu} - \check{\chi}{}_{\omega\mu\lambda\rho\nu\sigma} \\ \nonumber &&\hspace{59pt} - \check{\chi}{}_{\omega\nu\mu\rho\sigma\lambda} - \check{\chi}{}_{\omega\mu\nu\rho\sigma\lambda} - \check{\chi}{}_{\omega\mu\nu\rho\lambda\sigma} + \check{\chi}{}_{\omega\nu\rho\sigma\lambda\mu} \\ \nonumber &&\hspace{59pt} + \check{\chi}{}_{\omega\nu\rho\sigma\mu\lambda} - \check{\chi}{}_{\omega\nu\lambda\rho\sigma\mu} - \check{\chi}{}_{\omega\nu\lambda\rho\mu\sigma} - \check{\chi}{}_{\omega\nu\mu\rho\lambda\sigma} \Big) ,
\end{eqnarray}
\begin{eqnarray}
\label{eq:cc-decomposition-3}
      {}^{(3)}\check{\chi}{}_{\alpha\beta\mu\gamma\delta\nu} &=& \frac{1}{36} C{}^{\rho\sigma\omega\lambda}_{\alpha\beta\gamma\delta} \Big( -\check{\chi}{}_{\rho\sigma\omega\lambda\mu\nu} + \check{\chi}{}_{\rho\sigma\omega\lambda\nu\mu} + 2\check{\chi}{}_{\rho\sigma\omega\mu\nu\lambda}\\ \nonumber &&\hspace{59pt} + \check{\chi}{}_{\rho\sigma\mu\omega\lambda\nu} + \check{\chi}{}_{\rho\sigma\mu\omega\nu\lambda} - \check{\chi}{}_{\rho\sigma\nu\omega\lambda\mu}\\ \nonumber &&\hspace{59pt} - \check{\chi}{}_{\rho\sigma\nu\omega\mu\lambda} - \check{\chi}{}_{\rho\omega\sigma\lambda\mu\nu} + \check{\chi}{}_{\rho\omega\sigma\lambda\nu\mu}\\ \nonumber &&\hspace{59pt} + 2\check{\chi}{}_{\rho\omega\sigma\mu\nu\lambda} + \check{\chi}{}_{\rho\omega\lambda\sigma\mu\nu} - \check{\chi}{}_{\rho\omega\lambda\sigma\nu\mu}\\ \nonumber &&\hspace{59pt} - 2\check{\chi}{}_{\rho\omega\lambda\mu\nu\sigma} + 2\check{\chi}{}_{\rho\omega\mu\sigma\lambda\nu} + \check{\chi}{}_{\rho\omega\mu\sigma\nu\lambda}\\ \nonumber &&\hspace{59pt} - \check{\chi}{}_{\rho\omega\mu\lambda\nu\sigma} - 2\check{\chi}{}_{\rho\omega\nu\sigma\lambda\mu} - \check{\chi}{}_{\rho\omega\nu\sigma\mu\lambda}\\ \nonumber &&\hspace{59pt} + \check{\chi}{}_{\rho\omega\nu\lambda\mu\sigma} + \check{\chi}{}_{\rho\mu\sigma\omega\lambda\nu} + \check{\chi}{}_{\rho\mu\sigma\omega\nu\lambda}\\ \nonumber &&\hspace{59pt} + \check{\chi}{}_{\rho\mu\omega\sigma\lambda\nu} + 2\check{\chi}{}_{\rho\mu\omega\sigma\nu\lambda} + \check{\chi}{}_{\rho\mu\omega\lambda\nu\sigma}\\ \nonumber &&\hspace{59pt} + \check{\chi}{}_{\rho\mu\nu\sigma\omega\lambda} - \check{\chi}{}_{\rho\mu\nu\omega\lambda\sigma} - \check{\chi}{}_{\rho\nu\sigma\omega\lambda\mu}\\ \nonumber &&\hspace{59pt} - \check{\chi}{}_{\rho\nu\sigma\omega\mu\lambda} - \check{\chi}{}_{\rho\nu\omega\sigma\lambda\mu} - 2\check{\chi}{}_{\rho\nu\omega\sigma\mu\lambda} - \check{\chi}{}_{\rho\nu\omega\lambda\mu\sigma}\\ \nonumber &&\hspace{59pt} - \check{\chi}{}_{\omega\lambda\rho\sigma\mu\nu} + \check{\chi}{}_{\omega\lambda\rho\sigma\nu\mu} - \check{\chi}{}_{\rho\nu\mu\sigma\omega\lambda} + \check{\chi}{}_{\rho\nu\mu\omega\lambda\sigma} \\ \nonumber &&\hspace{59pt} + 2\check{\chi}{}_{\omega\lambda\rho\mu\nu\sigma} + \check{\chi}{}_{\omega\lambda\mu\rho\sigma\nu} + \check{\chi}{}_{\omega\lambda\mu\rho\nu\sigma} - \check{\chi}{}_{\omega\lambda\nu\rho\sigma\mu}\\ \nonumber &&\hspace{59pt} - \check{\chi}{}_{\omega\lambda\nu\rho\mu\sigma} - \check{\chi}{}_{\omega\mu\rho\sigma\lambda\nu} + \check{\chi}{}_{\omega\mu\rho\sigma\nu\lambda} + 2\check{\chi}{}_{\omega\mu\rho\lambda\nu\sigma}\\ \nonumber &&\hspace{59pt} - \check{\chi}{}_{\omega\mu\nu\rho\sigma\lambda} - \check{\chi}{}_{\omega\mu\nu\rho\lambda\sigma}  + \check{\chi}{}_{\omega\mu\lambda\rho\sigma\nu} + \check{\chi}{}_{\omega\mu\lambda\rho\nu\sigma} \\ \nonumber &&\hspace{59pt} + \check{\chi}{}_{\omega\nu\rho\sigma\lambda\mu} - \check{\chi}{}_{\omega\nu\rho\sigma\mu\lambda} - 2\check{\chi}{}_{\omega\nu\rho\lambda\mu\sigma} - \check{\chi}{}_{\omega\nu\lambda\rho\sigma\mu}\\ \nonumber &&\hspace{59pt} - \check{\chi}{}_{\omega\nu\lambda\rho\mu\sigma} + \check{\chi}{}_{\omega\nu\mu\rho\sigma\lambda} + \check{\chi}{}_{\omega\nu\mu\rho\lambda\sigma}\\ \nonumber &&\hspace{59pt} - 2\check{\chi}{}_{\mu\nu\rho\sigma\omega\lambda} + 2\check{\chi}{}_{\mu\nu\rho\omega\lambda\sigma} + 2\check{\chi}{}_{\mu\nu\omega\rho\sigma\lambda} + 2\check{\chi}{}_{\mu\nu\omega\rho\lambda\sigma} \Big) ,
\end{eqnarray}
\begin{eqnarray}
  \label{eq:cc-decomposition-4}
 {}^{(4)}\check{\chi}{}_{\alpha\beta\mu\gamma\delta\nu} &=& \frac{4}{45} C{}^{\rho\sigma\omega\lambda}_{\alpha\beta\gamma\delta} \Big( -\check{\chi}{}_{\rho\sigma\omega\lambda\mu\nu} + \check{\chi}{}_{\rho\sigma\omega\lambda\nu\mu} + 4\check{\chi}{}_{\rho\sigma\mu\omega\lambda\nu} \\ \nonumber &&\hspace{59pt} + \check{\chi}{}_{\rho\sigma\mu\omega\nu\lambda} - \check{\chi}{}_{\rho\sigma\nu\omega\mu\lambda} + \check{\chi}{}_{\rho\omega\sigma\lambda\mu\nu}\\ \nonumber &&\hspace{59pt} + 2\check{\chi}{}_{\rho\omega\sigma\lambda\nu\mu} - \check{\chi}{}_{\rho\omega\sigma\mu\nu\lambda} + \check{\chi}{}_{\rho\omega\lambda\sigma\mu\nu}\\ \nonumber &&\hspace{59pt} + \check{\chi}{}_{\rho\omega\lambda\sigma\nu\mu} + \check{\chi}{}_{\rho\omega\lambda\mu\nu\sigma} + 4\check{\chi}{}_{\rho\omega\mu\sigma\lambda\nu}\\ \nonumber &&\hspace{59pt} + \check{\chi}{}_{\rho\omega\mu\sigma\nu\lambda} + \check{\chi}{}_{\rho\omega\mu\lambda\nu\sigma} + \check{\chi}{}_{\rho\omega\nu\sigma\mu\lambda}\\ \nonumber &&\hspace{59pt} + 2\check{\chi}{}_{\rho\omega\nu\lambda\mu\sigma} + \check{\chi}{}_{\rho\mu\sigma\omega\lambda\nu} - 2\check{\chi}{}_{\rho\mu\sigma\omega\nu\lambda}\\ \nonumber &&\hspace{59pt} - \check{\chi}{}_{\rho\mu\omega\sigma\lambda\nu} - \check{\chi}{}_{\rho\mu\omega\sigma\nu\lambda} + 2\check{\chi}{}_{\rho\mu\omega\lambda\nu\sigma}\\ \nonumber &&\hspace{59pt} + 2\check{\chi}{}_{\rho\mu\nu\sigma\omega\lambda} - \check{\chi}{}_{\rho\mu\nu\omega\lambda\sigma} - \check{\chi}{}_{\rho\nu\sigma\omega\lambda\mu}\\ \nonumber &&\hspace{59pt} - 2\check{\chi}{}_{\rho\nu\sigma\omega\mu\lambda} - 2\check{\chi}{}_{\rho\nu\omega\sigma\lambda\mu} + \check{\chi}{}_{\rho\nu\omega\sigma\mu\lambda}\\ \nonumber &&\hspace{59pt} + 2\check{\chi}{}_{\rho\nu\omega\lambda\mu\sigma} + \check{\chi}{}_{\rho\nu\mu\sigma\omega\lambda} + \check{\chi}{}_{\rho\nu\mu\omega\lambda\sigma}\\ \nonumber &&\hspace{59pt} + \check{\chi}{}_{\omega\lambda\rho\sigma\mu\nu} - 2\check{\chi}{}_{\omega\lambda\rho\mu\nu\sigma} - \check{\chi}{}_{\omega\lambda\mu\rho\nu\sigma} - \check{\chi}{}_{\omega\mu\rho\sigma\lambda\nu}\\ \nonumber &&\hspace{59pt} - \check{\chi}{}_{\omega\mu\rho\lambda\nu\sigma} - \check{\chi}{}_{\omega\mu\lambda\rho\sigma\nu} - 2\check{\chi}{}_{\omega\mu\lambda\rho\nu\sigma} + \check{\chi}{}_{\omega\mu\nu\rho\lambda\sigma}\\ \nonumber &&\hspace{59pt} - \check{\chi}{}_{\omega\nu\rho\sigma\lambda\mu} + 2\check{\chi}{}_{\omega\nu\rho\sigma\mu\lambda} + \check{\chi}{}_{\omega\nu\rho\lambda\mu\sigma} - \check{\chi}{}_{\omega\nu\mu\rho\sigma\lambda}\\ \nonumber &&\hspace{59pt} + \check{\chi}{}_{\omega\nu\mu\rho\lambda\sigma} + \check{\chi}{}_{\mu\nu\rho\sigma\omega\lambda} - 2\check{\chi}{}_{\mu\nu\omega\rho\sigma\lambda} - \check{\chi}{}_{\mu\nu\omega\rho\lambda\sigma} \Big) ,
\end{eqnarray}
\begin{eqnarray}
\label{eq:cc-decomposition-5}
      {}^{(5)}\check{\chi}{}_{\alpha\beta\mu\gamma\delta\nu} &=& \frac{1}{36} C{}^{\rho\sigma\omega\lambda}_{\alpha\beta\gamma\delta} \Big( \check{\chi}{}_{\rho\sigma\omega\lambda\mu\nu} + \check{\chi}{}_{\rho\sigma\omega\lambda\nu\mu} - 2\check{\chi}{}_{\rho\sigma\omega\mu\nu\lambda}\\ \nonumber &&\hspace{59pt} + 2\check{\chi}{}_{\rho\sigma\mu\omega\lambda\nu} - \check{\chi}{}_{\rho\sigma\mu\omega\nu\lambda} - \check{\chi}{}_{\rho\sigma\nu\omega\mu\lambda}\\ \nonumber &&\hspace{59pt} + 3\check{\chi}{}_{\rho\omega\sigma\lambda\mu\nu} - \check{\chi}{}_{\rho\omega\sigma\lambda\nu\mu} + \check{\chi}{}_{\rho\omega\lambda\sigma\mu\nu}\\ \nonumber &&\hspace{59pt} - 3\check{\chi}{}_{\rho\omega\lambda\sigma\nu\mu} + 2\check{\chi}{}_{\rho\omega\mu\sigma\lambda\nu} + \check{\chi}{}_{\rho\omega\mu\sigma\nu\lambda}\\ \nonumber &&\hspace{59pt} + 3\check{\chi}{}_{\rho\omega\mu\lambda\nu\sigma} + 2\check{\chi}{}_{\rho\omega\nu\sigma\lambda\mu} - 3\check{\chi}{}_{\rho\omega\nu\sigma\mu\lambda}\\ \nonumber &&\hspace{59pt} - \check{\chi}{}_{\rho\omega\nu\lambda\mu\sigma} - \check{\chi}{}_{\rho\mu\sigma\omega\lambda\nu} + 5\check{\chi}{}_{\rho\mu\sigma\omega\nu\lambda}\\ \nonumber &&\hspace{59pt} - 3\check{\chi}{}_{\rho\mu\omega\sigma\lambda\nu} - \check{\chi}{}_{\rho\mu\omega\lambda\nu\sigma} - \check{\chi}{}_{\rho\mu\nu\sigma\omega\lambda}\\ \nonumber &&\hspace{59pt} - \check{\chi}{}_{\rho\mu\nu\omega\lambda\sigma} - \check{\chi}{}_{\rho\nu\sigma\omega\lambda\mu} + \check{\chi}{}_{\rho\nu\sigma\omega\mu\lambda}\\ \nonumber &&\hspace{59pt} + \check{\chi}{}_{\rho\nu\omega\sigma\lambda\mu} - 5\check{\chi}{}_{\rho\nu\omega\lambda\mu\sigma} + 3\check{\chi}{}_{\rho\nu\mu\sigma\omega\lambda}\\ \nonumber &&\hspace{59pt} - \check{\chi}{}_{\rho\nu\mu\omega\lambda\sigma} + \check{\chi}{}_{\omega\lambda\rho\sigma\mu\nu} + \check{\chi}{}_{\omega\lambda\rho\sigma\nu\mu}\\ \nonumber &&\hspace{59pt} + 2\check{\chi}{}_{\omega\lambda\rho\mu\nu\sigma} - \check{\chi}{}_{\omega\lambda\mu\rho\nu\sigma} + 2\check{\chi}{}_{\omega\lambda\nu\rho\sigma\mu}\\ \nonumber &&\hspace{59pt} - \check{\chi}{}_{\omega\lambda\nu\rho\mu\sigma} - \check{\chi}{}_{\omega\mu\rho\sigma\lambda\nu} - 5\check{\chi}{}_{\omega\mu\rho\sigma\nu\lambda}\\ \nonumber &&\hspace{59pt} - \check{\chi}{}_{\omega\mu\lambda\rho\sigma\nu} + \check{\chi}{}_{\omega\mu\lambda\rho\nu\sigma} - \check{\chi}{}_{\omega\mu\nu\rho\sigma\lambda}\\ \nonumber &&\hspace{59pt} - 3\check{\chi}{}_{\omega\mu\nu\rho\lambda\sigma} + 3\check{\chi}{}_{\omega\nu\rho\sigma\lambda\mu} - \check{\chi}{}_{\omega\nu\rho\sigma\mu\lambda}\\ \nonumber &&\hspace{59pt} - \check{\chi}{}_{\omega\nu\lambda\rho\sigma\mu} + 5\check{\chi}{}_{\omega\nu\lambda\rho\mu\sigma} - \check{\chi}{}_{\omega\nu\mu\rho\sigma\lambda}\\ \nonumber &&\hspace{59pt} + \check{\chi}{}_{\omega\nu\mu\rho\lambda\sigma} - 2\check{\chi}{}_{\mu\nu\rho\omega\lambda\sigma} + 2\check{\chi}{}_{\mu\nu\omega\rho\sigma\lambda} \Big) ,
\end{eqnarray}
\begin{eqnarray}
\label{eq:cc-decomposition-6}
{}^{(6)}\check{\chi}{}_{\alpha\beta\mu\gamma\delta\nu} &=& \frac{1}{18}
C{}^{\rho\sigma\omega\lambda}_{\alpha\beta\gamma\delta}\Big(
-2\check{\chi}{}_{\rho\sigma\omega\mu\nu\lambda}
+3\check{\chi}{}_{\rho\sigma\mu\omega\lambda\nu}
+\check{\chi}{}_{\rho\sigma\nu\omega\lambda\mu}
\\ \nonumber &&\hspace{59pt} -\check{\chi}{}_{\rho\omega\sigma\lambda\mu\nu}
+\check{\chi}{}_{\rho\omega\sigma\lambda\nu\mu}
-3\check{\chi}{}_{\rho\omega\lambda\sigma\mu\nu}\\
  \nonumber &&\hspace{59pt} -
              \check{\chi}{}_{\rho\omega\lambda\sigma\nu\mu} -
              4\check{\chi}{}_{\rho\omega\mu\sigma\lambda\nu} -
              3\check{\chi}{}_{\rho\omega\mu\sigma\nu\lambda}\\ \nonumber &&\hspace{59pt} -
              \check{\chi}{}_{\rho\omega\mu\lambda\nu\sigma} -
              4\check{\chi}{}_{\rho\omega\nu\sigma\lambda\mu} -
              \check{\chi}{}_{\rho\omega\nu\sigma\mu\lambda}\\
  \nonumber &&\hspace{59pt} +
              \check{\chi}{}_{\rho\omega\nu\lambda\mu\sigma} -
              3\check{\chi}{}_{\rho\mu\sigma\omega\nu\lambda} +
              \check{\chi}{}_{\rho\mu\omega\sigma\lambda\nu}\\ \nonumber &&\hspace{59pt} +
              \check{\chi}{}_{\rho\mu\omega\lambda\nu\sigma} +
              \check{\chi}{}_{\rho\mu\nu\sigma\omega\lambda} -
              \check{\chi}{}_{\rho\nu\sigma\omega\mu\lambda}\\
  \nonumber &&\hspace{59pt} -
              \check{\chi}{}_{\rho\nu\omega\sigma\lambda\mu} +
              3\check{\chi}{}_{\rho\nu\omega\lambda\mu\sigma} -
              \check{\chi}{}_{\rho\nu\mu\sigma\omega\lambda}\\ \nonumber &&\hspace{59pt} -
              \check{\chi}{}_{\omega\lambda\rho\sigma\mu\nu} -
              \check{\chi}{}_{\omega\lambda\rho\sigma\nu\mu} +
              2\check{\chi}{}_{\omega\lambda\rho\mu\nu\sigma}\\ \nonumber &&\hspace{59pt} -
              \check{\chi}{}_{\omega\lambda\mu\rho\sigma\nu} +
              \check{\chi}{}_{\omega\lambda\mu\rho\nu\sigma} +
              \check{\chi}{}_{\omega\lambda\nu\rho\sigma\mu}\\ \nonumber &&\hspace{59pt} +
              \check{\chi}{}_{\omega\lambda\nu\rho\mu\sigma} +
              3\check{\chi}{}_{\omega\mu\rho\sigma\lambda\nu} -
              \check{\chi}{}_{\omega\mu\rho\sigma\nu\lambda}\\
  \nonumber &&\hspace{59pt} +
              \check{\chi}{}_{\omega\mu\lambda\rho\sigma\nu} +
              3\check{\chi}{}_{\omega\mu\lambda\rho\nu\sigma} +
              \check{\chi}{}_{\omega\mu\nu\rho\sigma\lambda}\\ \nonumber &&\hspace{59pt} -
              \check{\chi}{}_{\omega\mu\nu\rho\lambda\sigma} +
              \check{\chi}{}_{\omega\nu\rho\sigma\lambda\mu} -
              3\check{\chi}{}_{\omega\nu\rho\sigma\mu\lambda}\\
  \nonumber &&\hspace{59pt} +
              \check{\chi}{}_{\omega\nu\lambda\rho\sigma\mu} +
              \check{\chi}{}_{\omega\nu\lambda\rho\mu\sigma} +
              \check{\chi}{}_{\omega\nu\mu\rho\sigma\lambda}\\ \nonumber &&\hspace{59pt} -
              3\check{\chi}{}_{\omega\nu\mu\rho\lambda\sigma} -
              2\check{\chi}{}_{\mu\nu\rho\omega\lambda\sigma} +
              2\check{\chi}{}_{\mu\nu\omega\rho\sigma\lambda} \Big) ,
\end{eqnarray}
\begin{eqnarray}
  \label{eq:cc-decomposition-7}
  {}^{(7)}\check{\chi}{}_{\alpha\beta\mu\gamma\delta\nu} &=& \frac{1}{20} C{}^{\rho\sigma\omega\lambda}_{\alpha\beta\gamma\delta} \Big( 2\check{\chi}{}_{\rho\sigma\omega\lambda\mu\nu} + 2\check{\chi}{}_{\rho\sigma\omega\mu\nu\lambda} + 4\check{\chi}{}_{\rho\sigma\mu\omega\lambda\nu}\\ \nonumber &&\hspace{59pt} - 2\check{\chi}{}_{\rho\sigma\mu\omega\nu\lambda} - 2\check{\chi}{}_{\rho\sigma\nu\omega\lambda\mu} + 3\check{\chi}{}_{\rho\omega\sigma\lambda\mu\nu}\\ \nonumber &&\hspace{59pt} - 3\check{\chi}{}_{\rho\omega\sigma\lambda\nu\mu} - \check{\chi}{}_{\rho\omega\lambda\sigma\mu\nu} + \check{\chi}{}_{\rho\omega\lambda\sigma\nu\mu}\\ \nonumber &&\hspace{59pt} - 2\check{\chi}{}_{\rho\omega\mu\sigma\lambda\nu} - \check{\chi}{}_{\rho\omega\mu\sigma\nu\lambda} + 3\check{\chi}{}_{\rho\omega\mu\lambda\nu\sigma}\\ \nonumber &&\hspace{59pt} + 2\check{\chi}{}_{\rho\omega\nu\sigma\lambda\mu} + \check{\chi}{}_{\rho\omega\nu\sigma\mu\lambda} - 3\check{\chi}{}_{\rho\omega\nu\lambda\mu\sigma}\\ \nonumber &&\hspace{59pt} - 2\check{\chi}{}_{\rho\mu\sigma\omega\lambda\nu} + \check{\chi}{}_{\rho\mu\sigma\omega\nu\lambda} - 3\check{\chi}{}_{\rho\mu\omega\sigma\lambda\nu}\\ \nonumber &&\hspace{59pt} - 3\check{\chi}{}_{\rho\mu\omega\lambda\nu\sigma} - 3\check{\chi}{}_{\rho\mu\nu\sigma\omega\lambda} + 3\check{\chi}{}_{\rho\nu\sigma\omega\mu\lambda}\\ \nonumber &&\hspace{59pt} + 3\check{\chi}{}_{\rho\nu\omega\sigma\lambda\mu} - \check{\chi}{}_{\rho\nu\omega\lambda\mu\sigma} + 3\check{\chi}{}_{\rho\nu\mu\sigma\omega\lambda}\\ \nonumber &&\hspace{59pt} - 2\check{\chi}{}_{\rho\nu\mu\omega\lambda\sigma} - \check{\chi}{}_{\omega\lambda\rho\sigma\mu\nu} - \check{\chi}{}_{\omega\lambda\rho\sigma\nu\mu}\\ \nonumber &&\hspace{59pt} + \check{\chi}{}_{\omega\lambda\mu\rho\nu\sigma} - 2\check{\chi}{}_{\omega\lambda\nu\rho\sigma\mu} + \check{\chi}{}_{\omega\lambda\nu\rho\mu\sigma}\\ \nonumber &&\hspace{59pt} + \check{\chi}{}_{\omega\mu\rho\sigma\lambda\nu} + 5\check{\chi}{}_{\omega\mu\rho\sigma\nu\lambda} + \check{\chi}{}_{\omega\mu\lambda\rho\sigma\nu}\\ \nonumber &&\hspace{59pt} + \check{\chi}{}_{\omega\mu\lambda\rho\nu\sigma} + \check{\chi}{}_{\omega\mu\nu\rho\sigma\lambda} + \check{\chi}{}_{\omega\mu\nu\rho\lambda\sigma}\\ \nonumber &&\hspace{59pt} - \check{\chi}{}_{\omega\nu\rho\sigma\lambda\mu} - \check{\chi}{}_{\omega\nu\rho\sigma\mu\lambda} + \check{\chi}{}_{\omega\nu\lambda\rho\sigma\mu}\\ \nonumber &&\hspace{59pt} - 5\check{\chi}{}_{\omega\nu\lambda\rho\mu\sigma} + \check{\chi}{}_{\omega\nu\mu\rho\sigma\lambda} - \check{\chi}{}_{\omega\nu\mu\rho\lambda\sigma}\\ \nonumber &&\hspace{59pt} + 2\check{\chi}{}_{\mu\nu\rho\omega\lambda\sigma} \Big) .
\end{eqnarray}
It has been verified by computer algebra that the above expressions
indeed belong to orthogonal subspaces.
 


\end{document}